%% file: jpsi.tex
\documentclass[reprint,twocolumn,nofootinbib]{revtex4}

\usepackage[utf8]{inputenc}
\usepackage{graphicx}
\usepackage{amsmath}
\usepackage{url}
\usepackage{hyperref}
\usepackage{color}
\usepackage{tikz}  
\usepackage{lineno}  
\usepackage{draftcopy}  

\newcommand{\J}{$J/\psi$ }
\newcommand{\Js}{{$J/\psi$s}}
\newcommand{\Jm}{{J/\psi} }
\newcommand{\cc}{$c$ and $\bar c$ }
\newcommand{\Q}{$Q \bar Q$ }

\newcommand{\pv}{\vec p}
\newcommand{\Pv}{\vec P}
\newcommand{\n}{\nonumber \\}
\newcommand{\JA}[1]{{\color{black}#1}}

\def\be{\begin{equation}}
	\def\ee{\end{equation}}
\def\bea{\begin{eqnarray}}
	\def\eea{\end{eqnarray}}
\def\bfr{{\mathbf r}}
\def\bfp{{\mathbf p}}

\begin{document}
\title{A new Microscopic Model for $J/\psi$ Production in Heavy Ion Collisions}
\author{Denys Yen Arrebato Villar, Jiaxing Zhao, Joerg Aichelin, Pol Bernard Gossiaux}
\affiliation{SUBATECH, Nantes University, IMT Atlantique, IN2P3/CNRS, \\
4 rue Alfred Kastler, 44307 Nantes cedex 3, France}

\date{\today}

\begin{abstract} 
\noindent
We present a new model for the creation of \J mesons in ultrarelativistic heavy ion collisions, which allows to follow the individual heavy quarks from their creation until the detector through the Quark Gluon Plasma (QGP), which is formed in these collisions and described by the EPOS2 event generator. The \cc quarks interact via a potential, based on results of lattice gauge calculations. The annihilation and creation of \J is described by a density matrix approach whose time evolution is studied in the expanding system. The comparison with PbPb data at $\sqrt{s}$=5.02 TeV shows that this model can describe simultaneously the nuclear modification factor $R_{AA}$ and the elliptic flow $v_2$ of the \J at low transverse momentum. Perspectives for further improvement are discussed.
\end{abstract}


\maketitle

\section{Introduction}
There is overwhelming evidence that in ultrarelativistic heavy ion collisions a plasma of quarks and gluons (QGP) is created, which evolves in time and disintegrates at the end of its lifetime into  hadrons. The multiplicity of light and strange hadrons is well described by statistical model calculations \cite{Andronic:2017pug}. The consequence of this observation is that light and strange hadrons cannot provide direct information about the time evolution of the QGP from its creation to hadronization. To study this time evolution and to get insight into the early phase of the heavy ion collision one has to focus on probes which do not come to equilibrium with the expanding QGP. They include electromagnetic probes, jets as well as hadrons, which contain heavy quarks.

Among these probes, especially the hidden heavy flavour meson $J/\psi$, composed of a $c$ and a $\bar c$ quark has recently gained a lot of interest. This is due to two experimental results, which came as a surprise. 
\begin{itemize}
    \item 
The nuclear modification factor $R_{AA}=\frac{d\sigma_{AA}/dp_T}{N_{coll}d\sigma_{pp}/dp_T}$, where $N_{coll}$ is the number of initial binary collisions in the AA system, stays almost constant as a function of the centrality in heavy ion collisions at LHC \cite{ALICE:2016flj} whereas it decreases strongly at RHIC energies \cite{PHENIX:2011img}.
\item 
In some specific approaches, like for example in the color glass condensate approach, \J  as well as charmed mesons are produced in correlation to light flavors mesons. This could explain the elliptic flow of $J/\psi$ observed in pp and pA collisions. However, such correlations are local in space and do not add coherently in the case of AA collisions, while 
the $J/\psi$'s observed in experiment show a strong elliptic flow, which follows the systematics of the $v_2$ observed for light hadrons \cite{ALICE:2020pvw}. This can only be explained if one assumes that $v_2$ is transferred to the individual charm quarks. The observation of a $v_2$ of \J questions the idea that it traverses the QGP as a color-neutral, weakly interacting object.
\end{itemize}

In this paper we study how these observations can be understood and what we can learn from the $J/\psi$ about the properties of the QGP, created in heavy ion collisions. 

The idea to use \Js~ for such studies goes back to the seminal paper of Matsui and Satz \cite{Matsui:1986dk} who argued that in strongly interacting thermal matter the color charges of the $c$ and $\bar c$ are screened by color charges of the medium to the extent that the $J/\psi$ ceases to exist as a bound state if the density of these charges becomes high enough.  Later this melting has been confirmed by lattice gauge calculations \cite{Mocsy:2007jz,Digal:2001ue} but the exact dissociation temperature, $T_{\rm diss}$, is still subject of debate. 

The Wilson loop allows to determine the free energy between the c and $\bar c$ as a function of their distance. The lattice gauge results for the Wilson loop as a function of the temperature allowed to develop a static $c\bar c$ potential which can be employed in a Schrödinger equation and allows for studying how the ground state energy of the $c \bar c$ pair develops as a function of the temperature of the QGP\cite{Bala:2021fkm}. These calculations confirmed the conclusions of 
\cite{Matsui:1986dk} that there is a limiting temperature above which the {\J} becomes unstable. For a recent review we refer to Ref.\cite{Rothkopf:2019ipj}.

In ultrarelativistic heavy ion collisions the situation is more complex than in a static medium. Shortly after the initial binary collisions of the nucleons of projectile and target a high temperature QGP is formed in which a \J  cannot survive. It can only be produced when the temperature of the expanding system gets lower than $T_{\rm diss}$. Therefore, the c and $\bar c$ of the final \J and those $c$ and $\bar c$ quarks, which are finally part of open heavy flavour hadrons, traverse initially the same QPG. Hence the knowledge, which one has acquired in the last years about open heavy flavour mesons, is also of use also for the study of \J.

Open heavy flavour hadrons, produced in heavy ion collisions,  have been extensively studied in the last years, experimentally and theoretically. Recently the theoretical models, which differ in details, have been compared \cite{Xu:2018gux,Cao:2018ews, Rapp:2018qla}.
This comparison suggests that the initial c and $\bar c$ quarks are created in elementary baryon-baryon collisions at the beginning of the heavy ion reaction and that their initial transverse momentum distribution is well described by FONLL (first order next to leading log) calculations \cite{Cacciari:2008gp,Cacciari:2011ma}. The heavy quarks interact subsequently with the QGP constituents, the light quarks and gluons, in elementary collisions, which are described by pQCD (perturbative QCD) Born diagrams, and pick up by these collisions a finite elliptic flow. Finally, they convert into heavy hadrons when the QGP hadronizes. The last process is usually described by a combination of coalescence and fragmentation. A modification of the \J distribution by hadronic rescattering is also possible, see ref. \cite{Linnyk:2008hp,Song:2015sfa}, but beyond the scope of the present article.

When the local temperature of the QGP gets lower than $T_{\rm diss}$ of the \J, it can be formed but also destroyed by an elastic or an inelastic collision of one of its constituents with a QGP parton. 
The difference between formation and collisional decay determines the \J spectrum at the end of the QGP phase. 

Recently it has been shown that in central collisions at LHC energies also in the statistical hadronization model the relative abundance of charmed hadrons but not the total multiplicity of charmed hadrons can be understood assuming that all the charmed hadrons are formed at chemical freeze out, when also the light hadrons are produced \cite{Andronic:2021erx}.

Transport models have also been advanced to study the dynamical production of \J \cite{Du:2015wha,Zhou:2014kka}. The model of Du et al.\cite{Du:2015wha} describes the \J production in central PbPb collisions at $\sqrt{s} = 2.76$ by a kinetic rate equation applied in an expanding fireball. The rate for \J production in the fireball is based on many body quantum mechanics using as main ingredient a potential $V$, which is calibrated to lattice results such as the free energy and the quarkonium correlators \cite{PhysRevC.97.034918}. It is only dependent on the local temperature of the system. 
While the main part of the quarkonium production happens in the fireball, it is also supplemented by a significant regeneration contribution in the expanding hadron gas after the fireball has been disintegrated into hadrons. Being in quasi-equilibrium\footnote{A reduction of the equilibrium limit is account for through a thermal relaxation factor.} with the expanding fireball, the c and $\bar c$  acquire a finite elliptical flow when the geometrical anisotropy in coordinate space is converted into an anisotropy in momentum space. The absolute value of the elliptic flow is underestimated. It can, however, be increased \cite{He:2021zej} by introducing, in the $c\bar{c}\to J/\psi$ hadronization process, off-equilibrium $c$  and $\bar{c}$ distributions from the Langevin dynamics as well as some space-momentum correlations. 

Zhou et al.\cite{Zhou:2014kka} have advanced a dynamical semi-classical model for dissociation and regeneration of \J when the c and $\bar c$ pass through the QGP, which is modeled by hydrodynamics. Dissociation and regeneration are calculated via the $\sigma_{gJ/\psi}$ cross section,  assuming that the charm quarks are in equilibrium in the QPG.  

There are also two more recent and interesting approaches, which have not yet yield quantitative predictions. The one is the treatment of the \J production under the aspect of an open quantum system \cite{Akamatsu:2021dot,Delorme:2022hoo} whose time evolution is given by the Lindblad equation. The other is the description of the \J by the time evolution of a reduced density matrix \cite{Blaizot:2018oev}. Both treat the
\cc pairs as quantum systems, a challenging as well as complex task.

In this paper we advance a microscopic model for the \J production which follows the c and $\bar c$ s from the initial creation until hadronization. By this we avoid one-body transport approaches like Boltzmann or Fokker-Planck equations. These equations are not appropriate to study two-body correlations, which are at the origin of the \J formation. While travelling through the QGP, the heavy quarks have energy and momentum conserving collisions with the constituents of the QGP and interact among themselves by a potential derived from lattice QCD. The Lagrangian, which we employ for the potential interaction, includes relativistic corrections in the center of mass system up to the order $\gamma -1$ where $\gamma = 1/\sqrt{1-\beta^2}$. Below $T_{\rm diss}$ the \Js~ are described by a Wigner density in relative coordinates with a root mean square (rms) radius, which depends on the temperature of the QGP, while above $T_{\rm diss}$ a \J cannot be produced.
The rate of production and dissociation is obtained by solving the von Neumann equation for the two-body $c\bar c$ system in the expanding medium, following a formalism which has been developed by Remler et al. \cite{Remler:1981du,Gyulassy:1982pe,Aichelin:1987rh} for the production of deuterons in heavy ion collisions. It has also been  employed in the study of quarkonia production in pp collisions within the PHSD approach  \cite{Song:2017phm}. 

The paper is organized as follows: In section II we present our model. We introduce the density matrix formalism introduced by Remler and study the rate of \J production for time independent $Q \bar{Q}$ Wigner densities, where Q stands for a heavy quark. This is followed by a description of the interaction of heavy quarks with the QGP partons. Finally we discuss the non relativistic $Q \bar{Q}$ Wigner density and its relativistic extension. In section III we extend our formalism 
to the case that the $Q\bar Q$ Wigner density gets time dependent.
Section IV is devoted to the potential interaction between $Q$ and $\bar{Q}$.
In section V we report about the initial distribution of the heavy (anti)quarks. In section VI we present numerical details of our approach and study the consequences of the different ingredients on the observables. In section VII we compare our results with experimental data before we draw our conclusions in section VIII.  
In this initial study we limit ourself to the charmonium ground state, knowing that feeding from B decay and excited charmonia gets
important at LHC energies. Here it is the primary goal to understand the global trends associated to such a microscopic approach. This limits also our possibility to compare our results with experimental data. Feed down, a more careful treatment of the color structure, a possible color screening of the cross section of a \J in the QGP, hadronic \J interactions and including the directly produced \J (means those which do not pass the QGP) will be subject for a later publication).

\section{The Model}
We start out with an outline of the approach of Remler, which we employ, adapted to the problem of heavy quarks: In the initial collisions between projectile and target  nucleons heavy (anti)quarks $Q(\bar Q)$ are created, which we assume to be uncorrelated in momentum space. Their individual transverse momenta reproduce the distribution of FONLL calculations \cite{Cacciari:2008gp,Cacciari:2011ma}. These heavy quarks then enter the QGP, which is created after a thermalization time of $t_0=0.35$ fm/c, and modelled by EPOS2 \cite{Drescher:2000ha,Werner:2010aa} or vHLLE hydrodynamics \cite{Karpenko:2013wva}.  While traversing the QGP the heavy quarks interact with the plasma constituents according to MC@sHQ \cite{Gossiaux:2008jv,Gossiaux:2009mk}. At the same time \Q pairs interact among themselves via a chromoelectical potential, a new feature, which is based on lattice results. It yields correlated $Q\bar{Q}$ trajectories. When the QGP has cooled down locally to $T_{\rm diss}^\Phi$, the dissociation temperature of a heavy \Q meson of type $\Phi$, these mesons can be created but also destroyed. We employ the Remler formalism to describe their creation and annihilation rates. These processes cease when the heavy quarks hadronize to open heavy flavour hadrons. The Remler formalism predicts the final momentum distribution of the quarkonia. 

 \subsection{The Remler density matrix formalism}
 The Remler formalism assumes that all information about a N-particle system is encoded in the N-body density operator, $\rho_{N}(t)$, of the system. Among the N particles there may be one or several $c\bar c$ pairs. Because the relative motion of heavy quarks in bound heavy quark systems is small compared to the heavy quark mass, we use here non relativistic kinematics and discuss the extension towards a relativistic treatment later. 
 
 The density operator obeys the von Neumann equation \cite{Gyulassy:1982pe}:
\begin{equation}
	\partial\rho_N/\partial t=-\frac{i}{\hbar}[H,\rho_N] \label{eq:vN} 
\end{equation}
where $H$ is the Hamiltonian of the full system 
\begin{equation}
	H=\Sigma_{i} K_{i} +\Sigma_{i>j}V_{ij}.
\end{equation}
$K_{i}$ is the kinetic energy operator of the particle $i$ and $V_{ij}$ is the interaction between the particles $i$ and $j$. Quarkonia, like a $J/\psi$, are two-body objects described by the  two-body density operator  $\rho^{\Phi} =|\Phi><\Phi|$. $\Phi$ is the wave function of the eigen state $\Phi$ of the two-body \Q system.
Thus 
\begin{equation}
	P^{\Phi}(t)={\rm Tr}[\rho^{\Phi}\rho_{N}(t)]
	\label{eq:rho2},
\end{equation}
where the trace is taken over all N-body coordinates (which include the $Q$ and $\bar{Q}$ degrees of freedom), measures the probability of finding the $Q$ and the $\bar Q$ at time t in the eigen state $|\Phi>$. In the case where several $Q\bar{Q}$ pairs are present in the system, this definition extends to the average number of $\Phi$-states which can be measured at the time of the projection, including possible interferences and taking into account the rare cases where several $\Phi$-states could be measured simultaneously.
We are in particular interested in the value of 
$P^\Phi(t\to +\infty)$, as it corresponds to experimental measurements. From the viewpoint of heavy quarks, our standard EPOS2+MC@sHQ is quite similar to an intranuclear cascade model to which the Remler algorithm was originally applied: In heavy ion reactions the QGP expands until hadronization. Propagating Q and $\bar Q$ as classical particles without potential, the distance between the $Q$ and $\bar Q$ quarks increases and at the end of the QGP expansion it is large with respect to the radius of the eigen state $\Phi$. Therefore $P^{\Phi}(t\to\infty)$ tends to zero. To circumvent this issue of semiclassical transport approaches, we resort to the method of Remler's original work: We express the probability to observe $Q\bar Q$ pairs in the eigen state $\Phi$ at a time $\tilde{t}$ as the integral of the rate of decay and formation of pairs in the eigen state $\Phi$, $\Gamma^\Phi(t)$: 

\bea
P^\Phi(\tilde t)&=& P^\Phi(0)+\int_0^{\tilde t}\Gamma^\Phi(t) dt
\label{eq:rho2bis}
\eea
with the rate $\Gamma^{\Phi}$ defined as
\begin{equation}
\Gamma^{\Phi}(t) =
\frac{dP^{\Phi}}{dt}=
\frac{d}{dt}{\rm Tr}[\rho^{\Phi}\rho_{N}(t)].
\label{eq:rateint}
\end{equation}
In our numerical scheme, we have introduced an attractive potential acting between $Q$ quarks and $\bar{Q}$ antiquarks, see section \ref{sectionIV}. Consequently, for some pairs the relative distance between the $Q$ and $\bar{Q}$ remains finite when $t\to \infty$. However, in our semi-classical modelling, which is best suited when many momentum exchanges occur but less reliable to describe the long time dynamics of this quantity, the formulation based on the rate, eq.~\ref{eq:rho2bis}, is more accurate than the direct projection eq.~\ref{eq:rho2}. In a full quantum evolution of $\rho_N$ both methods would give identical results.

Proceeding with the time derivative inside the ${\rm Tr[\cdots]}$, assuming that $\rho^{\Phi}$ is time independent and using the von Neumann equation (\ref{eq:vN}), one gets 
\bea
P^{\Phi}(\tilde t)&=& P^\Phi(0)+
  \int_0^{\tilde t} Tr[\rho^{\Phi},\frac{\partial \rho_N}{\partial t}]dt\nonumber \\
  &=& P^\Phi(0) -\frac{i}{\hbar}\int_0^{\tilde t} Tr[\rho^{\Phi},[H,\rho_N]] dt. 
 \label{eq:defGamma}
\eea  
We first focus on the case that among the N particles we find only a single $Q\bar{Q}$ pair. We assign to this \Q pair the indices 1 and 2 and  decompose the total Hamiltonian as
\begin{equation}
	H=H_{1,2}+H_{N-2}+U_{1,2}
	\label{fullhamiltonianremler}
\end{equation}
where 
\begin{equation}
H_{1,2}=K_{1}+K_{2}+V_{12}
\label{eq:2bodyhamilt}
\end{equation}
is the two particle Hamiltonian of the \Q pair, $H_{N-2}=\Sigma_{i}K_{i}+\Sigma_{j>i\geq 3}V_{ji}$ is the Hamiltonian of the remaining N-2 body system and $U_{1,2}$ 
is the interaction of the heavy quarks 1 and 2 with the rest of the system
\begin{equation}
	U_{1,2}=\Sigma_{j}V_{1j}+\Sigma_{j}V_{2j}.
\end{equation}
We replace in eq. \ref{eq:defGamma} the full Hamiltonian of the system by this decomposition and profit from the relations
\be
[\rho^{\Phi},H_{1,2}] = 0
\label{eq:vanishing commutator}
\ee 
because $|\Phi\rangle$ is an eigenstate of $H_{1,2}$ and
\be
[\rho^{\Phi}, H_{N-2}] =0
\ee
because $H_{N-2}$ does act only on the remaining N-2 particles due to the cyclic property of the trace. Therefore, we can write
\begin{equation}
	\frac{dP^{\Phi}(t)}{dt}=\Gamma^{\Phi}(t)=\frac{-i}{\hbar}Tr[\rho^{\Phi}[U_{1,2},\rho_{N}(t)]].
	\label{eq:rate}
\end{equation}
This is the starting point of our approach. With eq.~\ref{eq:rate} we calculate the probability $P^\Phi(\tilde t)$ that a $Q$ and a $\bar Q$ are in a bound state $\Phi$ at $t=\tilde t$ by integrating the rate from $t=0$ to $t=\tilde t$. $P^\Phi(\tilde t\to \infty)$ is then the probability that at the end of the heavy ion reaction a meson of type $\Phi$ is observed. To make calculations possible we have to know $\rho_{N}(t)$. A full quantum treatment of the evolution of $\rho_{N}$ or of the equivalent N-body Wigner density, $W_{N}$, defined as
\bea
		&&W_{N}(\{\mathbf{r_i}\},\{\mathbf{p_i}\},t)=\frac{1}{h^{3N}}\int d^3 y_{1}...d^3y_{N}(e^{i\frac{\mathbf{p}_{1}\cdot \mathbf{y}_{1}}{\hbar}} ... e^{i\frac{\mathbf{p}_{N}\cdot \mathbf{y}_{N}}{\hbar}}) \nonumber \\ &&
		\langle \mathbf{r}_{1}+\frac{\mathbf{y}_{1}}{2},...,\mathbf{r}_{N}+\frac{\mathbf{y}_{N}}{2}\vert \rho(t) \vert \mathbf{r}_{1}-\frac{\mathbf{y}_{1}}{2},...,\mathbf{r}_{N}-\frac{\mathbf{y}_{N}}{2}\rangle, 
	\label{eq:Weyloperator}
\eea
where $\mathbf{r}_{i}$ and $\mathbf{p}_{i}$ are the coordinates and momentum of the particles in the Wigner representation, is out of reach but in the past it turned out that many observables in heavy ion collisions can be well described if one replaces the N-body Wigner density by an average over classical N-body phase space densities $W^{c}_N$ 
\begin{equation}
	W_{N} \approx \langle W^{c}_{N}\rangle
	\label{WDcl}
\end{equation}
with
\be
W_{N}^c(\{\mathbf{r_i}\},\{\mathbf{p_i}\},t)=\prod_i^N\delta(\mathbf{r_i}-\mathbf{r}_{i0}(t))
\delta(\mathbf{p_i}-\mathbf{p}_{i0}(t)).
\label{eq:phc}
\ee
$W_{N}^c(\{\mathbf{r_i}\},\{\mathbf{p_i}\},t)$ as well as $W_{N}(\{\mathbf{r_i}\},\{\mathbf{p_i}\},t)$ are normalized to 1.
\bea
\int \prod_{i=1}^N  d^3r_id^3p_i W_{N}^c(\{\mathbf{r_i}\},\{\mathbf{p_i}\},t)=1\nonumber \\
\int\prod_{i=1}^N d^3r_id^3p_i W_{N}(\{\mathbf{r_i}\},\{\mathbf{p_i}\},t)=1.
\eea

\subsection{The rate for time independent $Q\bar Q$ Wigner densities}
In this section we assume that, as in the original Remler formalism, 
$W^\Phi$, the Wigner density of density matrix of the eigenstates of the \Q Hamiltonian, $|\Phi \rangle \langle \Phi|$,
is time independent. The extension to a time dependent $W^\Phi(t)$ will be discussed in section \ref{sc:tbasis}.

Employing Wigner densities we can rewrite the rate, eq.~\ref{eq:rateint}. We assume again that among the N particles there is only one heavy quark Q which carries the index 1 and one heavy antiquark $\bar Q$ with the index 2. Then we find 
\bea
	\frac{dP^{\Phi}(t)}{dt} &=& \Gamma^{\Phi}(t)=h^3\frac{d}{dt}
	\int \prod_j^N d^{3}{r}_{j}d^{3}{p}_{j} W^\Phi_{12}W^c_{N}(t) \nonumber \\
	&=&h^3\int \prod_j^N   d^{3}\mathbf{r}_{j}d^{3}\mathbf{p}_{j}\ W^\Phi_{12} \frac{\partial}{\partial t} W^c_{N}(t).
	\label{eq:WDprob}
\eea
where $W^\Phi_{12}=W^\Phi(\mathbf{r_1}-\mathbf{r_2},\mathbf{p_1}-\mathbf{p_2})$. 
The form of $W^\Phi$, the Wigner density of the quarkonium $\Phi$, will be discussed in section \ref{sc:wig}. 

The interaction between the $N$ partons is of short range (as compared to the mean free path). This means that we consider that the QGP partons and the heavy quarks move on straight line trajectories between the collisions whose strength is given by cross sections. 

We can number the collisions between a given couple of scattering partners $i$ and $j$ by $n$, up to $n^{\rm max}_{ij}$. We define as $t_{ij}(n)$ the time at which the $n^{\rm th}$ collision between the partons $i$ and $j$ takes place. This allows to calculate the momentum of particle $i$ at time $t$ as 
\begin{equation}
	p_i(t)=p_i(0)+
	\Sigma_{j\neq i}
	\Sigma_{n=1}\Theta(t-t_{ij}(n))\Delta p_{ij}(n)
\end{equation}
where $\Delta p_{ij}(n)$ is the momentum transfer in the $n-$th collision and where the sum on $n$ runs from 1 $\to n^{\rm max}_{ij}$. This notation will be used implicitly from now on. $\Delta p_{ij}(n)$ is equal
$-\Delta p_{ji}(n)$.
With this choice of time-dependent momenta in the Wigner density $W_N^c$ (eq.~\ref{eq:phc}) we can calculate the time evolution of the N-body Wigner density, (eq.~\ref{eq:WDprob})
\bea
		\frac{\partial}{\partial t} W_{N}^c(t)&=&\Sigma_{i} v_{i}\cdot \partial_{r_{i}} W_{N}^c(\{\mathbf{r}\},\{\mathbf{p}\},t)
\label{eq:partialrhoRemler}
\\ &+& 
\Sigma_{j\geq i}
\Sigma_{n}\delta(t-t_{ij}(n))\nonumber \nonumber \\ &\cdot& 
		(W_{N}^c(\{\mathbf{r}\},
		\{\mathbf{p}\},t+\epsilon)-W_{N}^c(\{\mathbf{r}\},\{\mathbf{p}\},t-\epsilon)).
		\nonumber
\eea
The first term arises from the straight line motion of the particles between the collisions while the second is due to the impulse received at the time $t_{ij}(n)$ when the $n^{\rm th}$ collision between particle $i$ and $j$ takes place. The $\delta(t-t_{ij}(n))$ assures that a momentum transfer takes place exactly at the time of collisions. Besides, we can separate the change of $\rho_N$ due to kinetic terms (straight line motion ) and potential ones (collisions) by writing
\begin{equation}
	\partial \rho_{N}(t)/\partial t=-\frac{i}{\hbar}\Sigma_{j}[K_{j},\rho_{N}(t)]-\frac{i}{\hbar}\Sigma_{k>j}[V_{jk},\rho_{N}(t)].
	\label{eq:partialrho}
\end{equation}
From the comparison between the equations (\ref{eq:partialrhoRemler}) and (\ref{eq:partialrho})  we find that
\bea
	-\frac{i}{\hbar}\Sigma_{j}[K_{j},\rho_{N}(t)] \equiv \langle
\Sigma_{i} 
v_{i}\cdot \partial_{r_{i}}
	 W_{N}^c(\{\mathbf{r}\},\{\mathbf{p}\},t)\rangle
\eea
and (renaming indices)
\bea
	&&-\frac{i}{\hbar}\Sigma_{k>j}[V_{jk},\rho_{N}(t)]\equiv \langle \Sigma_{k > j}\Sigma_{n}\delta(t-t_{jk}(n)) \\ &\cdot& (W_{N}^c(\{\mathbf{r}\},\{\mathbf{p}\},t+\epsilon)-W_{N}^c(\{\mathbf{r}\},\{\mathbf{p}\},t-\epsilon))\rangle \nonumber.
\eea
Strictly speaking we assume that the equivalence holds for each term of the sum separately. This means that like in cascade calculations the interaction range is small as compared to the mean free path. Substituting in eq.~\ref{eq:rate} the square bracket by the r.h.s of eq.~\ref{eq:partialrhoRemler} and passing globally to the Wigner representation,
we find

\bea
		\Gamma^\Phi(t)&=&\sum_{i=1}^{2}\sum_{j\ge 3}^N
		\sum_n \delta(t-t_{ij}(n))\int \prod^{N}_{k=1} d^{3}\mathbf{r}_{k}d^{3}\mathbf{p}_{k} 
		\label{eq:rate1}\\ 
		&\cdot&h^3 W^{\Phi}(\mathbf{r}_{1},\mathbf{r}_{2},\mathbf{p}_{1},\mathbf{
		p}_{2})
\nonumber \\ &\cdot&
		[W_{N}^c(\{\mathbf{r}\},\{\mathbf{p}\};t+\epsilon)-W_{N}^c(\{\mathbf{r}\},\{\mathbf{p}\};t-\epsilon)],\nonumber
\eea
where $j$ sums over the light quarks and gluons of the QGP.  
Hence, in the Remler formalism, collisions of the $Q$ and $\bar{Q}$ with the QGP medium determine the rate of creation and destruction of quarkonia. 
\begin{figure}[h]
	\centering
	\includegraphics[width=0.70\linewidth]{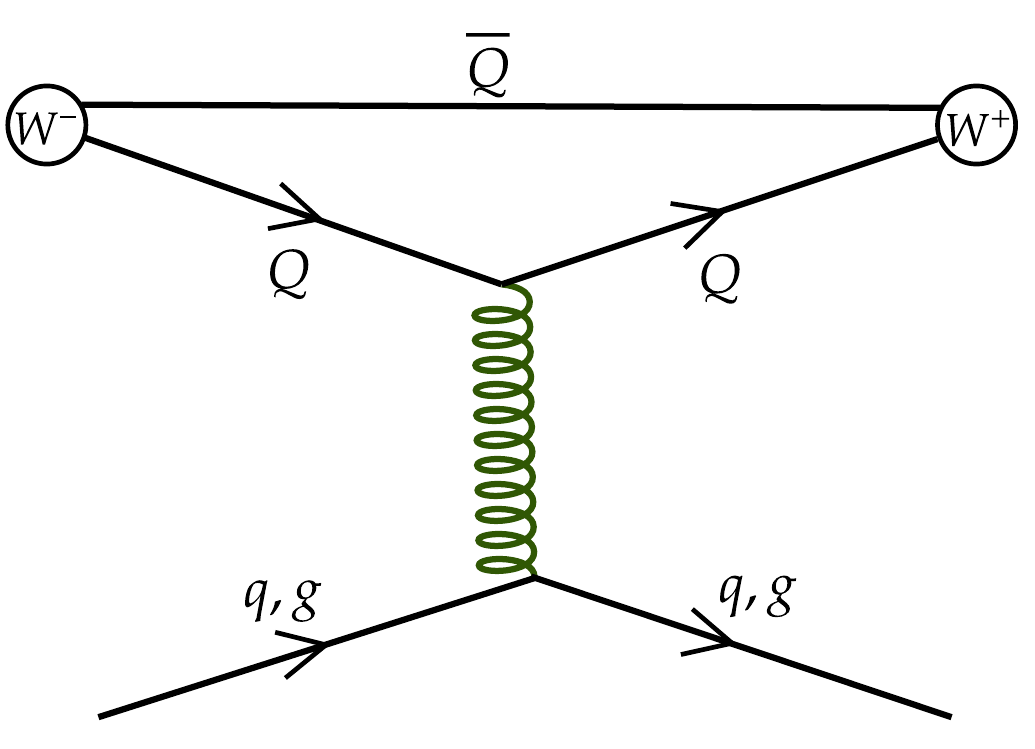}
	\caption{Visualization of eq.~\ref{eq:rate1}. A quark (gluon) from the QGP collides with a heavy quark Q causing a momentum change of the heavy quark. The probability that the heavy quark formed a quarkonia $\Phi$ with the $\bar Q$  before the collision ($W-=W^\Phi W_N^c(t-\epsilon)$) hence differs from the probability that it forms a $\Phi$ after the collision ($W+=W^\Phi W_N^c(t+\epsilon)$). The difference $W+\ $-$\ W-$ is therefore the change of the \J multiplicity
	due to this collision.}
	\label{remler}
\end{figure}
Fig. \ref{remler} visualizes eq.~\ref{eq:rate1}.
A parton from the QGP collides with a heavy quark Q leading to a momentum change of Q. 
We calculate
$W^\Phi$ before ($W_-$,  at $t-\epsilon$, where $\epsilon$ is an infinitesimal time) and after ($W_+$, at $t+\epsilon$) the collision for the $Q \bar Q$ pairs, which the $Q$ can form.

It is useful to explain this equation a bit more. At time $t=t_{1j}(n)$ a heavy quark $Q$, to which we assign the coordinates $\mathbf{r}_{1},\mathbf{p}_{1}$, has a collision with a QGP parton $j\ge$ 3. To the heavy antiquark we assign the coordinates 
$\mathbf{r}_{2},\mathbf{p}_{2}$, 
and define the two-body Wigner density of the $Q\bar Q$ pair as
\be
W_2(\mathbf{r}_{1},\mathbf{r}_{2},\mathbf{p}_{1},\mathbf{p}_{2},t)= \prod_{k=3}^N \int d^{3}r_k d^{3}p_k W_N^c(\{\mathbf{r}\},\{\mathbf{p}\},t).
\label{eq:psint}
\ee 
We can calculate the contribution of this collision to the yield of the state $\Phi$.
For this we define the relative and center of mass coordinates of the $Q \bar Q$ pair $\mathbf{q}=\frac{\mathbf{p}_{1}-\mathbf{p}_{2}}{2}\ (\mathbf{r}=\mathbf{r}_{1}-\mathbf{r}_{2})$
and
 $\mathbf{P}=\mathbf{p}_{1}+\mathbf{p}_{2}\
(\mathbf{R}=\frac{\mathbf{r}_{1}+\mathbf{r}_{2}}{2})$.  
The contribution of this Q-parton $n^{\rm th}$ collision to the $\Phi$ production rate can then be expressed as 
\bea
&&\Gamma^{\Phi}_{1,2;j}(n;t)=
h^3 \delta(t-t_{1j}(n))\int d^3P d^3R d^{3}rd^{3}q
W{^\Phi}(\mathbf{r},\mathbf{q}) \nonumber \\
	&\cdot& \left(W_2(\mathbf{R},\mathbf{r},\mathbf{P},\mathbf{q},t+\epsilon)-(W_2(\mathbf{R},\mathbf{r},\mathbf{P},\mathbf{q},t-\epsilon)\right). \label{eq:defnPhi} 
\eea
This allows for expressing the total rate in a form suitable for Monte Carlo implementations
\begin{equation}
\Gamma_{1,2}^\Phi(t) \equiv  \langle   \sum_{i=1,2}\sum_{j\ge 3}\sum_n \Gamma_{i,3-i;j}^{\Phi}(n;t)\rangle, 
\label{eq:gamma12MC}
\end{equation}
where $\Gamma_{2,1;j}^{\Phi}(n;y)$ is the equivalent quantity to  $\Gamma_{1,2;j}^{\Phi}$ when parton $j$ collides with the $\bar{Q}$.

Let us now consider the general situation with $N_Q$ $Q$-quarks as well as $N_{\bar Q}$ $\bar Q$-quarks in the N-body system, and let us assign indices $i\in[1,N_Q]$ for $Q$ and $j\in[N_{Q }+1,  N_Q + N_{\bar Q}]$ for $\bar{Q}$. The total rate of
quarkonia formation then writes
\bea
		\Gamma^\Phi(t)&=&\sum_i \sum_j 
		\sum_{k> N_Q + N_{\bar{Q}}}
		\sum_n \left[\delta(t-t_{ik}(n))+\delta(t-t_{jk}(n))\right] \nonumber \\
		&& \int \prod^{N}_{l=1} d^{3}\mathbf{r}_{l}d^{3}\mathbf{p}_{l} \,h^3 W^{\Phi}(\mathbf{r}_{i},\mathbf{r}_{j},\mathbf{p}_{i},\mathbf{
		p}_{j}) \label{eq:rateseveral} \\ &\cdot&
		[W_{N}^c(\{\mathbf{r}\},\{\mathbf{p}\};t+\epsilon)-W_{N}^c(\{\mathbf{r}\},\{\mathbf{p}\};t-\epsilon)],\nonumber
\label{eq:rate1several}
\eea
where collisions between heavy quarks are neglected as they are rare. We have to sum over all possible $Q\bar Q$ pairs because they can all lead to the formation of a $\Phi$ meson after the scattering of either the $Q$ ($\delta(t-t_{ik}(n))$) or the $\bar{Q}$ ($\delta(t-t_{jk}(n))$) with light particles. One can then generalize eq.~\ref{eq:psint} to
\be
W_2(\mathbf{r}_{i},\mathbf{r}_{j},\mathbf{p}_{i},\mathbf{p}_{j},t)= \prod_{\underset{\underset{\scriptstyle l\neq j}{\scriptstyle l\neq i}}{l=1}}^N \int d^{3}r_l d^{3}p_l W_N^c(\{\mathbf{r}\},\{\mathbf{p}\},t).
\label{eq:psintgen}
\ee 
and eq.~\ref{eq:defnPhi} to
\bea
&&\Gamma^{\Phi}_{ij;k}(n;t)=
h^3 \delta(t-t_{ik}(n))\int d^3P d^3R d^{3}rd^{3}q  W{^\Phi}(\mathbf{r},\mathbf{q})  \nonumber \\
	&\cdot& (W_2(\mathbf{R},\mathbf{r},\mathbf{P},\mathbf{q},t+\epsilon)-(W_2(\mathbf{R},\mathbf{r},\mathbf{P},\mathbf{q},t-\epsilon)
\label{eq:Gammaijk}
\eea
as well as $\Gamma^{\Phi}_{2,1;k}(n;t)$ to $\Gamma^{\Phi}_{ji;k}(n;t)$.
One thus obtains
\begin{equation}
\Gamma^\Phi(t)  \equiv  \langle   \sum_i\sum_j\sum_{k > N_Q+N_{\bar{Q}}}\sum_n( \Gamma_{ij;k}^{\Phi}(n;t)+\Gamma_{ji;k}^{\Phi}(n;t))\rangle,
\label{eq:ratefullMC}
\end{equation}
where $i$ runs from 1 to $N_Q$ and $j$ from $N_Q+1$ to $N_Q+N_{\bar{Q}}$, allowing to take into account all possible $Q\bar Q$ pairs, independent of whether the entrained heavy quarks come originally from the same vertex or from different vertices. With this approach, we are thus able to treat consistently the primordial and the regenerated components introduced in usual transport models.

\subsection{Heavy Quark - parton interactions}
Our approach for the \J production, derived in the last section, is based on the collisions of heavy quarks with partons from the QGP. The study of these collisions was already presented quite a while ago \cite{Gossiaux:2008jv,Gossiaux:2009mk,Nahrgang:2016lst} to investigate the production of open heavy flavour mesons. In this study we calculate the interaction rate for heavy quark - parton interactions and determine whether a collision takes place by a Monte Carlo procedure. If a collision is taking place we determine randomly, from the local equilibrium distribution, the momentum of the light parton. 
The interaction of the gluons and quarks with the heavy quarks is then described by Born-type matrix elements. These matrix elements have two inputs: The running coupling constant and the infrared regulator. The running coupling constant remains finite at zero momentum transfer and agrees with the analysis of $\tau$ decays and $e^+e^-$ scattering \cite{Gossiaux:2008jv}. The infrared regulator has been chosen to make the result independent of the scale which separates the hard thermal loop dominated low energy behaviour and the high momentum transfer region, which is described by Born terms. This approach has been successfully used to describe the open heavy flavor meson production in ultrarelativistic heavy ion collisions \cite{Gossiaux:2009hr}.
For the calculations presented here we limit us to elastic collisions and employ a $K$ scaling factor of 1.5 for the collision probabilities of heavy quarks with light partons. As shown such a scaling factor compensates for radiative collisions, which are not considered here \cite{Nahrgang:2014vza}.

\subsection{Wigner density of quarkonia}
\label{sc:wig}
To make use of eq.~\ref{eq:rate1} we need to know
the Wigner density of the quarkonia. For quarkonia, which are created in the vacuum in pp collision, the Wigner density has been discussed and employed in ref \cite{Song:2017phm}. Here we follow this approach. Considering first quarkonia in non relativistic motion, the center of mass ($\mathbf{R},\mathbf{P}$)  motion of the state is given by a plane wave. Due to the large mass of the heavy quark we assume that the relative wave functions of the different eigenstates $\Phi$, $|\Phi\rangle$, of the $Q\bar Q$ pair in vacuum, as well as its Wigner density, $W^\Phi(\mathbf{r},\mathbf{p})$, can be calculated by solving the Schr\"odinger equation for the relative motion in a Cornell potential \cite{Das:2015bta}. The calculation of the \J production in heavy ion collisions becomes more convenient if we replace $W^\Phi(\mathbf{r},\mathbf{p})$ for the s-wave states by a Gaussian Wigner density 
\begin{equation}
	W^{\Phi}_{1s}(\mathbf{r},\mathbf{p})=\frac{8 g}{h^3} e^{-\frac{r^{2}}{\sigma_{1s}^2}-p^2\frac{\sigma_{1s}^2}{\hbar ^2}}   \label{eq:WignerMC}
\end{equation}
with
$$\frac{3}{2}\sigma_{1s}^2=\langle r^2_{1s}\rangle,$$
where $\sqrt{\langle r^2_{1s}\rangle}\approx 0.4$ fm for \J.
$g=\frac{3}{4}$ is the  spin factor of a vector meson. 
The factor $\frac{8}{h^3}$ is due to the normalization of the Wigner density. To simplify the calculation we assign initially to each heavy quark-antiquark pair whether it is in a color singlet (probability = 1/9)  or in a color octet state (probability = 8/9) and stick then to this assignment. Thus we do not follow the color flow. This is foreseen as a future project.

\subsection{Operational summary}
We come back now to the relation between the probability
and the rate. Following eq.~\ref{eq:rho2bis}, the total probability that a $Q\bar Q$ pair, which has the coordinates $\{1,2\}$,  forms a quarkonium state at time $t$ is given by a time integration of the rate (eq.~\ref{eq:rate})
\begin{equation}
	P^\Phi_{1,2}(\tilde t)=P^{\rm prim}_{1,2}(0)+\int^{\tilde t}_{t_{0}}\Gamma_{1,2}(t)d t
	\label{eq:Remlerform}
\end{equation}
with $\tilde t<t_{\rm hadr}$, the time when the QGP fully hadronizes. $P^{\rm prim}_{1,2}$ is the probability that at the moment of their creation the $Q\bar Q$ pair forms a quarkonium (see \cite{Song:2017phm}, as well as section V), $t_0$ is the time when the QGP is formed\footnote{Assuming no contribution to quarkonia production for the interval $[0,t_0]$.}. The rate (eq.~\ref{eq:rate}) is treated in a Monte Carlo approach, adopting the Remler method, leading to eq.~\ref{eq:gamma12MC}. The time integral of the rate thus accumulates the change of the probability that the $Q\bar Q$ pairs form a quarkonium caused by all collisions with plasma partons suffered either by the $Q$ or the $\bar Q$ during the time evolution until the time $\tilde{t}$. 

In the Remler formalism, equation (\ref{eq:Remlerform}), which refers only to a single $Q\bar Q$ pair, naturally extends to many $Q\bar Q$ pairs that are in the QGP at a given time $\tilde{t}$. Using eq.~\ref{eq:Gammaijk}, the later can be expressed by summing over all possible pair combinations at a given time. As exhibited in eq.~\ref{eq:ratefullMC}, the total rate of quarkonium formation, the sum of the rate due to the scattering of the heavy quark and that of the heavy antiquark,  at a given time can be expressed as 
\begin{equation}
	\Gamma(t)=\sum^{N_{Q}}_{i=1}\sum^{N_{Q}+N_{\bar Q}}_{j=N_{Q}+1} \left(\Gamma_{i,j}(t)+
	\Gamma_{j,i}(t)\right).   
\end{equation}

In practice, this sum over the rates is performed in the numerical program according to eq.~\ref{eq:ratefullMC}. 

We would like to stress that in the numerical implementation of our approach the $\Phi$ mesons are not represented by pseudo particles, produced and destroyed by $2\to 2$, like $c\bar c \leftrightarrow \Jm + g$, or $3\to 2$, like $Xc\bar c \leftrightarrow \Jm X$, processes, as done in standard cascade approaches. 
Instead we sum coherently the contributions to the rate of the different $Q\bar Q$ pairs, which offers the advantage to add coherently all possible contributions, what is not possible in standard MC approach based on pseudo-particles. The non-trivial effect of adding the diagonal and off diagonal components for the primordial contribution --$\sum_{i} \sum_j P^{\rm prim}_{i,j}(0)$ -- has already been discussed in \cite{Song:2017phm}.

Finally, it should be noted that the Monte Carlo implementation of the rate can be formulated locally -- see eq.~\ref{eq:Gammaijk} -- as a sum of a gain and a loss term. If one bins the phase space along any variable (f.i. transverse momentum ${\bf P}_T$), one can thus reformulate the Monte Carlo process as a depletion of some ${\bf P}_T$ bin and the population of a ${\bf P}_T +\Delta {\bf P}_T$ bin, where $\Delta {\bf P}_T$ is the transverse momentum transferred from the light parton to the heavy quark, its scattering partner. This opens the possibility to evaluate differential $\Phi$-spectra by book-keeping these gain and loss terms.

\subsection{Generalization for relativistic quarkonia}

Up to now we have formulated the Wigner density in a nonrelativistic approach. As shown in the appendix the corresponding relativistic Wigner density can be written as
\begin{eqnarray}
	W_i^\Phi(y,\mathbf{u}_{T},\mathbf{r}^{\rm cm},\mathbf{q}^{\rm cm})&=&\frac{\delta (y-y_{\Phi})}{(2\pi)^{3}}\delta^{2}(\mathbf{u}_{T,\Phi}-\mathbf{u}_{T}) \times \nonumber\\ 
	&& W_{i,\rm NR}(\mathbf{r}^{\rm cm},\mathbf{q}^{\rm cm}).
	\label{eq:boostWigner1}
\end{eqnarray}
In this expression,  $u_{T}$ is the transverse  component of the 4-velocity 
\begin{equation}
	\mathbf{u}_{T,\Phi}=\frac{\mathbf{P}_{T}}{m_{\Phi}}
	\label{eq:velocitytransv}
\end{equation}
where $\mathbf{P}_{T}$ is the total transverse momentum of the $Q \bar Q$ center of mass, $y_{\Phi}$ is the  rapidity of the quarkonium, while $\mathbf{r}^{\rm cm}$ and $\mathbf{q}^{\rm cm}$ are the (relative) coordinates in the center of mass frame. $\Phi$ and the index NR indicates that the Wigner density of the relative coordinates are evaluated in a non relativistic framework (see  eq.~\ref{eq:WignerMC}). This is justified because $Q\bar Q$ pairs with a large relative momentum do not form quarkonia.
Here it is important to  mention  that only for those states, for which we impose a well-defined center of mass 4-velocity and a well-defined relative momentum with respect to the center of mass, we have been able to successfully derive a prescription which allows to evaluate the Wigner density in any system of reference as a function of the Wigner density in the center of mass frame. The later condition comes from the fact that, even if we can always define a total momentum for the center of mass, due to the on-shell condition.the mass of the quarkonium state $m_{\Phi}$ depends on the relative momentum of the pair $\mathbf{q}$, as shown in the equation (\ref{eq:onshell}).  This implies that in our construction one cannot impose both, a fixed total momentum AND a fixed velocity for the two-body state. To overcome this problem one has to solve the Bethe Salpeter equation what is beyond the scope of the present approach.

The result obtained in the eq.\ref{eq:boostWigner1} allows us to study the formation of \J in the center of mass of the $c\bar c$ pair and at the same time to be able to evaluate the Wigner density at any time in any other system of reference. The latter is rather important because our multi-particle dynamics requires to adopt a common computational frame, as can be seen from the definition of the global rate, see eq.~\ref{eq:ratefullMC}. The standard computational frame is the center of mass frame of the heavy ion collision. Benefiting from the boost invariance of the phase space, it is nevertheless possible to define the equivalent Wigner density in this center of mass frame, called lab frame to distinguish it from the center of mass frame of the $c\bar c$ pair, (see equivalence between eq.~\ref{eqntransversevelocity}  and eq.~\ref{eq:densitylabcm}) by expressing 
$\mathbf{q}^{\rm cm}$ as a function of $\mathbf{q}^{\rm lab}$ as well as $\mathbf{r}^{\rm cm}$ as a function of $\mathbf{r}^{\rm lab}$ (while taking ${x^{\rm lab}}^0=0$ in the Lorentz transform). This leads to a Wigner density in the lab frame:
\bea
		W_{i}(y,\mathbf{u}_{T},\mathbf{q}^{\rm lab},\mathbf{p}^{\rm lab})&=&\frac{1}{(2\pi)^{3}}\delta(y-y_{\Phi})\delta^{2}(\mathbf{u}_{T,\Phi}-\mathbf{u}_{T}) \nonumber \\ 
		&\cdot& W_{i,\rm NR}(\mathbf{r}^{\rm cm}(\mathbf{r}^{\rm lab}),
		\mathbf{q}^{\rm cm}(\mathbf{q}^{\rm lab})).    
\label{eq:Wignerboostedreadc}
\eea

\section{Appropriate Basis for the Quarkonium States in the QGP }
\label{sc:tbasis}

\begin{figure}[h]
   \centering
   \includegraphics[width=1.2\linewidth]{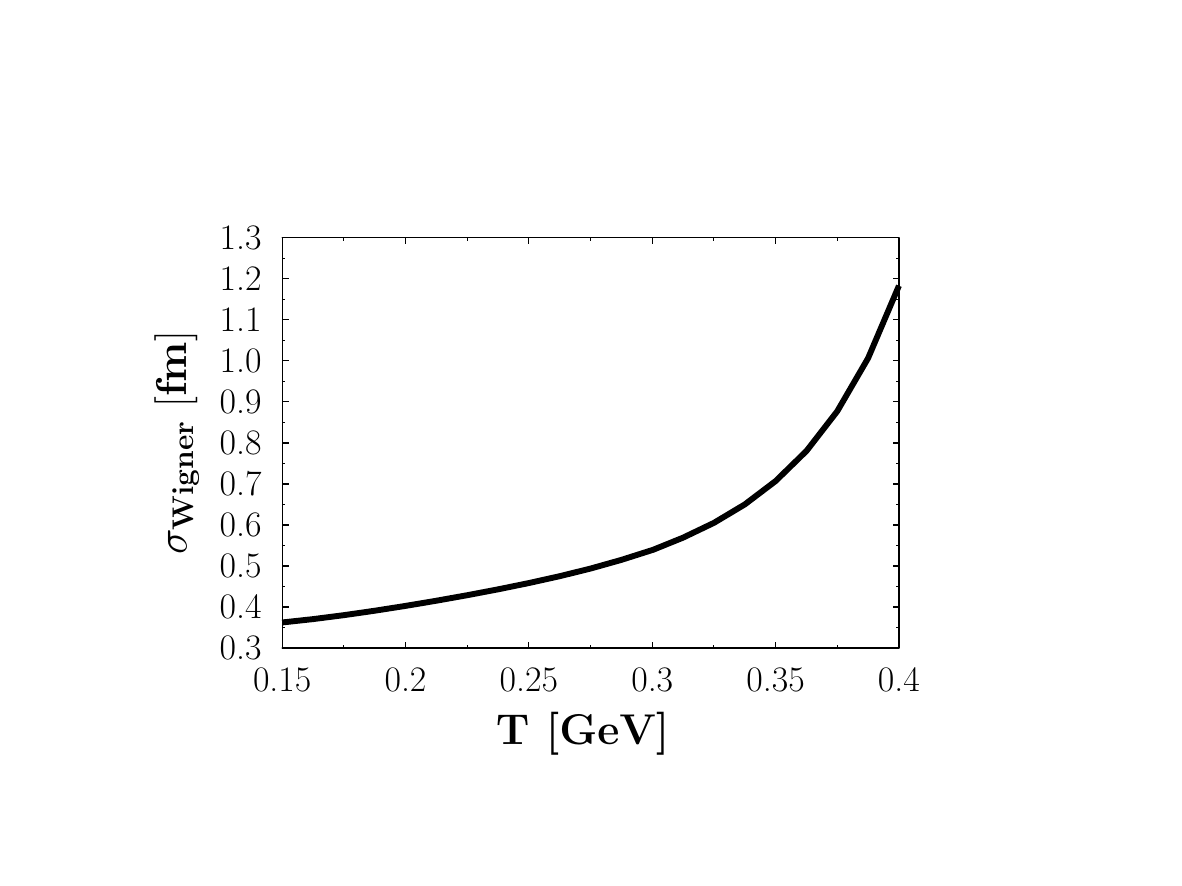} 
      \caption{Width of the Gaussian Wigner density, $\sigma(T)$, as function of the temperature, obtained by solving the Schr\"odinger equation with a potential taken from  \cite{Lafferty:2019jpr} in the interval 0.15 GeV $<T<$ 0.4 GeV. Above $T$= 0.4 GeV = $T_{\rm diss}$, the \J is unstable, below $T=0.15$ a QGP does not exist.}
         \label{Gaussianwidthvstemp}
\end{figure}

The Remler formalism was originally developed for two-body
systems for which the vacuum  eigenstates provide the appropriate basis. In this case the density operator $\rho^\Phi (\mathbf{r}_{1}, \mathbf{r'}_{1}, \mathbf{r}_{2}, \mathbf{r'}_{2})$  in 
eq.~\ref{eq:vanishing commutator} corresponds to the two-body vacuum density operator.
Lattice results \cite{Mocsy:2007jz} show that the potential between the $Q$ and $\bar Q$  changes with temperature and at high temperatures the quarkonia melt. To cope with these results we introduce a temperature dependent potential between the $Q$ and $\bar Q$, taken from  \cite{Lafferty:2019jpr}. This renders the 
two-body Hamiltonian temperature dependent and the eigenstates of the relative motion of the quarkonia need to be chosen accordingly in order to fulfil eq.~\ref{eq:vanishing commutator}.
We assume that also at finite temperature the \J wave function can be approximated by a Gaussian. To obtain the temperature dependence of the Gaussian width we solve the two-body Schr\"odinger equation with a Lafferty-Rothkopf potential \cite{Lafferty:2019jpr} and determine the rms radius of the \J wave function \cite{Katz:2015edb,Katz:2013rpa}. The rms radius of the \J wave function is related to the Gaussian width by $\sigma^{2}(T)=\frac{2}{3}\langle r^{2}(T)\rangle$. This calculation shows as well that the \J melts at $T_{\rm diss} = 0.4\,{\rm GeV}$.
For $T\to 0$ the temperature-dependent potential becomes the potential in vacuum and therefore we recover eq.~\ref{eq:WignerMC}.
The dependence of the Gaussian width $\sigma$ on the local temperature $T$ is displayed in Fig. \ref{Gaussianwidthvstemp}.
In an expanding QGP, the temperature changes rapidly as a function of time. Therefore the temperature dependence of the width is equivalent to a time dependence.

Introducing a temperature-dependent potential creates an additional term in eq.~\ref{eq:defGamma} when replacing $\rho_{\Phi}(\mathbf{r}_{1},\mathbf{r'}_{1},\mathbf{r}_{2},\mathbf{r'}_{2}$) by $\rho_{\Phi}(\mathbf{r}_{1},\mathbf{r'}_{1},\mathbf{r}_{2},\mathbf{r'}_{2},T(t))$. We call this term "local rate". This leads to 
\bea
	\Gamma_{\rm eff} &=& 	\Gamma+ \Gamma_{\rm loc}\\ &=& Tr[\rho^{\Phi}(\mathbf{r},\mathbf{r}^{'},T(t))
	\dot{\rho}_{N}(t)]+Tr[\dot{\rho}^{\Phi}(\mathbf{r},
	\mathbf{r}^{'},T(t))\rho_{N}(t)]\nonumber.
	\label{eq:totalrate}
\eea    
The first term is the rate from the Remler formalism eq.~\ref{eq:rateint} and the second is the new $\Gamma_{\rm loc}$. 

Performing the trace integral, we obtain
\begin{equation}
	\Gamma_{\rm loc}=\sum \int d^{3}\mathbf{r}d^{3}\mathbf{r}^{'}\dot{\rho}^{\Phi}
	(\mathbf{r},\mathbf{r}^{'},T(t))\rho_{Q\bar{Q}}
	(\mathbf{r},\mathbf{r}^{'},t)
\end{equation} 
where $\rho_{Q\overline{Q}}(\mathbf{r},\mathbf{r}^{'},t)$ is the density operator of the N-body system integrated over the positions of the remaining $N-2$ particles which are not part of the pair, while $\rho^{\Phi}$ is the density operator of the bound quarkonium states. The sum runs over all possible $Q\bar Q$ pairs. Converting $\rho^\Phi$ into the corresponding Wigner density 
\begin{equation}
	\rho^{\Phi}(\mathbf{r}+\frac{\mathbf{r'}}{2},\mathbf{r}-\frac{\mathbf{r}'}{2})=\int d^{3}p e^{-i\frac{\mathbf{p}\cdot \mathbf{r'}}{\hbar}}W^{\Phi}(\mathbf{r},\mathbf{p})
\end{equation}
we obtain
\be
\Gamma_{\rm loc}=(2\pi\hbar)^{3}\int d^{3}rd^{3}p\ \dot{W}^{\Phi}(\mathbf{r},\mathbf{p},T(t)) W_{Q\bar Q}(\mathbf{r},\mathbf{p},t).
\ee
For classical phase space densities,  eq.~\ref{eq:phc}, we arrive at
\begin{eqnarray}
	\Gamma_{\rm loc}&=&8\dot{\sigma}(T(t))\partial_{\sigma}e^{-(\frac{\mathbf{r}^{2}}{\sigma^{2}}+\frac{\sigma^{2}\mathbf{p}^{2}}{\hbar^2})} \\
	&=&16\dot{\sigma}(T(t))\left(\frac{\mathbf{r}^{2}}{\sigma^{3}(T)}-\frac{\sigma(T)\mathbf{p}^{2}}{\hbar^{2}}\right)e^{-(\frac{\mathbf{r}^{2}}{\sigma^{2}}+\frac{\sigma^{2}\mathbf{p}^{2}}{\hbar^2})}.\nonumber
	\label{eq:localrate}
\end{eqnarray}
where $\dot{\sigma}(T(t))=\dot{T}(t)\sigma'(T)$.
This local rate is non zero if the temperature changes with time and therefore the temperature dependent Gaussian width becomes time dependent. 
Including the local rate, the probability that  a quarkonium state is formed at time $t$ from a single $Q\bar{Q}$ pair now reads 
\begin{equation}
	P^{\Phi}_{Q\bar Q}(t)=P^{\rm init}(t^{Q,\bar Q}_{\rm init})+\int^{t}_{t^{Q,\bar Q}_{\rm init}}(\Gamma_{{\rm coll},Q\bar Q}(t^{'})+\Gamma_{{\rm loc},Q\bar Q}(t^{'}))dt^{'}.
	\label{eq:modifyremler}
\end{equation}
$\Gamma_{\rm coll}$ is the contribution of the collisions to the rate and $P^{\rm init}$ is the initial probability for the production of a quarkonium once the local $Q\bar Q$ temperature falls below $T_{\rm diss}$. Contrary to eq.~\ref{eq:Remlerform}, obtained from the projection on the vacuum basis, $t^{Q,\bar Q}_{\rm init}$, the time when the \Q passes $T_{\rm diss}$, is dependent on the environment of the $Q\bar{Q}$ pair and on the quarkonium state considered. 

\JA{Let us finally mention that the existence of a local rate is not specific to the Remler approach. It should appear in all transport approaches in which medium modified bound states are produced.}

\section{In-medium $Q\bar Q$ propagation}
\label{sectionIV}
Besides the usual short distance collisions between heavy quarks and light QGP partons, we have updated our  MC@sHQ scheme by implementing a long distance $Q\bar{Q}$ interaction. This is detailed in the present section.

\subsection{Global strategy}
First, let us recall that the implementation of the Remler method, described in section II, requires trajectory calculations in Minkowski time steps $\Delta t$. The EPOS2 event generator, which we use to simulate the time evolution of the QGP and to obtain the initial interaction points at which the \cc quarks are produced, uses,
like almost all hydrodynamical calculations, Milne coordinates (Bjorken time $\tau^{\rm Bjorken}=\sqrt{t^2-z^2}$ as well as the space time rapidity $\eta=\frac{1}{2}\ln\frac{t+z}{t-z}$). The coordinate and momentum space variables are updated employing a finite time step $\Delta \tau^{\rm Bjorken}$.
Therefore, before being able to apply the Remler algorithm, we have to construct the trajectories of the $Q$ and $\bar{Q}$ quarks in Minkowski space from the Milne coordinates. If $\Delta \tau^{\rm Bjorken}$ is too large and, as a consequence, the differences in momentum  of the heavy quarks between two Bjorken time steps becomes sizable, these trajectories become kinky.

Solving the relativistic dynamics of an ensemble or particles in mutual potential-interaction is a much involved problem and nowadays still represents a challenge. A first option consists in considering retarded potentials, generalizing the Li\'enard-Wiechert potential from classical electrodynamics. This approach suffers from the non-conservation of energy and angular momentum associated to radiative field emission. Such non-conserving features can be cured at the price of adding the retarded and advanced propagator \cite{Wheeler:1949hn} which, however, leads to advanced interactions from the future to the past. A second choice is
the constrained Hamilton dynamics which reduces the 8N dimensional phase space to a 6+1 dimensional by imposing time and energy constraints \cite{Sorge:1989dy,Marty:2012vs}. This
is, however, only possible if the potential has a form, which is invariant under a Lorentz transformation.

In the absence of an exact way of solving the multi-body dynamics, we have adopted the following strategy: We proceed to the evolution of each quark $Q$ from $\tau^{\rm Bjorken}$ to $\tau^{\rm Bjorken}+\Delta \tau^{\rm Bjorken}$ by considering the closest $\bar{Q}$ partners. For each such pair, we first transform the $Q\bar{Q}$ coordinates into the CM system of the pair, where the evolution can be performed exactly -- see next subsection  -- until the quark $Q$ reaches time $\tau^{\rm Bjorken}+\Delta \tau^{\rm Bjorken}$. Such an evolution leads to a variation $\delta x_Q$ with respect to the free propagation.

The total evolution of the $Q$ under consideration is thus defined as the sum of the various $\delta x_Q$ from the interaction with the different $\bar Q$ superposed to the free motion. Such a linearized algorithm leads to acceptable results when $\Delta \tau^{\rm Bjorken}$ is small with respect to the revolution time of the $Q\bar{Q}$ pairs.

If the distance between the quarks is large, the potential does not affect the trajectories. Only the trajectories of neighboring \Q pairs are therefore concerned and usually one finds for each heavy quark $Q$ not more than one $\bar Q$ (or two, in the early stage), which is sufficiently close that the potential has an influence on the trajectory. 

 The drawback of such a sequence of Lorentz transformations is that small shifts in space-time coordinates are introduced at each step due to Lorentz transformations\footnote{Equal time $t(Q)=t(\bar{Q})$ in the CM does generally not correspond to a unique time in the computational frame.}. As those shifts are proportional to the ${Q\bar{Q}}$ CM velocity, this approach is not suited for the calculation of high $p_{T}$ pairs. 
 
\subsection{The $Q\bar Q$ potential interaction}
\label{sectionQQbarinteraction}
The simplest relativistic modification to the movement of a particle in a central potential field, $V(r)$,  is described by the Lagrangian
\cite{Lemmon:2010ya}.

\begin{equation}
\mathcal{L} = -\gamma^{-1} m c^2 - V(r)
\label{eq:lag}
\end{equation}
with $\gamma^{-1}=\sqrt{1-v^2/c^2}$ and 
$m$ being the particle mass. With
\be
\frac{\partial \mathcal{L}}{\partial v_i}=p_i = \gamma m v_i
\ee
and
\bea
H=p_iv_i -\mathcal{L} &=&\gamma m v_iv_i+\frac{m c^2}{\gamma}+ V(r)\nonumber = \gamma m c^2+ V(r) \\&=& \sqrt{m^2c^4+p^2c^2}+ V(r)=E.
\label{ener}
\eea
 $E$, the energy,  is a constant of motion. If we employ spherical coordinates $\dot r^2=\dot r^2_r+r^2 \dot\theta^2$, $ p^2= p^2_r+p_\theta^2/r^2$, we find the
corresponding momenta from the Euler - Lagrange equations (we employ now c=1)
\bea
p_\theta=\frac{\partial {\mathcal L}}{\partial \dot \theta}&=& \gamma m r^2 \dot \theta \nonumber \\ 
p_r=\frac{\partial {\mathcal L}}{\partial \dot r} &=& \gamma m \dot r.  
\eea
Expressing the Hamiltonian in terms of $p_r$ and $p_\theta$
\be
H= \sqrt{m^2+p_r^2+\frac{p^2_\theta}{r^2}} + V(r)
\label{ham}
\ee
we obtain the equations of motion from the Hamilton's equation
\bea
\dot r &=& \frac{\partial H}{\partial p_r}= \frac{p_r}{\sqrt{m^2+p_r^2+\frac{p^2_\theta}{r^2}}}\nonumber \\
\dot \theta &=& \frac{\partial H }{\partial  p_\theta}= \frac{p_\theta}{r^2\sqrt{m^2+p_r^2+
\frac{p^2_\theta}{r^2}}}
\nonumber \\
\dot p_r &=& -\frac{\partial H}{\partial r}= \frac{p_\theta^2  }{r^3\sqrt{m^2+p_r^2+\frac{p^2_\theta}{r^2}}}
- \frac{\partial V}{\partial r} \nonumber \\
&=& \frac{p_\theta \dot \theta}{r} - \frac{\partial V}{\partial r} \nonumber \\
\dot p_\theta &=&- \frac{\partial H}{\partial \theta}= 0 \to p_\theta = {\rm const} = L 
\label{tee}\eea
The last equation states that the generalized angular momentum
\be
L= \gamma m r^2 \dot \theta
\ee
is conserved in this ansatz. Thus one only needs to solve the radial equations of motion on $r$ and $p_r$ numerically and can then integrate the differential on $\dot \theta$ .

For two heavy quarks in their center of mass system we can formulate a spinless Hamiltonian, generalizing eq.~\ref{ener} 
\be
H_2=\sqrt{m_1^2+p_1^2}+\sqrt{m_2^2+p_2^2} - V(r_{12})=E
\label{ener2}
\ee
where $\bfp_1=-\bfp_2$. $m_1$ and $m_2$ are the masses of the heavy quarks and $\bfr_{12}$ is their relative distance. For the case we consider here, $m_1=m_2$, the two-body dynamics is directly mapped on the single one, discussed above, by taking $V(r_{12})= V(2\times r/2)$ in eq.~\ref{ener}. $V(r_{12})$ is here taken as the Lafferty-Rothkopf potential \cite{Lafferty:2019jpr}, which depends on the temperature. 

\subsection{Calculation of $\Gamma_{\rm coll}$}
\label{chIIIb}
At the boundaries of each time interval $[t,t+ \Delta t$] we
compare the momentum change of each $c(\bar c)$.
If the $c \bar c$ potential is not active, means switched off, and the heavy quark has changed its momentum in this time interval we know that this heavy quark had a collision with a QGP parton and we calculate the Wigner density of this quark with the antiquarks, which are in a hydro cell with $T < T_{\rm diss}$, to determine $\Delta W= W(t+ \Delta t)- W(t)$. $\Delta W$ is then the contribution of this $c\bar c$ pair to the \J multiplicity and the sum of $\Delta W$ over all $\bar c(c)$ is the contribution of this collision to the \J multiplicity.



\subsection{Consequences of the $Q\bar Q$ Interaction}
\begin{figure}[h]
	\centering
	\includegraphics[width=0.90\linewidth]{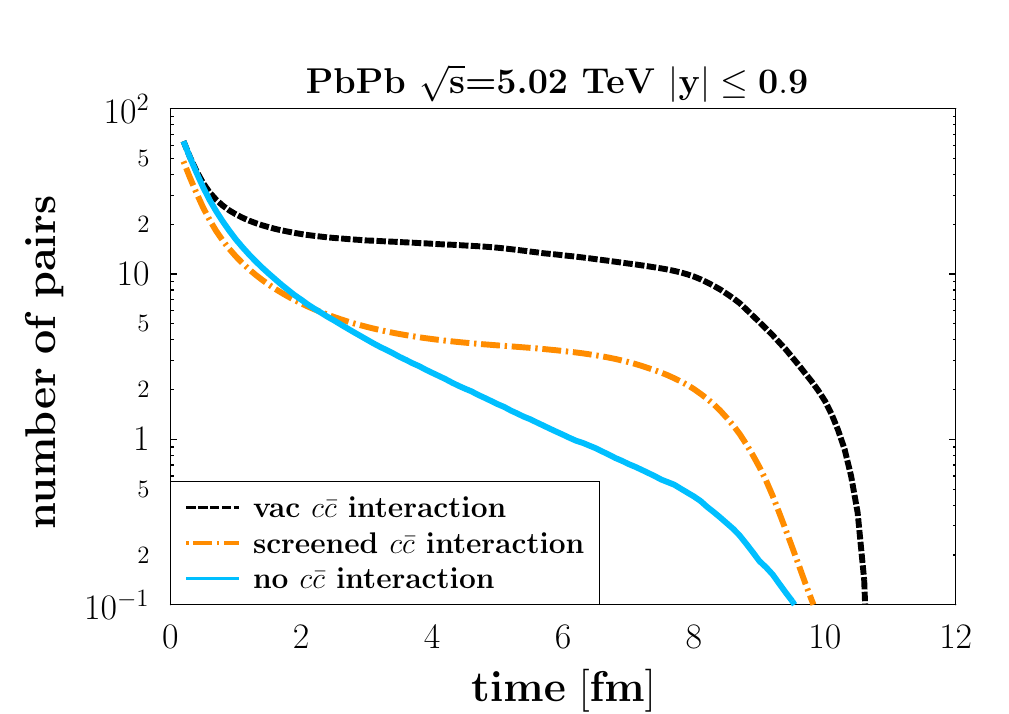}	
	\caption{ Number of $c\bar{c}$ pairs whose relative distance in the CM is lower than 1 fm as function of time under three conditions: without any $c\bar c$ interaction (blue line),
	with the interaction screened by the medium (dashed orange line) and with the vacuum interaction (dotted black line).}            
	\label{figureHQclosepotential}
\end{figure}

Fig. \ref{figureHQclosepotential} shows the influence of th \Q potential on the time evolution of the $c\bar{c}$ pairs. We display there the number of $c\bar{c}$ pairs whose constituents have an invariant distance (the relative distance between the $c$ and $\bar{c}$ quarks, measured in their CM) of $r \le 1$ fm as a function of the Minkowski time.
We show this quantity for three scenarios: a) without any $c\bar c$ interaction (blue line), b) with the interaction screened by the medium \cite{Lafferty:2019jpr} (dashed orange line)  and c) with the vacuum interaction (dotted black line).  Without potential the number of $c\bar{c}$ pairs that stay close indeed decreases strongly with time. We see that a medium screened $Q\bar Q$ potential keeps the $c \bar c$ pairs longer together, especially at the final stage of the evolution ($t\approx 4-8$ fm/c).  This is quite important because in a fast expanding medium the heavy quarks tend to move away from each other.  Adding an interaction potential enhances the recombination (regeneration) process, especially at the latest stage of QGP evolution. In the final stage $t\gtrsim 8$ fm/c, the $c$ and $\bar{c}$ quarks escape from the QGP as mesons and are not counted anymore in Fig. \ref{figureHQclosepotential}. 
We display in Fig.\ref{figureHQclosepotential} as well the (extreme and nonphysical) case of an unscreened potential, which maximizes the number of close pairs, to allow for a better judgement of the influence of the potential in the calculation presented here.

The influence of the $Q\bar Q$ potential, both for the collision rate and for the local rate,  will be studied in section \ref{section6}. 
\section{$J/\psi$ Production in pp and Initial $c\bar{c}$ State Calibration}
\input{pp}

\section{Results}
\input{results}
\section{Conclusions}
\input{Conclusions}

\section*{Acknowledgement}
\input{acknowledgement}
\section*{Appendix}
\input{appendix}

\bibliography{jpsi.bib}

\end{document}

%% file: pp.tex
Although there are some well established models  and formalisms 
available to deal with quarkonia formation in individual pp collisions, we have chosen, for consistency reasons, to evaluate this production in the same Wigner density coalescence approach as the one used for AA collisions, hence following \cite{Song:2017phm}. 

For this purpose, we start from the double differential spectrum eq.~\ref{eq:densitylabcm} or its Monte Carlo equivalent eq.~\ref{eq:densitylabcmMC}, considering that one single $Q\bar{Q}$ pair is produced, so normalizing $\frac{d^9N_{Q\bar Q}}{dY d^2u_{T}d^3q^{\rm lab}d^3x^{\rm lab}_{r}}$ to unity. In our present approach, we neglect momentum correlations between the initial $Q$ and $\bar{Q}$ quarks, although they can become significant at large $p_T$. Accordingly, we express more conveniently 
\begin{equation}
 \frac{d^9N_{Q\bar Q}}{dY d^2u_{T}d^3q^{\rm lab}d^3x^{\rm lab}_{r}} = 
\mathcal{J}\,e_{Q\bar{Q}}^{\rm lab}\,
\frac{d^3\tilde{N}_{Q\bar{Q}}}{d^3x_r^{\rm lab}} 
\times
\frac{d^3N_{Q}}{d^3p_Q^{\rm lab}}
\times
\frac{d^3N_{\bar{Q}}}{d^3p_{\bar{Q}}^{\rm lab}}
\label{eq:spectrum_base_uT}  
\end{equation}
where $\frac{d^3\tilde{N}_{Q\bar{Q}}}{d^3x_r^{\rm lab}}$ is the normalized distribution to observe a distance ${\bf x}_r^{\rm lab}$ between the $Q$ and $\bar Q$ quark. $\mathcal{J}=s+u_T^2\frac{\partial s}{\partial u_T^2}$ is the Jacobian of the variable transformation. As the longitudinal $z$-space is Lorentz contracted, we moreover assume that
\begin{equation}
\frac{d^3\tilde{N}_{Q\bar{Q}}}{d^3x_r^{\rm lab}} \propto
\delta(z)\,e^{-\left(\frac{x_{rT}^{\rm lab}}{\sigma_r}\right)^2}
\end{equation}
where $x_{rT}^{\rm lab}$ is the transverse initial distance between $Q$ and $\bar{Q}$ quark. A last assumption is to consider factorisation of the individual $Q$ and $\bar{Q}$ production along longitudinal and transverse direction according to
\begin{equation}
\frac{d^3N_{Q}}{d^3p_Q^{\rm lab}}=
\frac{d^2\tilde{N}_{Q}}{dp_{T,Q}^2} \frac{d\tilde{N}_{Q}}{dp_{L,Q}}
=\frac{1}{e_Q} \frac{d^2\tilde{N}_{Q}}{dp_{T,Q}^2} \frac{d\tilde{N}_{Q}}{dy_{Q}},
\end{equation}
where both distributions are normalized to unity and where the transverse spectrum is taken from the FONLL approach. From these hypothesis, it is possible to establish that
\begin{equation}
\frac{dN_{\Phi}}{dy} \approx P_{\Phi} \times \frac{d\tilde{N}_{Q}}{dy} \frac{d\tilde{N}_{\bar{Q}}}{dy}, 
\end{equation}
where all rapidity-distributions are evaluated at the same rapidity and where $P_{\Phi}$ represents a kind of conditional probability for a quarkonia $\Phi$ to be formed in the $Q\bar{Q}$ Wigner density coalescence, see eq.~\ref{eq:rho2} where $P(t)$ is in this case time independent. It can be evaluated semi-analytically once the width $\sigma_{1s}$ in eq.~\ref{eq:WignerMC}, $\sigma_r$  and the $p_T$ spectrum of the individual quarks are specified. It is important to realize that in our Wigner density coalescence model, the quarkonium production at a given rapidity scales quadratically with the local abundance of heavy quarks. Next, one obtains an equivalent relation for the production cross section in pp by noticing a) that the conditional probability to produce a $Q\bar{Q}$ pair in such collision is $\frac{\sigma_{Q}}{\sigma_{\rm tot}}$ (${\sigma_{\rm tot}}$ being the total pp cross section) and b) that the normalized distribution $\frac{d\tilde{N}_{\bar{Q}}}{dy}=
\frac{d\sigma_Q/dy}{\sigma_Q}$. The average distribution per pp collision is therefore 
\begin{equation}
\frac{dN_{\Phi}}{dy} \approx \frac{P_\Phi}{\sigma_{\rm tot} \sigma_Q} \times \frac{d\sigma_{Q}}{dy} \frac{d\sigma_{\bar{Q}}}{dy} 
\end{equation}
leading to 
\begin{equation}
\frac{d\sigma_{\Phi}}{dy} \approx \frac{P_\Phi}{\sigma_Q} \times \left( \frac{d\sigma_{Q}}{dy} \right)^2  \Leftrightarrow \frac{\frac{d\sigma_{\Phi}}{dy} }{\frac{d\sigma_{Q}}{dy} } = P_\Phi \times \frac{\frac{d\sigma_{Q}}{dy}}{\sigma_Q}, 
\label{eq:sigmaphiinpp}
\end{equation}
where the last factor can be seen as an inverse rapidity width. The last relation allows to calibrate the model using experimental results for $c$ quarks. In our case, we only adjust the $\sigma_r$ parameter because $\sigma_{1s}=\sigma_\Jm$ is constrained by the vacuum wave function to $\sigma_{1s}=0.35\,{\rm fm}$ (see eq.~\ref{eq:WignerMC}). For $y_{\rm CM}\approx 0$, one takes $\frac{\frac{d\sigma_{c}}{dy}}{\sigma_c}=0.125$, a value in agreement with NLO calculations \footnote{ http://www.lpthe.jussieu.fr/$\sim$cacciari/fonll/fonllform.html}, while $\frac{d\sigma_{J/\psi}}{dy}$ and  $\frac{d\sigma_{c}}{dy}$ were respectively taken as 6 $\mu$b and 1.165 mb following \cite{ALICE:2019pid} and \cite{ALICE:2021dhb}. Tuning $\sigma_r$ in our MC code, we obtain the corresponding $P_{J/\psi}$ for $\sigma_r=0.25\,{\rm fm}$, which is a reasonable value according to the $m_c$ scale. Once the parameters are fixed, the $p_T$ distribution of $J/\psi$ production in pp can be calculated without further assumption. It will be discussed in section \ref{section_scaled_pp}.

%% file: results.tex
\label{section6}
\subsection{Preliminary remarks }
\label{section6a}
As discussed in previous sections, the production and disintegration of \Js~  is a complex process. Therefore we start out with a short overview and some definitions.

Whereas in pp collisions the $c$ and $\bar c$ in the \J come almost exclusively from the same interaction vertex, in heavy ion reactions, when several $c\bar c$ pairs are produced, this is not necessarily the case. In our analysis we call those \J, which contain a $c\bar c$ pair
from the same vertex, {\it diagonal} \J, the others are called {\it off-diagonal}. In central heavy ion collisions the system forms a QGP. If its  temperature is higher than $T_{\rm diss}$ the \J are not stable and a bound state cannot be formed. At these high temperatures only unbound \cc exist, which interact, however, among each other and with the QGP constituents. We call the distribution of $c$ and $\bar c $  at the moment of their production in initial hard collisions {\it primordial distribution}. The distribution at $T_{\rm diss}$, when \J formation starts, is named {\it initial distribution}.  
The distribution of \cc at that moment differs considerably from their primordial distribution, due to collisions of the \cc  with the QGP constituents, due to the potential interaction between the \cc and due to free streaming of the \cc quarks. Therefore, when applying the same \J Wigner density to the primordial and to the initial  distribution of the heavy quark pairs, to determine the \J yield, we expect large differences. 

Below $T_{\rm diss}$ the \J rate has two contributions:
The {\it collision rate}, which is a consequence of the $c$ or $\bar c$ collision with the QGP partons, described by the Remler formalism, and the {\it local rate}, a consequence of the change of the width of the \J Wigner density with temperature and hence with time. The rates are non-zero until the QGP hadronizes. During the hadronization of the QGP no further \J will be produced. We neglect here also hadronic rescattering of the \J.

From a more theoretical viewpoint it is known (see for instance \cite{Blaizot:2018oev}) that compact white objects do not interact with the QGP. The interaction rate of those objects in a QGP increases quadratically with their size $r_{\rm sing}$, until $r_{\rm sing} \sim l_{\rm corr}$, the correlation length of gluon thermal fields. From this value on, both, $Q$ and $\bar{Q}$, interact independently with the QGP. 
In other words, an interaction of the \J with the QGP gluons is only possible if their wave length is smaller than $r_{\rm sing}$. Otherwise a gluon does not see the individual color of the color neutral $c \bar c$ dipole. 

One may consider that the traditional models, based on a non-interacting initial singlet component, and our approach explore the both extreme facets of the more involved reality. Presently only
the finite value of the elliptic flow, observed in experiments and discussed in section \ref{sec:flow}, presents strong evidence that the \J or its constituents interact with the partons of the QGP.

\subsection{Scaled Proton-Proton  Production}

\label{section_scaled_pp}
One of the key observables in the study of \J production in heavy ion collisions is the nuclear modification factor \be
R_{AA}^{\Jm}(p_T)= \frac{\frac{d\sigma^\Jm_{AA}}{dp_T}}{N_{\rm coll}\frac{d\sigma^\Jm_{pp}}{dp_T}}.
\label{eq:RAA}\ee 
$N_{\rm coll}$ is the number of the initial hard pp collisions. The $R_{AA}$ calculation requires the knowledge of $N_{\rm coll}\frac{\sigma^\Jm_{pp}}{dp_T}$. 
In practice, we do not simulate pp calculations separately. Instead, to obtain $N_{coll}\frac{d\sigma^{\Jm}_{pp}}{dp_T dy}$, we can use the initial $p_T$ and $y$ distribution of the $c$ and $\bar c$ quarks in AA collisions, neglecting all possible cold nuclear matter effects (as for instance shadowing). For diagonal pairs the distributions are then - up to a constant -  identical for pp and AA. 
\be
N_{coll}\frac{dN^\Jm_{pp}}{dp_Tdy} =
\frac{dN^{\Jm,{\rm diag}}_{AA}}{dp_T dy}
\ee
where the rhs is evaluated in the initial stage of the evolution with the help of eq.~\ref{eq:densitylabcmMC}, with $u_{T,J/\psi}=\frac{p_T}{M_{J/\psi}}$, selecting $c$ and $\bar c$ coming from the same vertex. We recall that in  eq.~\ref{eq:densitylabcmMC} $\bf{r}^{\rm cm}$ and $\bf{q}^{\rm cm}$ are the relative distance in coordinate (momentum) space of the $Q$ and $\bar Q$ in the system defined by $\{y_\Phi,{\bf u}_{T,\Phi}\}$, the rapidity and the transverse components of the 4-velocity of the quarkonium, while $W_{\rm NR}$ is the Wigner density defined in $Q\bar Q$ CM system .

\begin{figure}[h]
\includegraphics[width=1.30\linewidth]{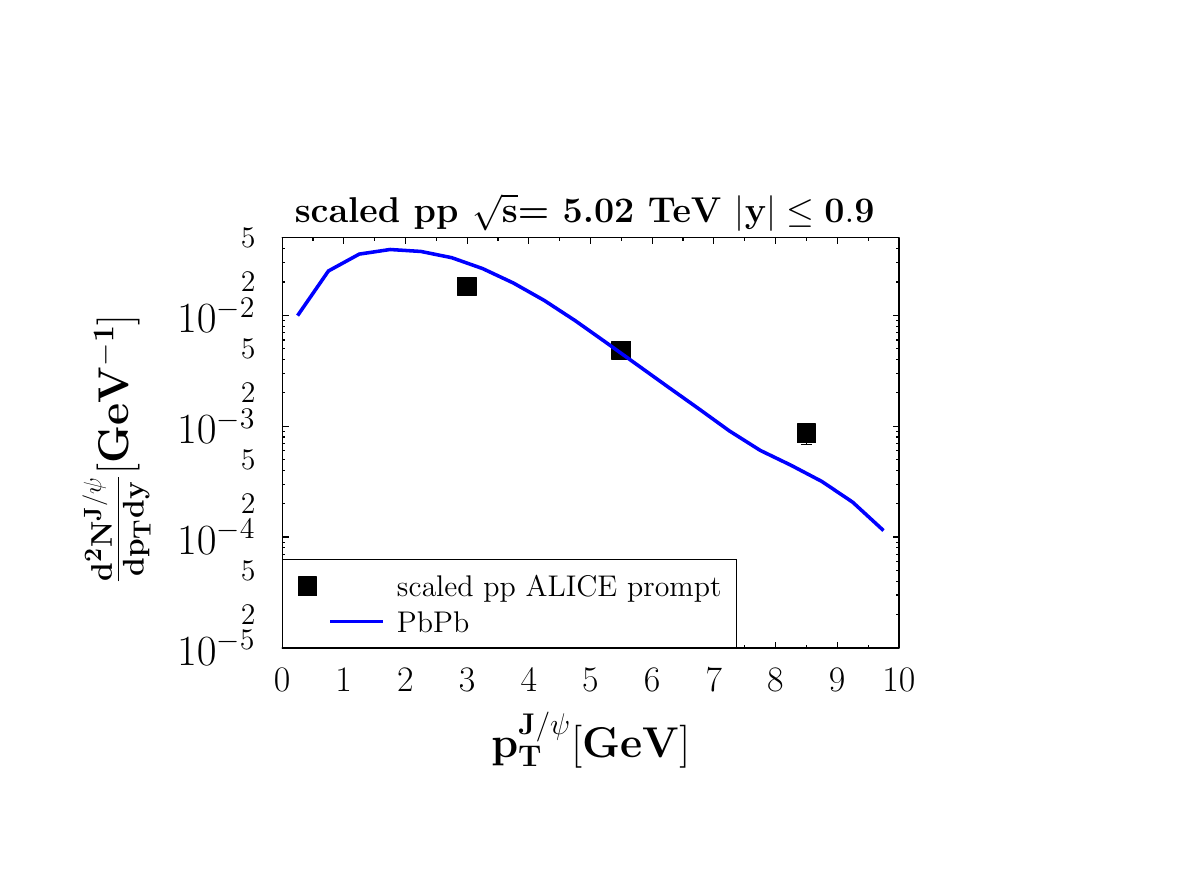}
      \caption{Comparison of our model prediction for the primordial $J/\Psi$ distribution with the prompt experimental pp cross section from the ALICE collaboration \cite{ALICE:2012vup}, scaled by  $T_{PbPb}= 20.55$ $mb^{-1}$. Cold nuclear matter effects are not considered here.}
      \label{Figppmidrap1}
\end{figure}

Neglecting cold nulcear matter effects, we can compare our primordial diagonal A+A distribution with the one obtained in the same conditions using the experimentally measured  pp cross section on prompt \J production 
\be
\frac{d\sigma ^\Jm_{pp}}{dp_Tdy} \cdot T_{{AA}} =
\frac{dN^{\Jm,{\rm diag}}_{AA}}{dp_T dy}.
\label{scale}
\ee
For the [0-20\%] centrality class, we display in Fig. \ref{Figppmidrap1} the midrapidity $p_T$ distribution of \Js~ in central PbPb collisions at $\sqrt{s} = 5.02\ {\rm TeV}$. The result is compared with the prompt pp data of the ALICE collaboration \cite{ALICE:2021edd}, scaled by eq.~\ref{scale} with an associated nuclear overlap function $T_{\rm PbPb}$ of 20.55 ${\rm mb}^{-1}$, compatible with the one extracted from EPOS2. The data points are marked as full squares, our MC results are given by a blue line.
We see that our approach reproduces quite nicely the experimental data. Deviations are seen at large $p_T$. This may be due to the fact that, in lack of having a better approach available,  the
$c$ and $\bar c$ are created uncorrelated, in $p_T$ as well as in the azimuthal angle. It is also important to emphasize  (as was mentioned in chapter 4) that our model calculates the direct $J/\psi$ production. Therefore, one has to be cautious to compare our results with the experimental prompt data since  decay from excited quarkonium states contribute to the spectra. For forward rapidities the ALICE results \cite{ALICE:2021qlw} show that this contribution is about 15\%.

\begin{figure}[h]
   \centering 
   \includegraphics[width=1.30\linewidth]{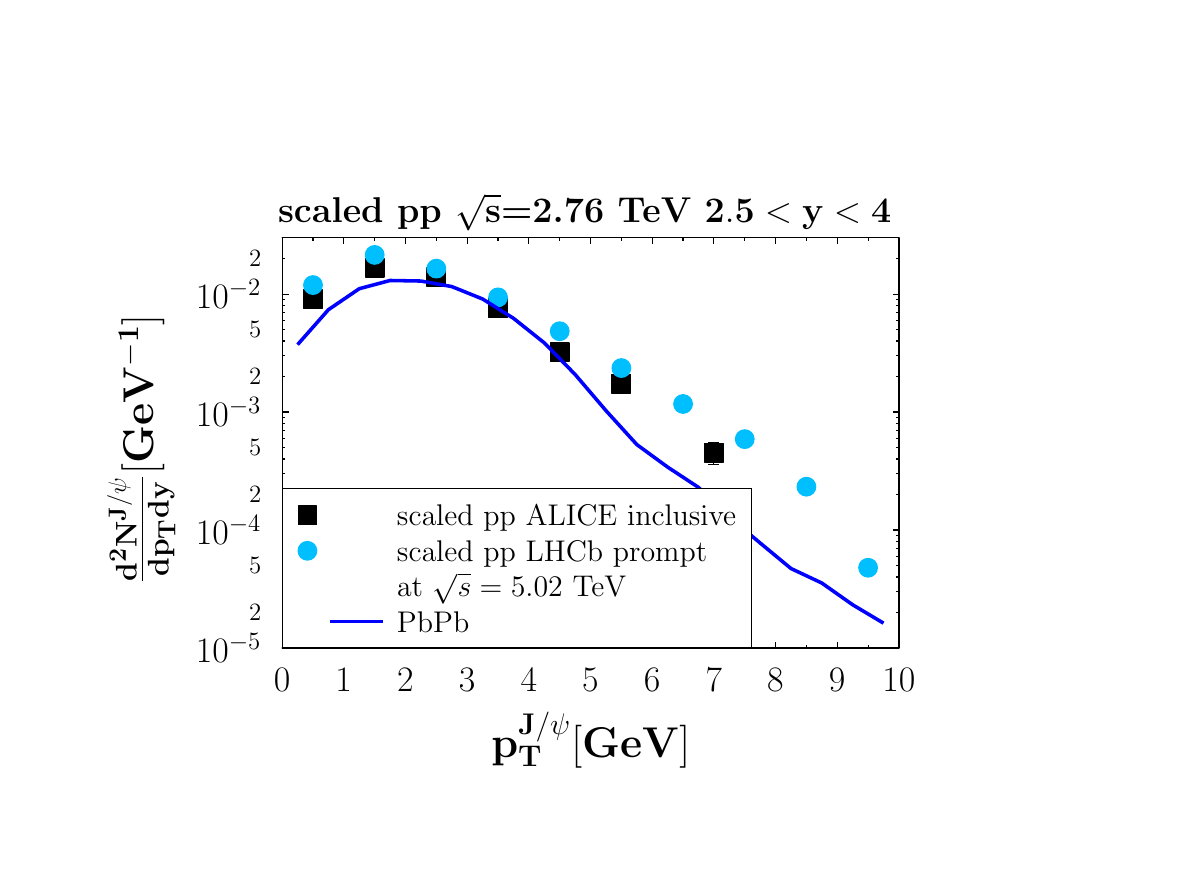}
      \caption{Comparison of our model prediction for the primordial $J/\Psi$ distribution in PbPb with the inclusive experimental pp cross section for forward production from the ALICE collaboration \cite{ALICE:2012vup} as well as with the prompt pp cross section, measured by the LHCb collaboration \cite{LHCb:2021pyk} (for $\sqrt{s}=5.02$ TeV), scaled by  $T_{PbPb}= 20.55$ $mb^{-1}$.} 
    \label{Figpforwardrap}
\end{figure}

In Fig. \ref{Figpforwardrap}, we display the same analysis for the forward rapidity ($2.5 \le y \le 4$) data at $\sqrt{s} = 2.76\ {\rm TeV}$. Since the non-prompt \J fraction increases from roughly 0.08 to 0.2 in the displayed $p_T$ interval \cite{LHCb:2011zfl} we expect deviations between our results for prompt \J and the inclusive experimental results at higher $p_T$ values. We display as well the prompt cross section, measured by the LHCb collaboration \cite{LHCb:2021pyk}, for $\sqrt{s}=5.02$ TeV.

We can conclude from this comparison that our formalism reproduces the \J production in elementary pp collisions. Thus we confirm the results obtained in \cite{Song:2017phm}, although the details of the modelling differ slightly.

\subsection{Heavy Ion Collisions}
\subsubsection{Initial Distribution}
\JA{In our simulations of heavy ion collisions, a global shadowing of 50\% is applied for most of the observables analyzed in this section. For the production of $c$ and $\bar{c}$ quarks at low $p_T$, such a value leads to a D-meson production compatible with the results stemming from the more sophisticated EPS09 shadowing combined with MC$@_s$HQ in \cite{Nahrgang:2016lst}. } The \Js~, which  would be produced in absence of a medium in the initial hard NN collisions, are not observed finally because they dissolve into \cc quarks when they pass the (high temperature) QGP. \cc can only form a stable \J when the QGP temperature falls below $T^{\Jm}_{\rm diss}$. In practice we apply  the following description: When a $c$-quark arrives for the first time in a region with $T<{T^{\Jm}_{\rm diss}}$ we calculate its probability to form a $J/\psi$ with all $\bar{c}$, which are already satisfying this condition:
\begin{equation}
N_k^{\Jm,{\rm init}}=\sum^{n_{\bar c}}_{l=1} W_{{\rm NR}}^\Jm(r^{cm}(k,l),q^{cm}(k,l)).
\label{eqninstantaneousprim}
\end{equation}
In this expression, $n_{\bar c}$ is the number of active charm anti-quarks (means from a region of the QGP with $T<{T^{\Jm}_{\rm diss}}$), $k$ is the index of the $c$-quark which has passed the dissociation temperature at time $t$, $l$ is the index of a $\bar c $ quark which is active. $r^{cm}(k,l)(q^{cm}(k,l))$ stand for the relative distance in coordinate  (momentum) space of the $\{k,l\}$ pair in the pair center of mass system.  The sum of all these contributions for all $c$-quarks (and analogously those for all $\bar c$ quarks) is the initial \J distribution 

We can also define the initial rapidity distribution of the \Js~, which contain a $c$ or $\bar c$ quark, which passed $T_{\rm diss}$ between $t$ and $t+\Delta t$ (the time-step used for the Remler algorithm):
\bea
&&\frac{dN^{\rm init}}{dy}(t,t+\Delta t)= \sum^{N_{\rm first}(t,t+\Delta t)}_{k=1}\sum^{n_{\bar c}(n_{c})}_{l=1} \n &\times&\int d^2u_T  W(y,u_T,r^{cm}(k,l),q^{cm}(k,l))
\label{eqntotalpriminstant}
\eea
where $N_{\rm first}(t,t+\Delta t)$ stands for the number of $c$ or $\bar c$ quarks, which passed the temperature threshold $T_{\rm diss}$ between t and $t+\Delta t$. $ n_{\bar c}(n_{c})$ is the number of $\bar c (c)$ quarks in cells below $T_{\rm diss}$. 

\begin{figure}
    \centering
     \includegraphics[width=1.3\linewidth]{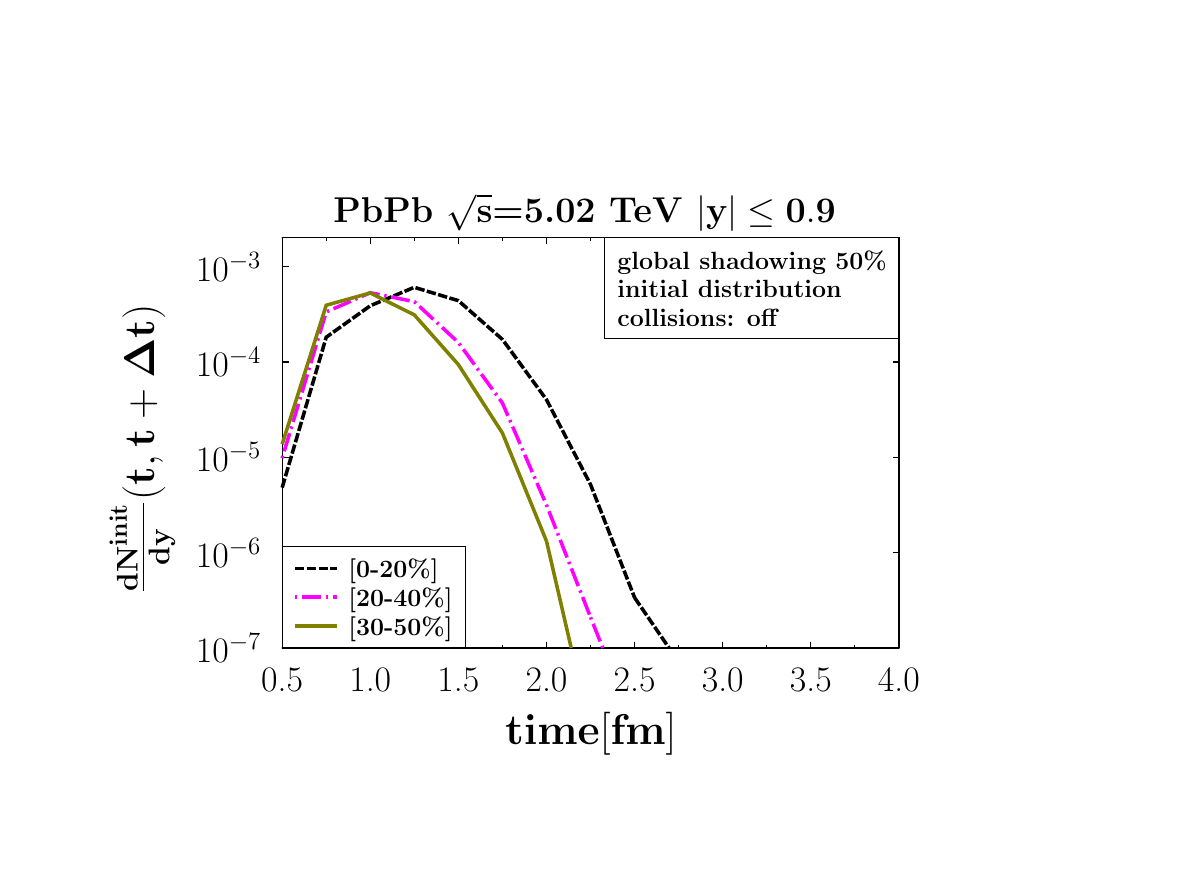}
     \includegraphics[width=1.3\linewidth]{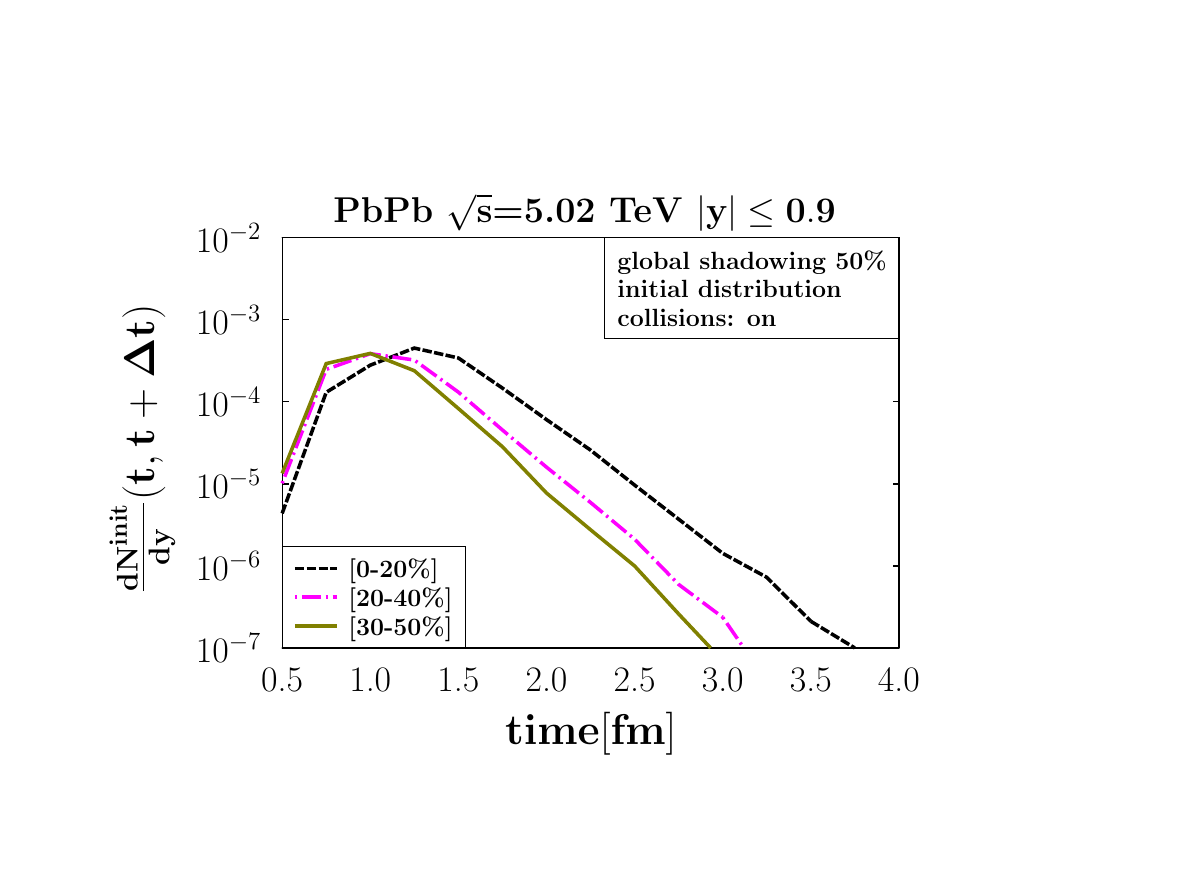}     
     \caption{Time evolution of the  $J/\psi$ initial production for different centralities and for two scenarios: without collisions of $c$ and $\bar c$ with QGP partons (top) and including these collisions (bottom). $\Delta t$ is taken as 0.25 fm/c.}
     \label{primordialweighted}
\end{figure}

The time evolution of the initial \J production at midrapidity, $\frac{dN^{\rm init}}{dy}$, is shown in Fig.\ref{primordialweighted}. \JA{We display this quantity, normalized to $(\frac{dN^{c}(t)}{dy})$ for different centrality intervals and for the reaction PbPb at $\sqrt{s} = 5.02\ {\rm TeV}$ (black dashed line for [0-20\%], red dashed-dotted line for [20-40\%], olive full line for [30-50\%]). Two well-defined limited cases can be identified: If \Js~ are produced in individual NN collisions, the production of charmonia scales with the total charm production and happens in the initial stage, while in a rate equation approach, assuming a system of fixed volume, the creation of \J would be proportional to $(\frac{dN^{c}(t)}{dy})^2$ and pretty much independent of time. }

\begin{figure}[h]
   \centering 
    \includegraphics[width=1.3\linewidth]{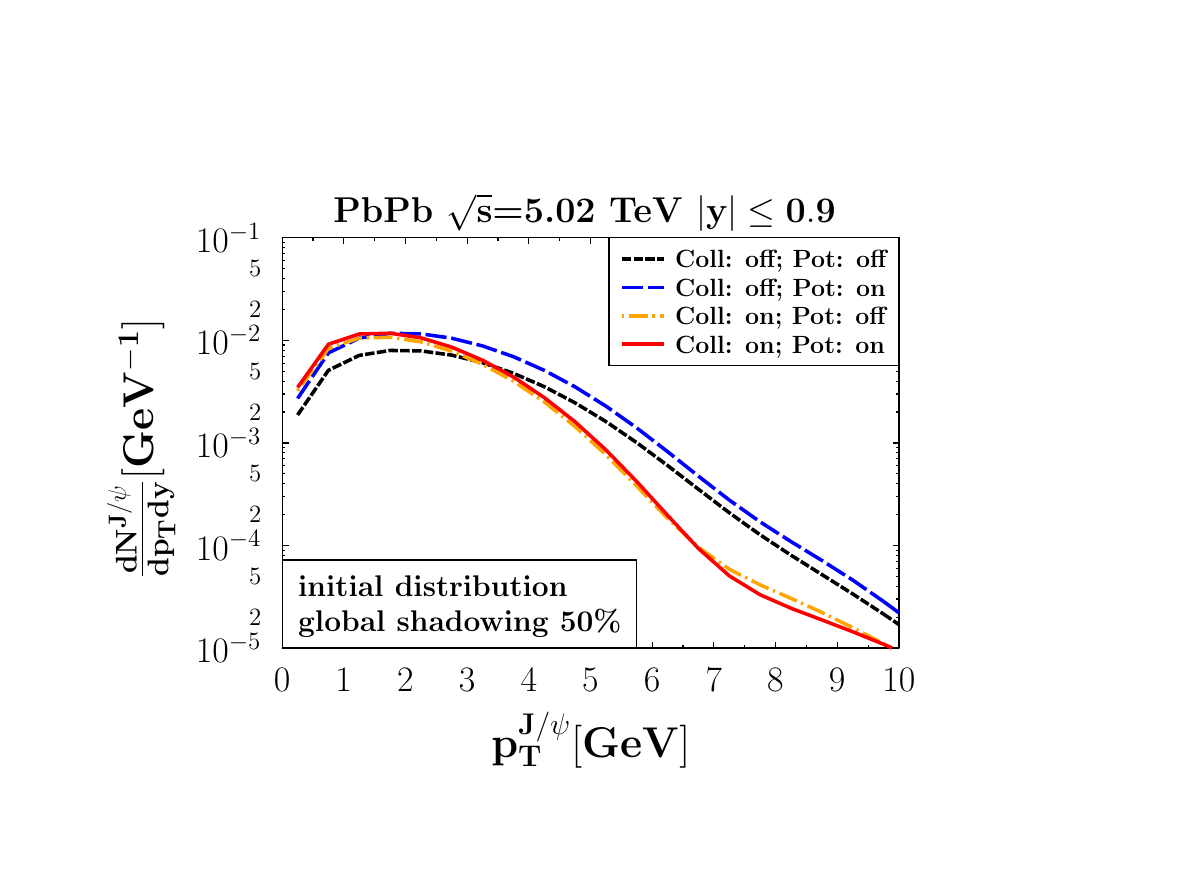} \caption{Time integrated initial $p_{T}$ distribution of $J/\psi$ in  midrapidity [0-20\%] central PbPb collisions at $\sqrt{s}=5.02$ TeV, obtained for the four different evolution conditions combining $c\bar c$ potential interaction ($ON$ and $OFF$) and elastic collisions ($ON$ and $OFF$) with QGP partons.}
    \label{figintegralprimordialpt}
\end{figure}
\JA{We see in Fig.\ref{primordialweighted} more the first type of production with however a distribution of times \JA{over a time interval of 1-2 fm } when a $c$ or $\bar c$ quark can form a \J for the first time, in mid-central as well as in central collisions. In central collisions the QGP is larger and therefore present for a longer time. Therefore the distribution is shifted to later times. All together, the necessary time to pass below $T_{\rm diss}$ is short enough that the $c$ and $\bar{c}$, which are produced far apart, do not have the chance to encounter during the first 2 fm/c, what explains the "canonical scaling" with $\frac{dN^{c}(t)}{dy}$ for the initial contribution.}
These distributions are rather independent on whether the potential interaction between the \cc is active or not but depend quite strongly on whether we admit collisions of the heavy quarks with the QGP partons. As we will see below, these collisions lower the $p_T$ momentum of the heavy quarks and therefore decelerate the expansion. As a consequence, they stay longer in the hot phase. In addition, heavy quarks with lower momenta have a higher chance to form a \J . 

We come now to the $p_T$ distribution of the initial \Js. It is displayed for $|y| \le 0.9$ and for [0-20\%] central  PbPb collisions at $\sqrt{s}=5.02\ {\rm TeV}$ in Fig. \ref{figintegralprimordialpt}. We display 4 different scenarios to show the consequences of the collisions of \cc with the QCD constituents and of the
presence of the potential between the \cc.
The black dashed line shows our result if neither collisions occur nor the $c\bar c$ potential is active. For the blue long dashed line we switched on the $c\bar c$ potential. The orange dashed dotted and the red full line -- which corresponds to our full model --  show the results without and with $c\bar c$ potential if collisions between the heavy quarks and QGP partons are admitted.

The potential has little influence on the initial $p_T$ spectrum but collisions shift the \J distribution to lower $p_T$ values, what corresponds to a $c$($\bar{c}$) quenching between the time when the $c$($\bar c$) enters the QGP and the "initial" time (when $T=T_{\rm diss})$. The knee in the calculations with collisions reflects the fact that c-quarks below 4 GeV/c are thermalized or in the process of thermalizing while those with a larger $p_T$ get decelerated but not thermalized.

\subsubsection{Impact of the potential on the correlations between the \cc quarks}
As discussed in section 3, the \cc quarks interact via a potential interaction whose parameters are taken from \cite{Lafferty:2019jpr}. We calculate the heavy quark trajectories in Minkowski space using the equations of motion of eq.~\ref{tee} after boosting the pair into their cm frame. 
\begin{figure}[h]
   \centering 
   \includegraphics[width=1.30\linewidth]{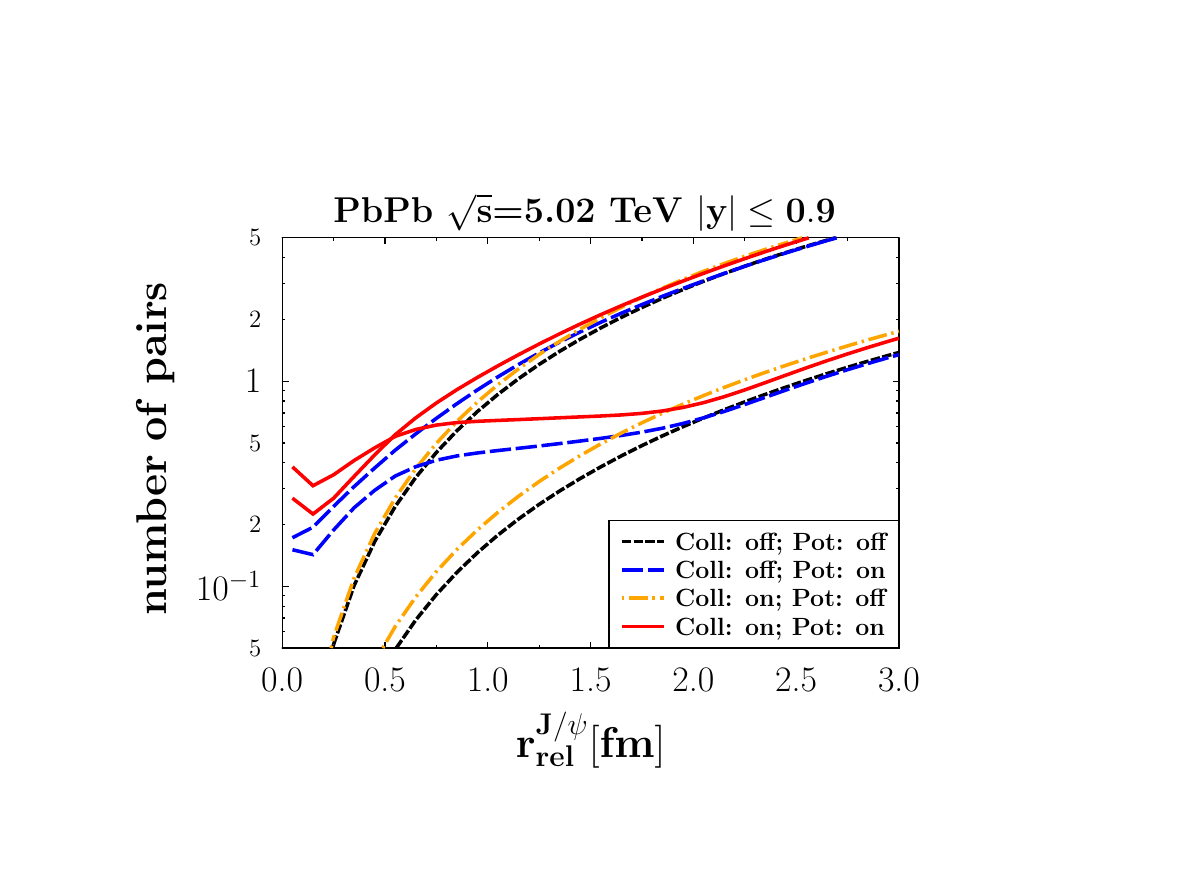}
   \caption{Number of $c\bar c$ pairs 
   as a function of $r_{rel}^{J/\psi}$, for t=4 fm/c (top lines) and t=8 fm/c (bottom lines). The color coding is the same as in Fig.\ref{figintegralprimordialpt}.}
    \label{HQclosephasespace}
\end{figure}
To show the influence of this potential we display in Fig. \ref{HQclosephasespace}, the number of $c \bar c$ pairs as function of their relative distance in their center of mass, $r_{\rm rel}^{J/\psi}$ for central reactions of PbPb at $\sqrt{s}= 5.02\ {\rm TeV}$ and at midrapidity, $|y|\le 0.9$. The line coding is the same as in Fig.\ref{figintegralprimordialpt}.
The 4 top lines show the number of pairs at 4 fm/c, the 4 bottom lines that at 8 fm/c. 
We expect larger $r_{rel}$ values at t=8 fm/c because the system is expanding. We see that especially at 8 fm/c the potential interaction leads to a much larger number of  $c\bar c$ pairs with a small relative distance, which are susceptible to form a \J.
The influence of collisions on the distributions is more subtle. For large distances, where the potential is weak, their influence is not strong and all 4 curves join. For distances smaller than 1.5~fm (the potential range), they enhance the correlations if the potential is active. Because such collisions enable the energy transfer from the $c\bar{c}$ internal motion to the medium, they lead to a lowering of the internal energy and thus to a reinforcement of the correlations.


\subsubsection{Impact of the local rate}
In an expanding QGP the temperature decreases as a function of time. Therefore the temperature dependent Wigner density for \J, which we employ, is time dependent. This leads in the Remler formalism, extended to temperature dependent eigenstates (see eq.~\ref{eq:totalrate}), to a local production rate for \J,
\be
\Gamma_{\rm loc}= \frac{d T }{d t}  
\frac{d \sigma}{d T}\frac{d W }{d \sigma}
\ee
where $\sigma$ is the width of the Gaussian Wigner density $W_{{\rm NR}}^\Jm(r^{cm}(k,l),q^{cm}(k,l))$ and $T$ is the temperature of the QGP region in which the \J is located at the time $t$. The \cc may be in regions of slightly different temperature although the most relevant $c\bar{c}$ contributions to $\Gamma_{\rm loc}$ are those for which $T_c \approx T_{\bar{c}}$. In our calculation we take the average value 
\begin{equation}
T_{c\bar c}=\frac{1}{2}(T_{c}+T_{\bar c}).    
\end{equation}
In the numerical program we use a fixed time step $\Delta t$. Therefore we replace
\be
\frac{d\sigma (t) }{dt} \approx \frac{\Delta \sigma (t)}{\Delta t/u^0}   
\ee
where $\Delta t/u^0$ is the computational Minkowski time step measured in the center of mass of the pair. This $c\bar c$ pair contributes to the final \J multiplicity with
\begin{equation}
N^{\rm loc}_{\rm pair} = \int^{\infty}_{t^{\rm first}} \Gamma^{c\bar c}_{\rm loc}(t)dt  \end{equation}
where $t^{\rm first}$ is the time in which the later of the two ($c$ or $\bar c$) passes $T_{\rm diss}$. For the multiplicity we sum over all $c \bar c$ pairs.
\begin{figure}[h]
   \includegraphics[width=1.30\linewidth]{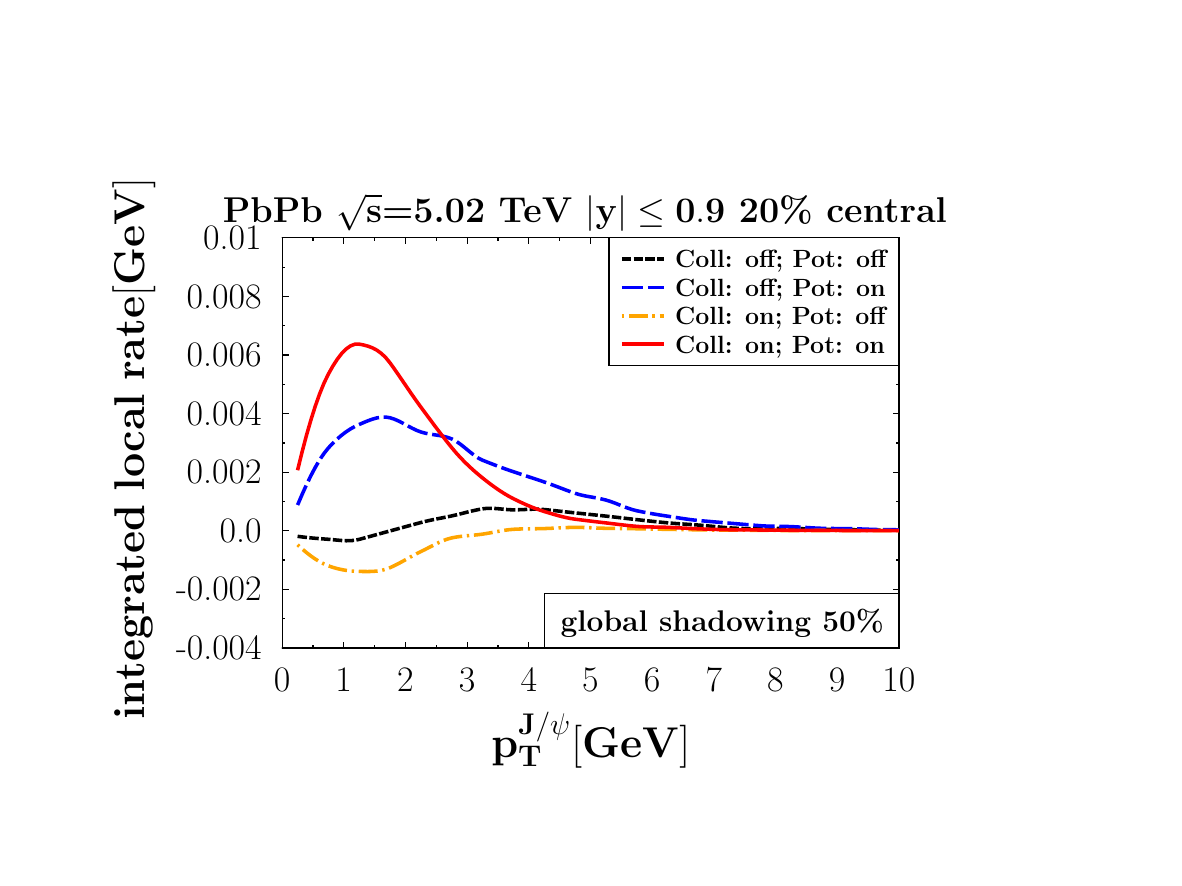}
   \caption{Integral (time integrated) local rate as a function of $p_{T}$ for $J/\psi$ production
   in PbPb in (0-20\%) central PbPb collisions at $\sqrt{s}$ = 5.02 TeV for four different evolution scenarios: $c\bar c$ potential interaction ($ON$ and $OFF$) and elastic collisions with QGP partons ($ON$ and $OFF$). 
   The color coding is the same as in Fig.\ref{figintegralprimordialpt}}. 
    \label{figintegralprimordial}
\end{figure}
In Fig. \ref{figintegralprimordial} we display the $p_T^{J/\psi}$ dependence of the integrated local rate for [0-20\%] central PbPb collisions at $\sqrt{s} =  5.02\ {\rm TeV}$ and for the different scenarios discussed in Fig. \ref{figintegralprimordialpt}.
The color coding is the same as in Fig.\ref{figintegralprimordialpt}.
The Wigner density gets larger with decreasing distance between the \cc. The potential interaction brings \cc quarks closer together (see Fig.\ref{HQclosephasespace})  and therefore it is expected that the local rate contribution gets larger when the potential interaction between the heavy quarks is active. This is indeed seen in the calculations. We see as well that the collisions of the \cc quarks with the QGP partons \JA {influence the local rate, however in a more complicated way. These collisions shift the transverse momentum distribution of the heavy quarks towards lower values (Fig. \ref{figintegralprimordialpt}) and therefore also the relative momentum between the \cc, which enters the Wigner density, gets smaller. If the potential is active collisions reinforces the rate at low $p_T$ whereas if the potential is switched off, collisions make the rate slightly negative.}
Comparing with Fig. \ref{figintegralprimordialpt}, one sees that the local rate is quantitatively of the same importance as the initial production, especially when both, potential and collisions with QGP, are active.


\subsubsection{Impact of the collision rate}
The collision rate $\Gamma_{\rm coll}$ is a central quantity in our approach, not only conceptually (it is responsible for the continuous suppression and production of quarkonia) but also for the numerical values of our results. This will be shown in this section.
Whenever a $c$ or $\bar c$ quark collides at time $t$ with a parton of the QGP, the heavy quark changes its momentum and therefore the overlap with the \J Wigner density is different before and after the collision. These collisions lead only to the production of \J for all $c\bar c$ pairs, which include that scattered heavy quark if the QGP around the heavy quark has a temperature $T<T_{\rm diss}$. In this case the difference $\Delta W = W(t+\epsilon)-W(t-\epsilon)$ is the contribution of this collision to the integrated collision rate of \Js. 

\begin{figure}[h]
   \centering 
   \includegraphics[width=1.30\linewidth]{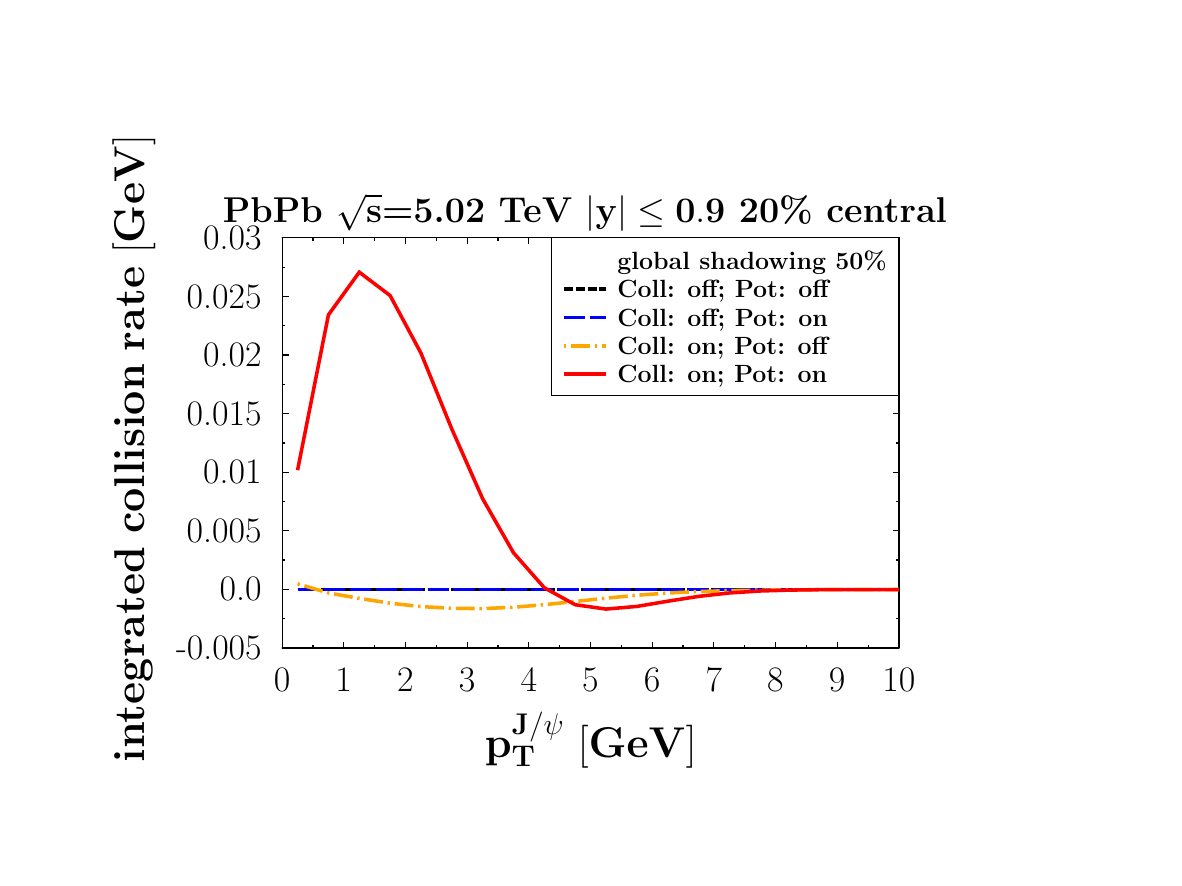}   
      \caption{ $\frac{dN_{\rm coll}}{dp_T}$,  integrated over the collision rate, 
      for the four different evolution scenarios combining $c\bar c$ potential interaction ($ON$ and $OFF$) and elastic collisions ($ON$ and $OFF$) with QGP partons.} 
    \label{collisionrate0020fullset}
\end{figure}

Fig. \ref{collisionrate0020fullset} shows how the integrated collision rate depends on the final center of mass momentum of the \J. We display this rate for central [0-20\%] PbPb collisions at $\sqrt{s}= 5.02 \ {\rm TeV}$ for all \J which have at the end of the reaction a rapidity $|y| \le 0.9$. We use the same coding as for Fig. \ref{figintegralprimordialpt}.
\JA{We see, first of all, that the integrated  collision rate is largely positive for small $p_T$ if collisions and potential are active, means that the multiplicity of \Js~increases. This is a clear sign of regeneration at low $p_T$, while the sign change for $p_T \gtrsim 4\,{\rm GeV}/c$ reflects the shrinking of the underlying $c$ and $\bar{c}$ quarks $p_T$-distributions. On the contrary, if the potential is inactive, the rate is negative in the full $p_T$ range, which may be considered as a consequence of the expansion of the $c$ and $\bar{c}$ spatial distributions. If the collisions are switched off, the collision rate is of course zero. 

For very small values of $p_T$ the momentum transfer is limited by kinematics and therefore the momentum of the heavy quark changes little. For very large \J momenta the cross section leads to a small angle scattering what also limits the possible momentum transfer. In both cases $\Delta W$ is small and so the contribution to the integrated rate is close to zero.}

\subsubsection{Total Multiplicity}
\begin{figure}[h]
   \begin{minipage}{0.48\textwidth}
     \centering
      \includegraphics[width=1.3\linewidth]{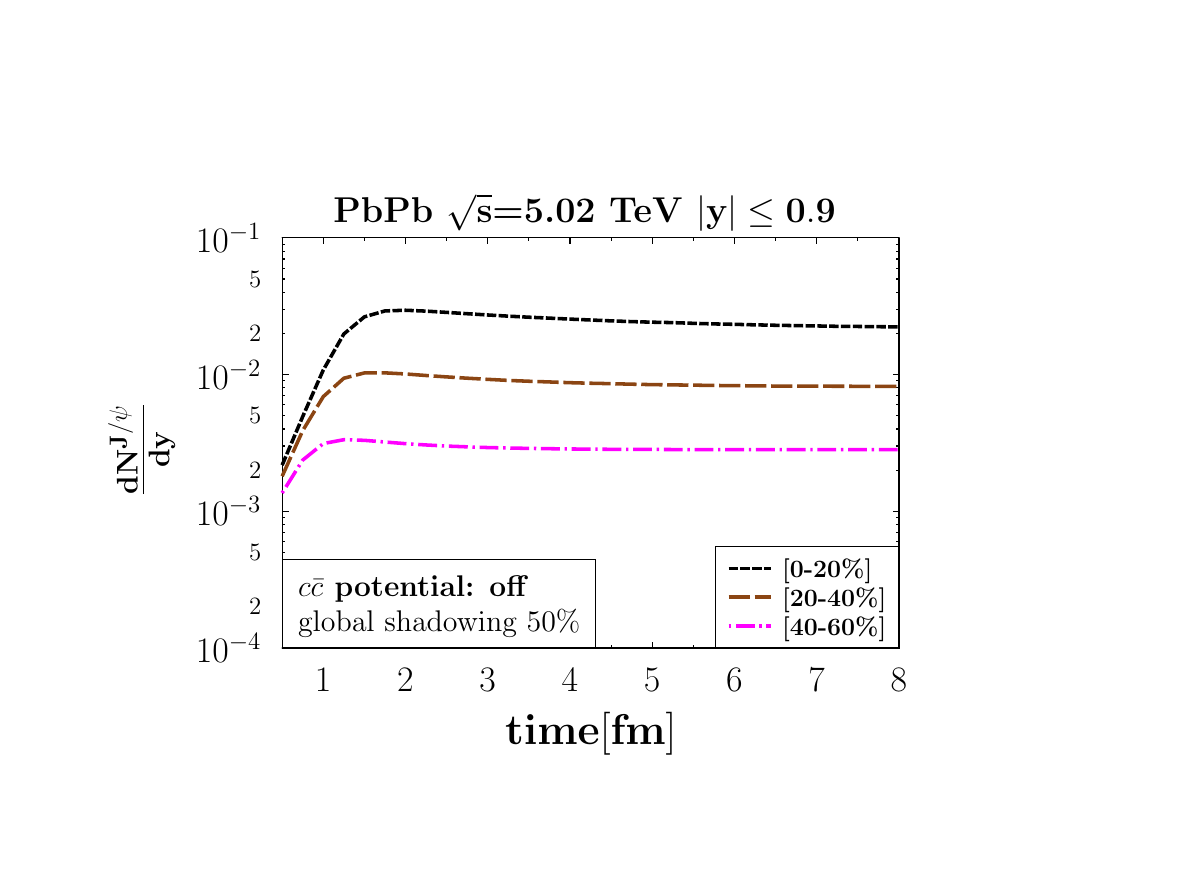}    
   \end{minipage}\hfill
   \begin{minipage}{0.48\textwidth}
     \centering
     \includegraphics[width=1.3\linewidth]{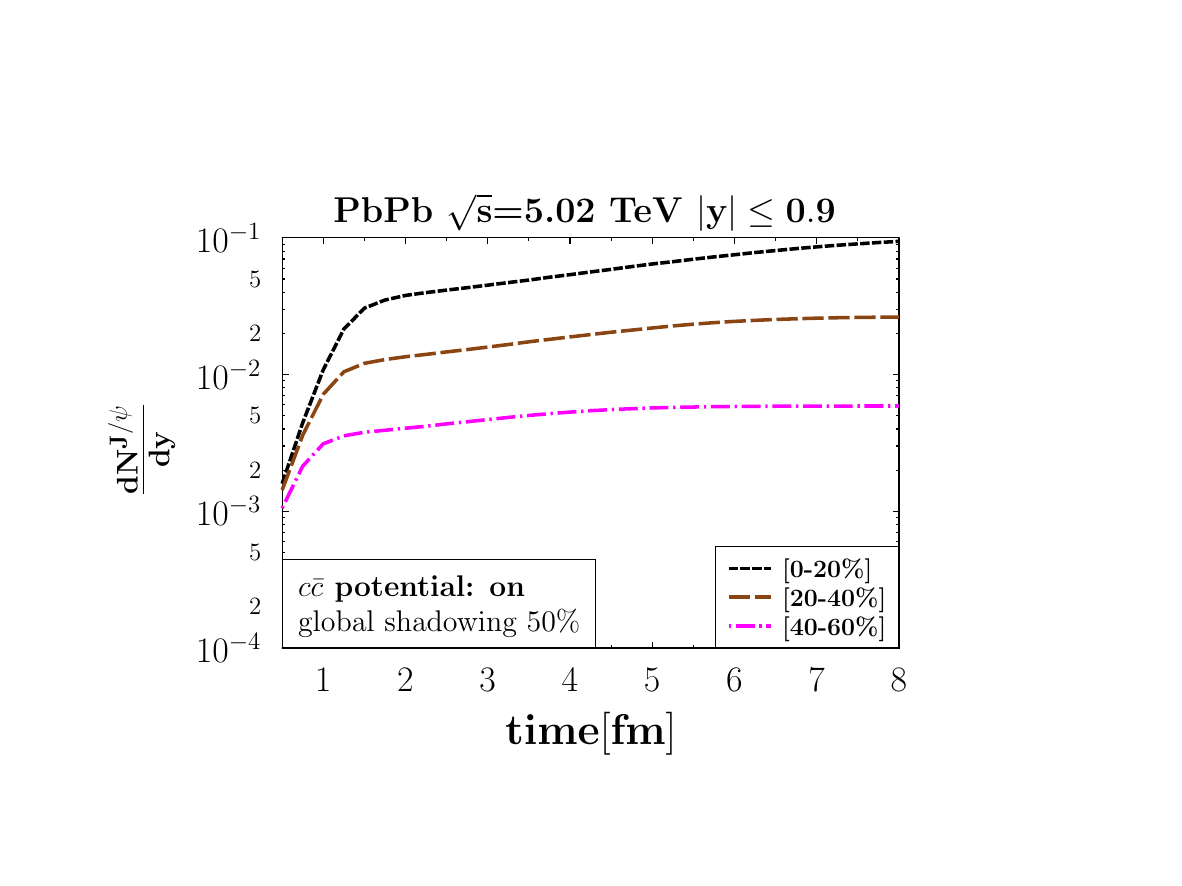}     
   \end{minipage}
   \caption{ Time evolution of the $J/\psi$  production $\frac{dN}{dy}$, integrated over  $p_{T}$, as a function of time for different centrality bins and for two different scenarios: $c\bar c$ interaction potential $ON$ (top) and $OFF$ (bottom) for 
Minkowski and Bjorken time steps 0.25 fm/c and 0.1 fm/c, respectively. } 
   \label{Fig:timeevolutiontotalprob}
\end{figure}

Adding the local and collisional rates as well as the initial production we can study the net-multiplicity, eq.~\ref{eq:modifyremler}, the difference between production and disintegration, of the \J par unit rapidity, $\frac{dN^{J/\psi}}{dy}$. In Fig.\ref{Fig:timeevolutiontotalprob} we display the net-multiplicity as a function of time for three different centralities [0-20\%], black dashed line, [20-40\%], brown long dashed line, and [40-60\%], magenta dashed dotted line.  On top we display this quantity if the potential interaction between \cc is not active, in the bottom figure it is included. We see that the production starts early, when the first \cc pass $T_{\rm diss}$ and increases strongly until around 1 fm/c. Without potential at later times there are more \cc disintegrating than produced, in agreement with the negative $\Delta W$, displayed in Fig.  
\ref{collisionrate0020fullset}. For an active $c\bar c$ potential the $c$ and $\bar c$ stay closer together (Fig. \ref{HQclosephasespace}) with the consequence of a steadily increasing yield until 8 fm/c, 
This increase is rather independent of the centrality of the collision.

Thus our results do not support an instantaneous coalescence mechanism during hadronization, which is the basis of statistical models at fixed volume and which is also frequently applied to describe the total or part of the \J yield  in heavy ion collisions. 
This is one of the key messages of our approach, that we expect to hold irrespective of the 
specific implementation of
the $c\bar{c}$ potential and of the collisions with the QGP partons in transport codes.

\section{Comparison with Experiment}
In this section we compare the production of \J , obtained from 
eq.~\ref{eq:modifyremler}, with the corresponding experimental heavy-ion results and make predictions where data have not been published yet. 
\subsection{$p_{T}$-Spectrum of $J/\psi$}
\begin{figure}[h]
     \centering
     \includegraphics[width=1.3\linewidth]{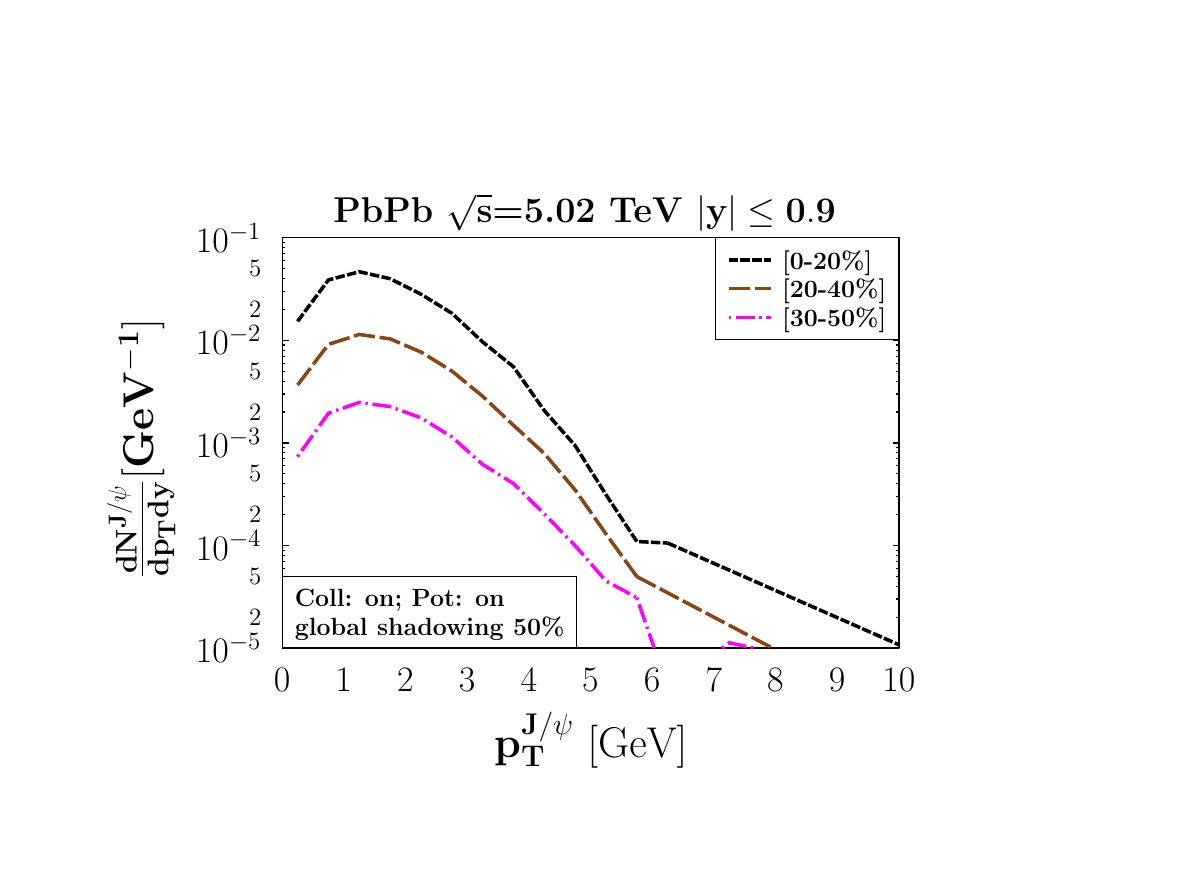}     
   \caption{Final $J/\psi$ $p_T$ spectrum,
   $\frac{dN^{J/\psi}}{dp_Tdy}$, at midrapidity obtained for different centrality bins.  The results were obtained with the standard values of the parameters: 
Minkowski time step value $\Delta t=$0.25 fm/c and Bjorken time step 0.1 fm/c.} 
   \label{Fig:finalprodptspectracent}   
\end{figure}

The calculated midrapidity $p_T$ spectra, for $|y| \le 0.9$, of \J produced in PbPb collisions at $\sqrt{s} = 5.07 \ {\rm TeV}$ is displayed in Fig. \ref{Fig:finalprodptspectracent} for three different centralities [0-20\%], [20-40\%] and [40-60\%].
 Here collisions with the QGP partons as well as the $Q\bar Q$ potential interactions are included. We observe that the form of the curves are rather similar. 
 
\begin{figure}[h]
   \centering 
   \includegraphics[width=1.3\linewidth]{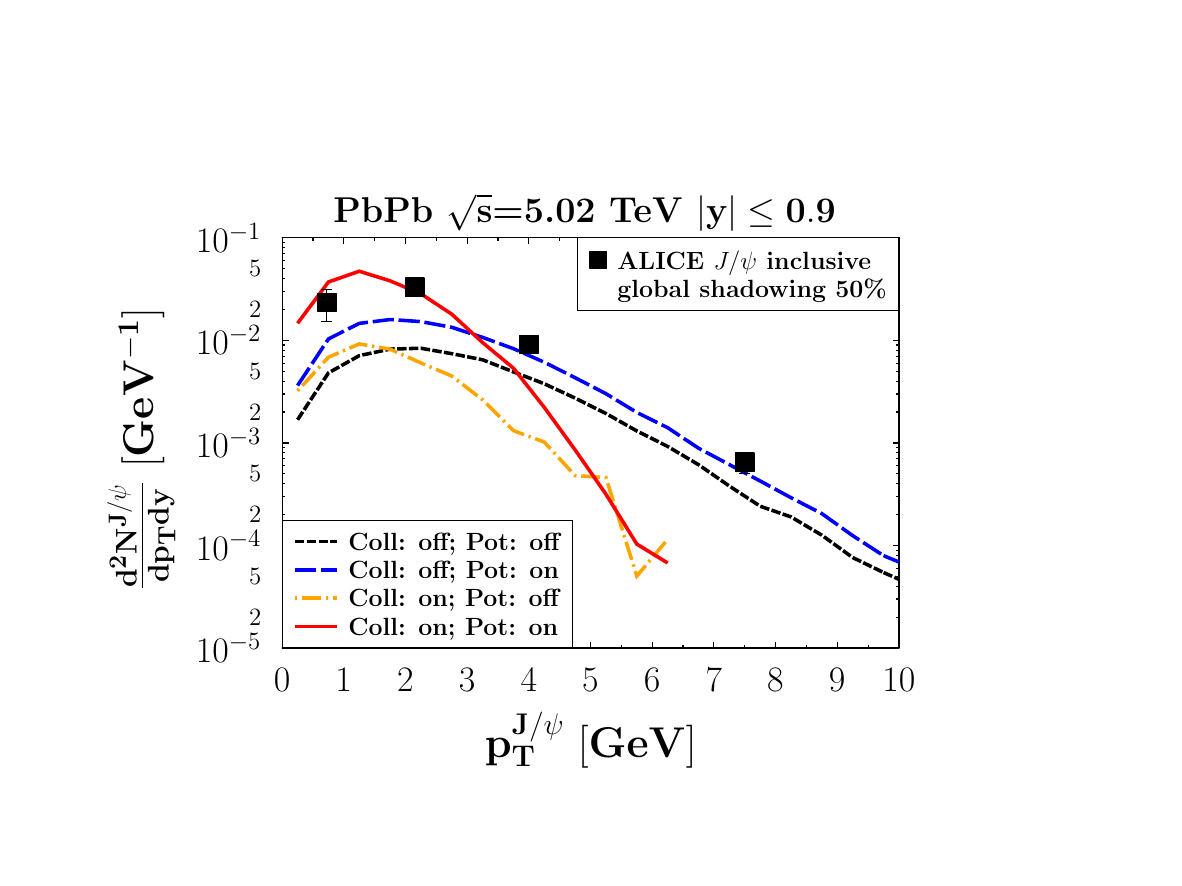} \caption{Final $p_{T}$-spectra at mid-rapidity for [0-20\%] central PbPb collisions, for the four scenarios (switching on/off interaction potential and in-medium elastic collision) and for 
Minkowski and Bjorken time steps of 0.25 fm/c and 0.1 fm/c, respectively. The full squares mark the ALICE experimental data \cite{ALICE:2019nrq}.}
    \label{Finalprobfullset}
\end{figure}

In Fig. \ref{Finalprobfullset} we investigate in detail how the different ingredients of our model influence the final $p_T$ distribution at midrapidity for [0-20\%] central PbPb collisions at $\sqrt{s} = 5.07 \ {\rm TeV}$. We present the results for the four possible combinations if we activate/deactivate collisions and potential. The color code corresponds to that in Fig. \ref{figintegralprimordialpt}. We observe that, if collisions are active, the maximum of the \J distribution is shifted to lower values of $p_T$ as a consequence of the shift of the heavy quark spectra under this condition. We see as well that collisions lead to a much steeper slope at large $p_T$. The $c\bar c$ potential enhances the yield without changing the high $p_T$ slope if there are no collisions.  
If collisions take place, the enhancement of the low $p_T$ yield due to the potential is of the order of a factor of 5 whereas at intermediate $p_T$ the yield changes little. In Fig. \ref{Finalprobfullset} we compare as well our results with the data of the ALICE collaboration \cite{ALICE:2019nrq}, which are shown as black squares. We see that for low $p_T$ our results are close to the experimental data, if both, potential and collisions, are active.
If collisions are active we see at high $p_T$ a much steeper slope than seen in experiments. 
Several features could be at the origin of this difference: the neglect of the feed-down from excited states, the absence of $c\bar{c}$ momentum correlations in the initial production, the disregard of \J production in the corona, the insufficiency  of the description of the potential interaction if the transverse 4-velocity $u_T$ is large,\ldots.  Some of these possible factors of disagreement will be reinvestigated in upcoming publications.

\subsection{$J/\psi$ nuclear modification factor} 
\begin{figure}[h]
   \begin{minipage}{0.48\textwidth}
     \centering
     \includegraphics[width=1.3\linewidth]{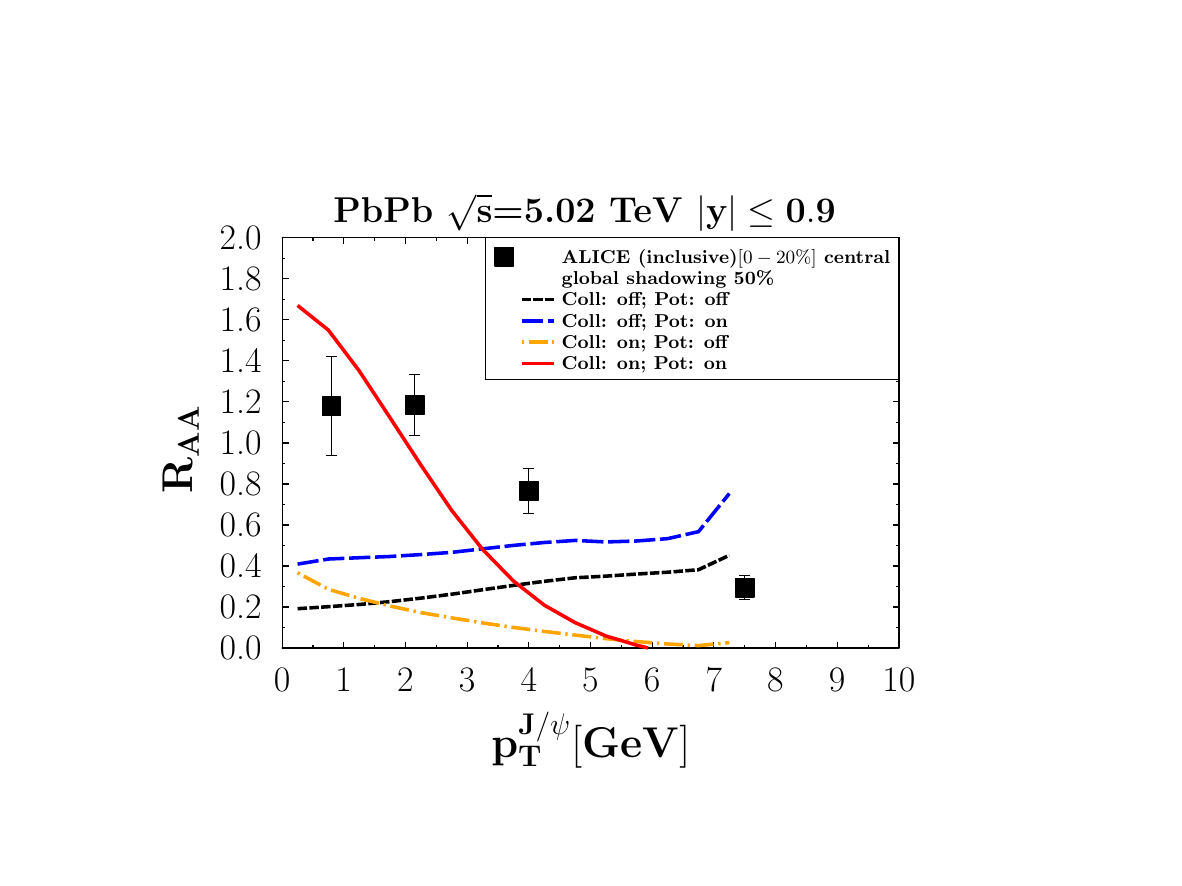}   \end{minipage}\hfill
   \begin{minipage}{0.48\textwidth}
     \centering
     \includegraphics[width=1.3\linewidth]{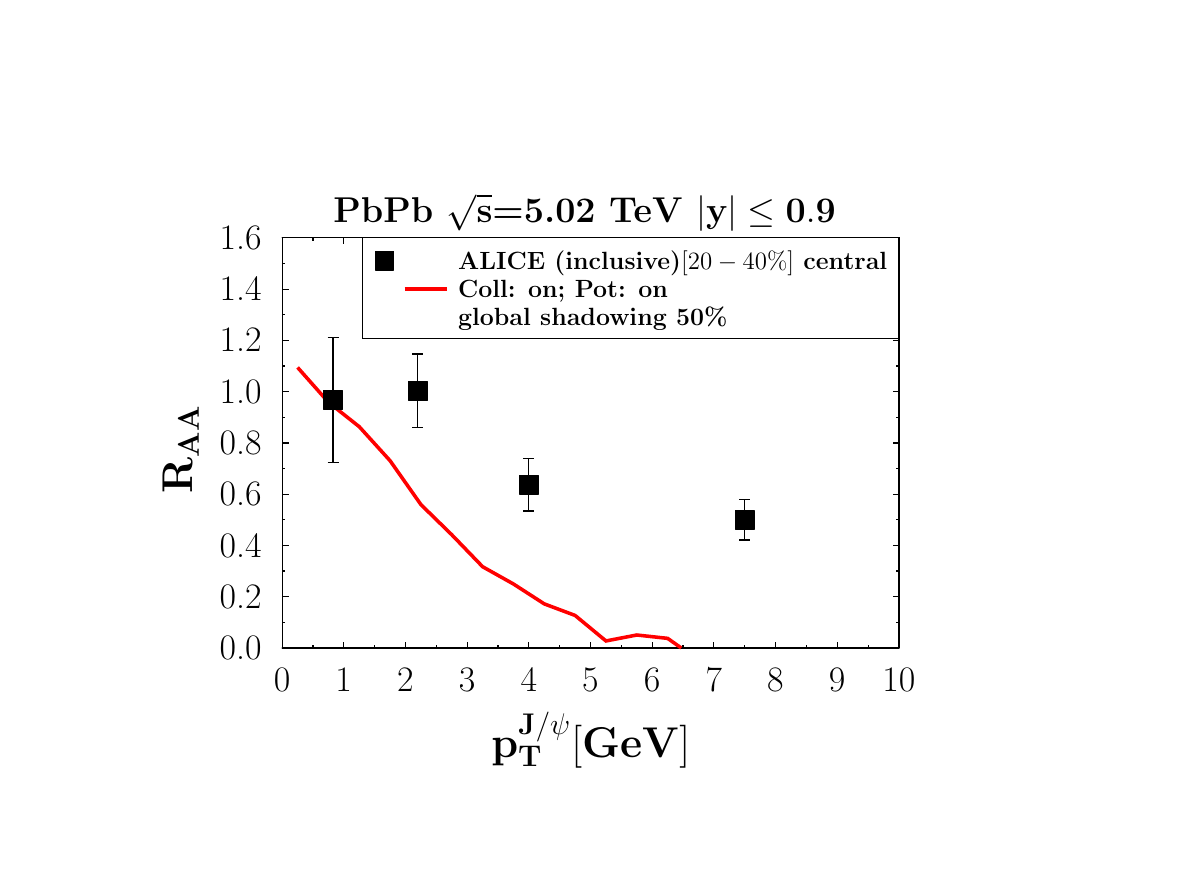}   \end{minipage}\hfill
   \caption{Comparison between our model prediction for the nuclear modification factor $R_{AA}$ for two different centrality ranges, $0-20\%$ (top) and $20-40\%$ (bottom), and the corresponding inclusive experimental data from the ALICE collaboration \cite{ALICE:2019nrq}.} 
      \label{Fig:primordialinst}
\end{figure}

One of the most interesting observables is the nuclear modification factor, $R_{AA}$, eq.~\ref{eq:RAA}. Its deviation from unity shows how the $p_T$ spectra are modified by nuclear effects. With the
proton reference spectrum discussed in subsection 5.A we present the results of our model in comparison with the ALICE data \cite{ALICE:2019nrq} in Fig. \ref{Fig:primordialinst} 
for PbPb collisions at $\sqrt{s} = 5.02\ {\rm TeV} $. The ALICE data present the results for inclusive \J whereas we calculate only the directly produced \J. B meson decay as well as the decay of excited  states, like the $\psi'$ and $\chi$, contribute to the experimental \J distribution and therefore the comparison between our results and the data has to be taken with caution.  
Fig. \ref{Fig:primordialinst} shows $R_{AA}(p_T) $, on top for central
([0-20\%]) on bottom for mid central [20-40\%] collisions. Our results
for the full model are presented as red lines, the ALICE data as black points. In the top figure we display as well the $R_{AA}$ values, which we obtain for the different combinations of switching on/off the $c\bar c$ potential and the collisions of the heavy quarks with the QGP partons. This exhibits the important role of the combined action of the $c\bar{c}$ potential and the collisions with the QGP partons for building up strong correlations. We see that our results, as the experimental data, show an enhancement at small $p_T$. For large $p_T$ $R_{AA}$ decreases but more in the calculations than in the data. The possible origins of this disagreement we discussed above.

It is remarkable that we obtain \JA{at low $p_T$} a $R_{AA}$ close to 1. As explained in the last section, in AA collisions we have off-diagonal contributions which dominate the primordial multiplicity (see \cite{Song:2017phm}), so naively one would expect a strong enhancement. The reason that this enhancement practically disappears is that in AA collisions the \J are created later (when $T< T_{\rm diss}$), where the average distance between $c$ and $\bar c$ is larger and therefore the overlap with the \J Wigner density is smaller. Despite collisions and $c \bar c$ potential, which enhance the yield, this almost compensates the primordial enhancement.

\begin{figure}[h]
     \centering
     \includegraphics[width=1.3\linewidth]{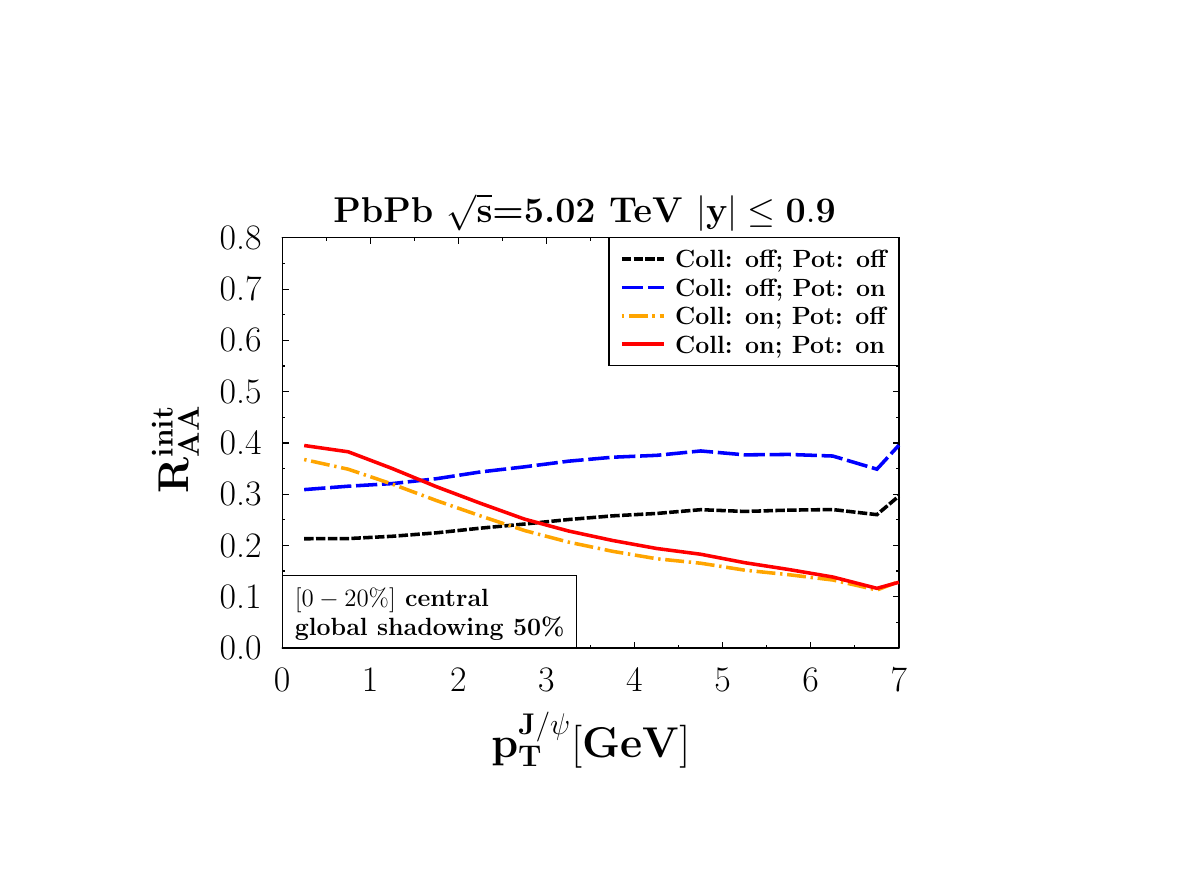}
   \caption{$R_{AA}^{\rm init}$ as a function of the transverse momentum at the initial time, when the charm quarks pass $T_{\rm diss}$, at midrapidity, $|y| \le 0.9$ for [0-20\%] central PbPb collisions at $\sqrt{s}$ = 5.02 TeV.
   We display this result for the 4 possible combinations of the set up. For details see text.} 
   \label{primandlocalraa}
\end{figure}

\begin{figure}[h]
   \centering 
   \includegraphics[width=1.3\linewidth]{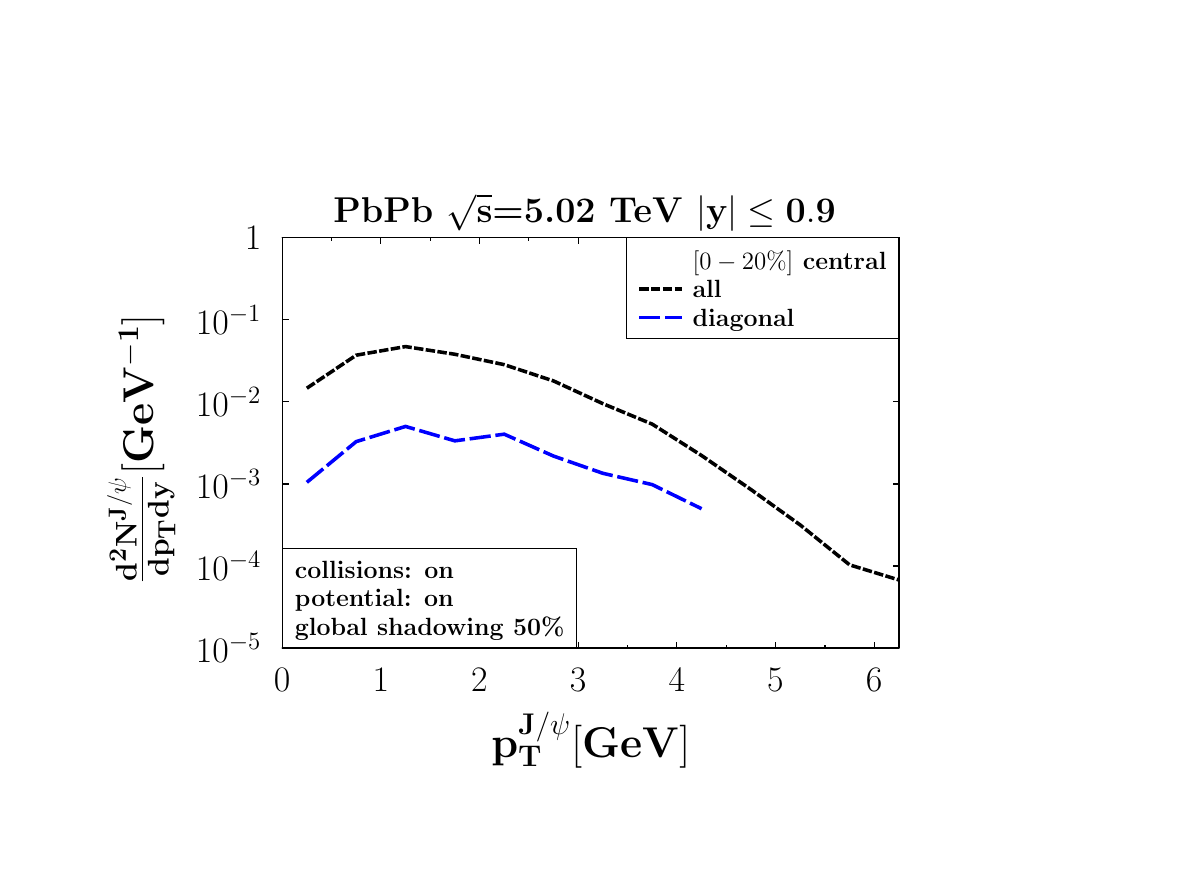}   
      \caption{Comparison of our model prediction for the final \ yield (short dashed black line)  at midrapidity, $|y| \le 0.9$, for [0-20\%] central PbPb collisions at $\sqrt{s}$ = 5.02 TeV. We display also our model prediction for the diagonal contribution (dashed blue line). For this calculation medium elastic collision and $c\bar c$ potential are included and we employ our standard parameter values: 
Minkowski and Bjorken time steps 0.25 fm/c and 0.1 fm/c, respectively.}
    \label{diagonalvstotal}
\end{figure}

This is demonstrated in Fig. \ref{primandlocalraa} where we display  the ratio -- called $R_{AA}^{\rm init}$ -- of \Js~ obtained from the $c\bar c$ which have passed $T_{\rm diss}$ and the diagonal \Js~ produced in the initial hard collisions. The ratio is displayed as a function of $p_T$ for the four different scenarios with the same color coding as in Fig. \ref{figintegralprimordialpt}.
The ratio is, despite of the off-diagonal contribution, smaller than one. Collisions enhance $R_{AA}^{init}$ at small $p_T$ and lower the ratio at large $p_T$, whereas the potential has little influence on the ratio at this stage. 
The importance of the off diagonal contribution to the \J yield is demonstrated in Fig.\ref{diagonalvstotal}. It shows for the same reaction and for central collisions the total yield (short dashed black line) and the diagonal contribution (dashed blue line).   
For low $p_T$ the diagonal part of the yield is up to one order of magnitude smaller than the total yield, so in most of the \Js~ the two heavy quarks come from different vertices. This is a consequence of the observation that there heavy quarks come to thermal equilibrium with the QGP \cite{Nahrgang:2013saa} at least in azimuthal direction and therefore the ratio of diagonal to off-diagonal contribution is determined by statistics. Therefore one could
call this low $p_T$ region "regeneration dominated".
\begin{figure}[h]
   \centering 
   \includegraphics[width=1.3\linewidth]{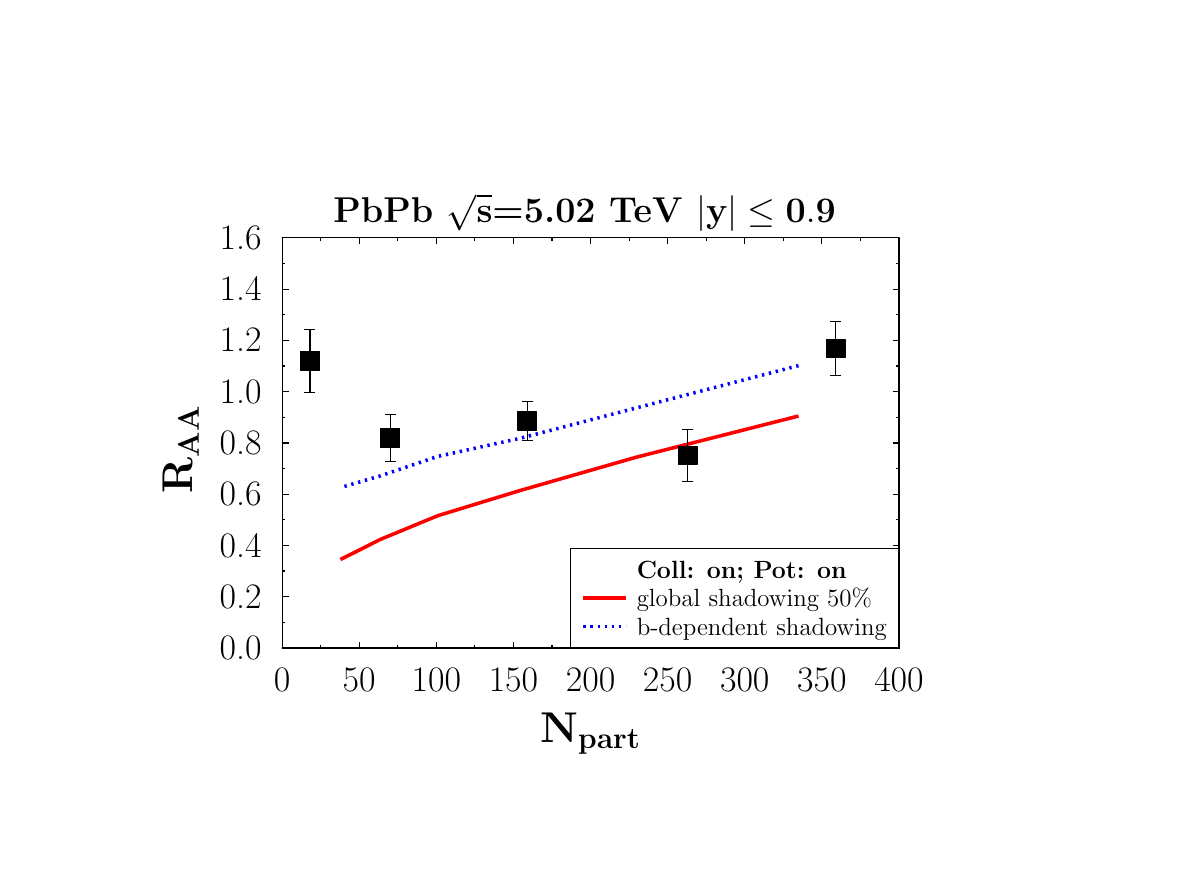}      \caption{Nuclear modification factor as a function of the average number of participants in PbPb obtained for active interaction potential and medium elastic collisions, employing  Bjorken and Minkwoski time steps of 0.1 fm/c and 0.25 fm/c, respectively.
\JA{We compare the results for a global shadowing of 50\% and of an impact parameter dependent shadowing \cite{Helenius:2012ny}} with ALICE data (black squares \cite{ALICE:2019nrq}.}
    \label{RAAvsnumberpast}
\end{figure}
The relative contribution of off-diagonal \J decreases but even at the largest $p_T$, investigated in this study, it does not fall below 50\% of the total yield.

Fig. \ref{RAAvsnumberpast} shows the centrality dependence of $R_{AA}$ for central PbPb collisions at midrapidity and at $\sqrt{s}=5.02 \ {\rm TeV}$. The ALICE data \cite{ALICE:2019nrq} are presented as black squares and the result of our calculation \JA{if we apply a global shadowing of 50\%} (what is only legitimate for central collisions) as a red line. To understand better the influence of the shadowing in the full $N_{\rm part}$ range, we also present calculations with an impact parameter dependent shadowing \cite{Helenius:2012ny}. The result is shown as a dotted blue line. We observe in theory as well as in experiment an enhancement of $R_{AA}$ for central collisions, where the number of produced \cc is large and therefore recombination is more probable, as well as a decrease with decreasing centrality. For peripheral reactions, which resemble pp collisions, in our calculation $R_{AA}$ is not equal one, as expected, because in this first version of the model, presented here, we neglect \J produced in the corona, which represent an increasing fraction of the yield when $N_{\rm part}$ becomes smaller.


\subsection{ Elliptic Flow of \J}
\label{sec:flow}
The azimuthal distribution of the transverse momentum distribution can be expanded in a Fourier series
\begin{equation}
\frac{d^2N}{d^2 p_T}=\frac{1}{2\pi p_T}\frac{dN}{dp_{T}}\left(1+2\sum^{\infty}_{n=1}v_{n}\cos(n(\phi -\Psi_{RP}))\right),    
\label{eq:azimdistrib}
\end{equation}
where $\phi$ is the azimuthal angle of the \J and $\Psi_{RP}$ is the angle of the reaction plane. 
The elliptic flow, $v_{2}$, the second coefficient
of the expansion, can be expressed (if the xz plane is the reaction plane) as 
\begin{equation}
v_{2}=\langle \frac{p^{2}_{x}-p^{2}_{y}}{p^{2}_{T}}\rangle.   \end{equation}
$v_2$ is another key observable in  heavy quark physics. The eccentricity of the almond shaped interaction region in coordinate space is converted, during the hydrodynamical expansion of the QGP, into an eccentricity in momentum space and hence into a finite $v_2$ value. The production of the initial heavy quarks in hard collisions is azimuthally isotropic and hence initially $v_2=0$. The $v_2$ value, observed for final \Js, is therefore a measure of their interaction (or that of their predecessors, the \cc quarks) with the QGP because only in collisions with the medium they can acquire a finite $v_2$. This is true for moderate $p_T$. At higher $p_T$ values the dependence of the path length in the medium on the azimuthal angle starts to play the leading role.
\begin{figure}[h]
   \begin{minipage}{0.48\textwidth}
     \centering
     \includegraphics[width=1.3\linewidth]
     {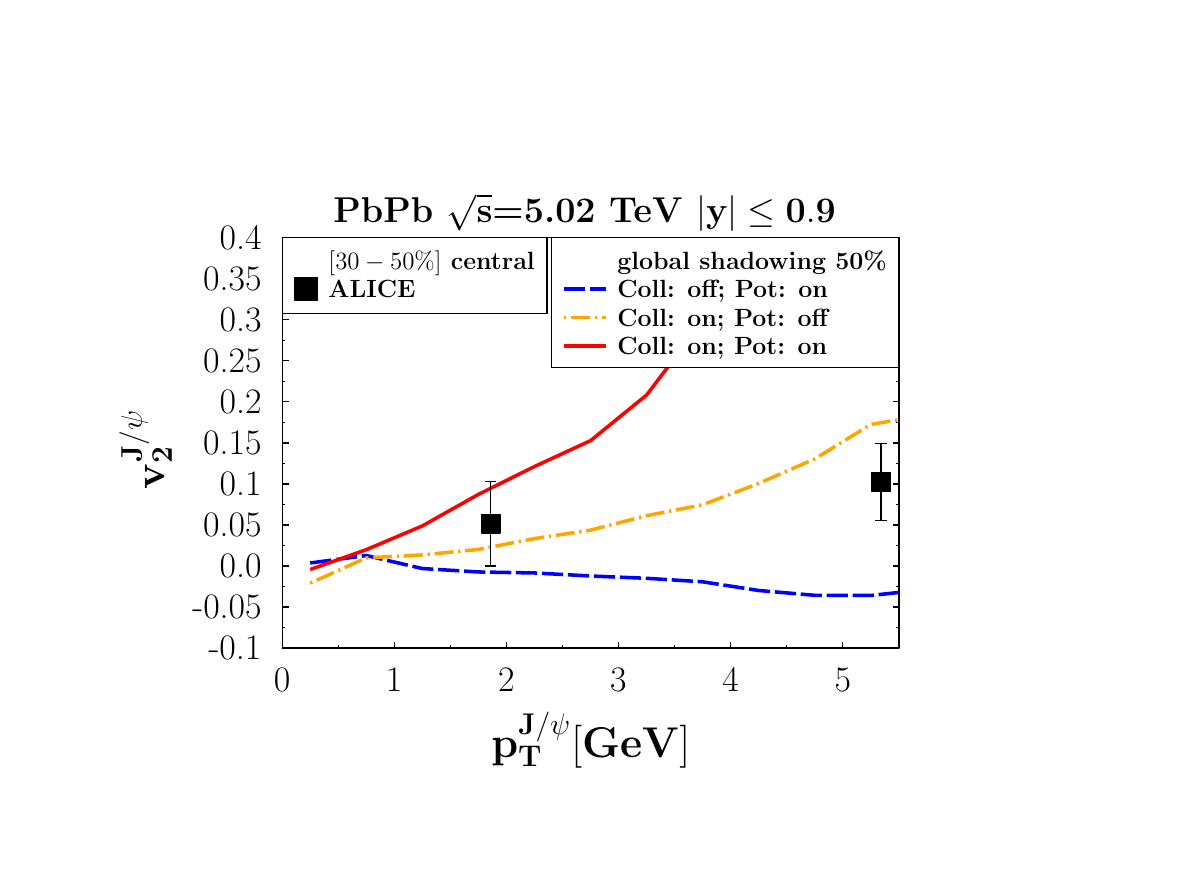}     
     \end{minipage}\hfill
   \caption{Elliptic flow, $v_{2}$, in the [30-50\%] centrality range between for PbPb at $\sqrt{s}$ = 5.02 TeV and for mid-rapidity ($\vert y \vert \le 0.9$. We compare our results,
   employing the standard parameters, with the data from the ALICE
   collaboration \cite{ALICE:2017quq}. Here for each of the  4.000 EPOS events we generated 20.000 MC@sHQ events.}
   \label{Fig:v23050}
\end{figure}     

In Fig. \ref{Fig:v23050}
we compare our results for the standard parametrization including collisions and potential (red line) for $|y|\le 0.9$ with
midrapidity data from the ALICE collaboration \cite{ALICE:2020pvw} (black squares) for the reaction PbPb at $\sqrt{s} = 5.02\ {\rm TeV}$ and for the [30-50\%] centrality interval. \JA{Our calculation shows a stronger increase of $v_2$ with $p_T$ in the standard version (collision and potential ON) than the experimental data. The origin of this large $v_2$ value, especially observed when the potential is ON, is the continuous production of \J during the expansion, which transfers the $v_2$ of the light partons to the heavy quarks. In standard transport approaches, a large fraction of the \Js~, observed for intermediate and large $p_T$, stem from the so-called primordial component, characterized by a small relative $c\bar{c}$ relative distance. As explained at the end of section \ref{section6a}, if the distance between heavy quarks is small, they do not scatter independently with the QGP partons but act, if the wave length of the exchanged gluons is smaller than this distance, as a color neutral object, explaining why $v_2^{\rm primordial}<v_2^{\rm regenerate}$ in these models. Such an interference mechanism, which could tame the $v_2$ at intermediate and high $p_T$ has not been considered yet in our model, neither the \J production in the corona, which could act in the same direction.}
\begin{figure}[h]
   \begin{minipage}{0.48\textwidth}
     \centering
   \end{minipage}\hfill
   \begin{minipage}{0.48\textwidth}
     \centering
     \includegraphics[width=1.3\linewidth]{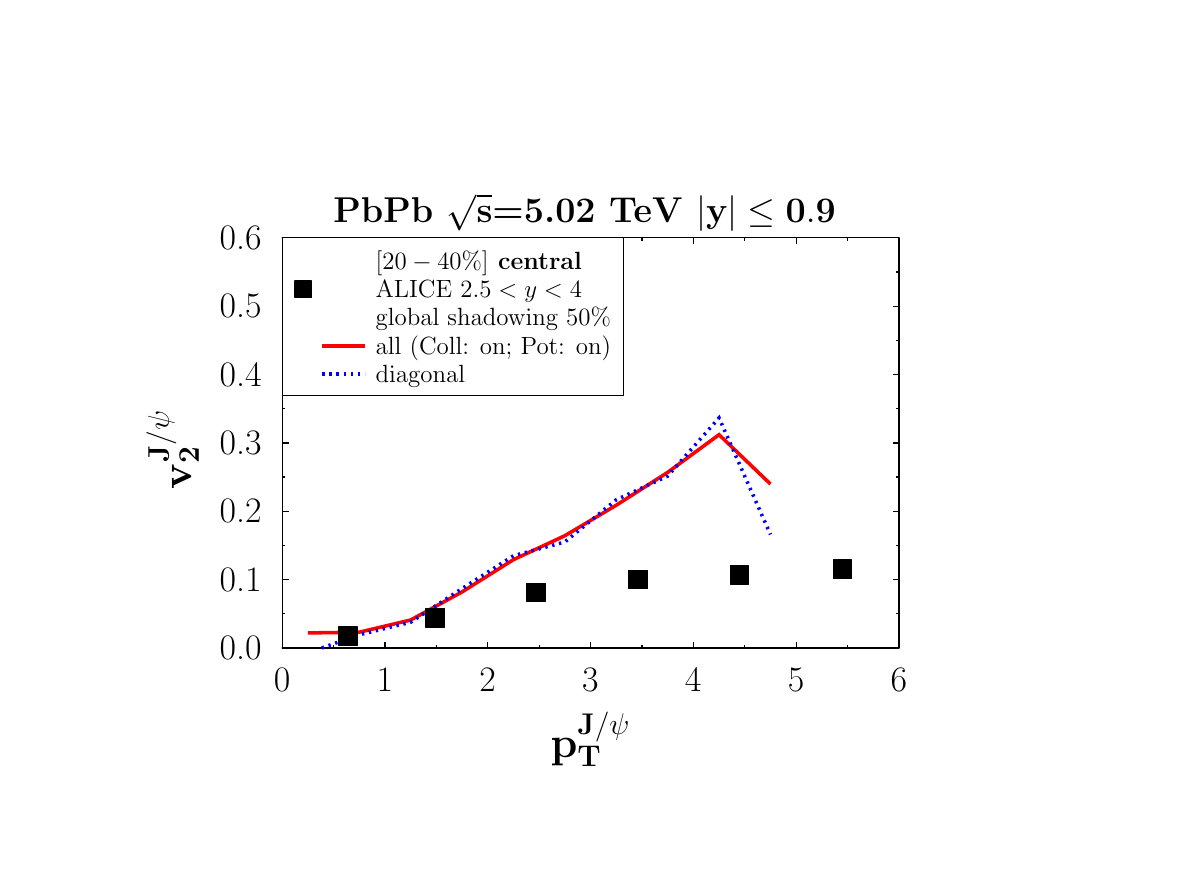}   
   \end{minipage}
   \begin{minipage}{0.48\textwidth}
     \centering
     \includegraphics[width=1.3\linewidth]
     {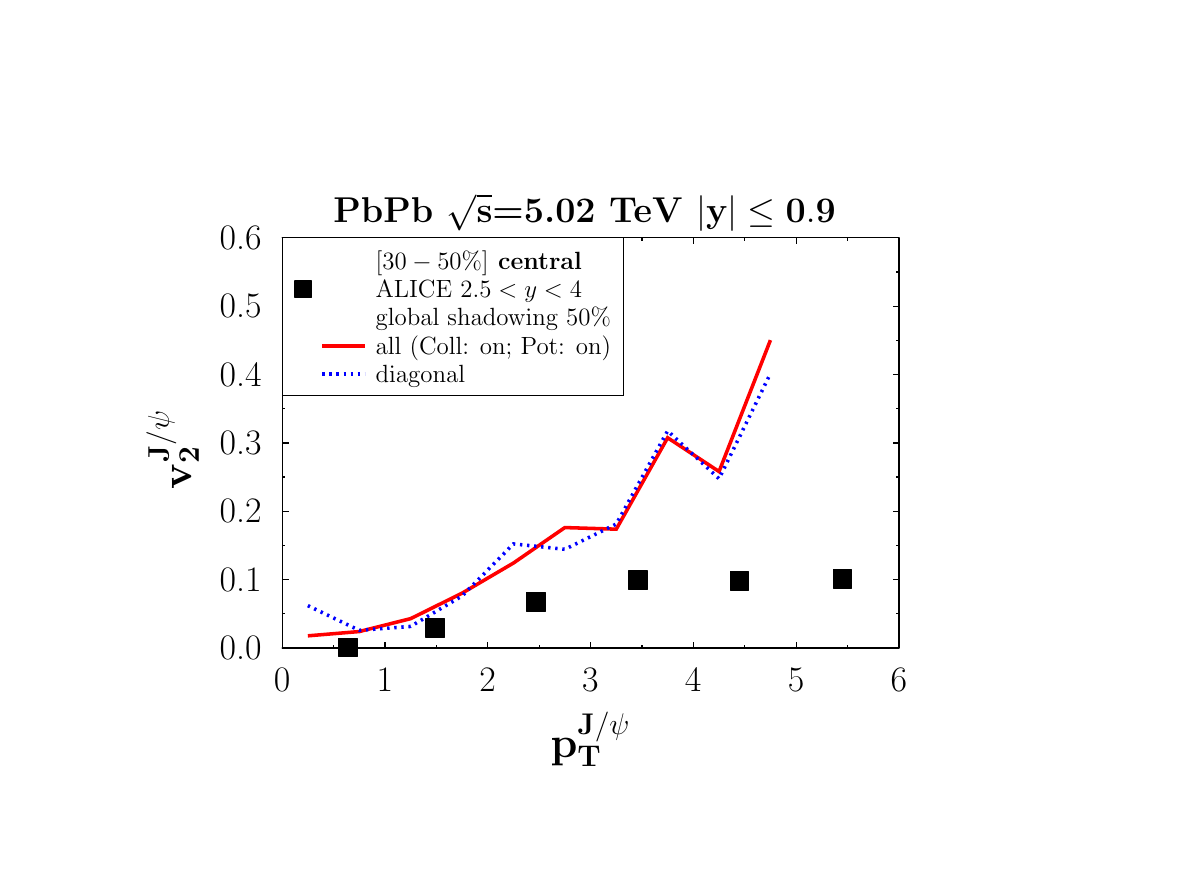}   \end{minipage}     
   \caption{Comparison between our model predictions for the elliptic flow, $v_{2}$, at mid-rapidity ($\vert y \vert \le 0.9$) with experimental data from the ALICE collaboration for the rapidity ($2.5< y < 4.0$) \cite{ALICE:2020pvw}. On top (bottom)
   we show the centrality ranges [20-40\%] ([30-50\%]).} 
\label{Fig:v21020}
\end{figure}  

In Fig. \ref{Fig:v21020} we compare our results for $|y| \le 0.9$ and for different centrality bins with the results of the ALICE collaboration for $2.5<y<4$.  The top figure shows the results for
[20-40\%], the bottom figure those for [30-50\%] centrality. This bin corresponds to the centrality bin shown in Fig. \ref{Fig:v23050}.
 Comparing both figures, we see that the experimentally measured $v_2$ at mid and forward rapidity is rather similar. Therefore we can profit from the better data available for the forward rapidity range (where our calculation is plagued from the large $\gamma_{cm}$ value).
 We include in this figure, as blue dotted line, $v_2$ of the diagonal \J, means from those where the \cc come from the same elementary vertex. We see that they have a similar $v_2$ and remark that one does indeed not recover the $v_2^{\rm primordial}<v_2^{\rm regenerate}$ observed in transport models, which would for us correspond to $v_2^{\rm diag}<v_2^{\rm off diag}$. This tension can be understood from the previous remark on neglecting the "dipole character" of the $c\bar{c}$ - QGP interactions
 and deserves further investigation.


%% file: Conclusions.tex
In this work, we presented a new theoretical approach to understand the experimental data on \J production in pp and AA collisions at LHC energies. Our main goal was to provide a microscopical model which allows to follow the individual \cc quarks from their creation in initial hard collisions until their final observation in the hidden heavy flavor mesons. Our treatment is thus in the same spirit as recent open quantum system approaches, which explicitly study the time evolution of the quarkonia density matrix under the influence of an approximated density matrix of the whole system. 

For pp collisions, as in ref. \cite{Song:2017phm}, the production of \Js~ is described by a sudden Wigner-coalescence approximation, which gives a good description of the experimental findings, not only for \J and $\psi'$ but also for $\chi_{c}$. In this approach
the primordial distribution of \cc is projected on the Wigner densities of the quarkonia states.

This primordial distribution of \J and its excited states
are of little relevance for the \J production in central heavy ion collisions at low and moderate $p_T$. There a QGP is produced. Lattice gauge calculations reveal that if its temperature is above $T=T_{diss}$, \Js~ are not stable and that below $T_{diss}$ but finite T, the \J wave function is quite different from the vacuum wave function due to the interactions of the \cc with the QGP environment. This renders it more complicated to model the dynamics of \Js~ in heavy ion collisions. 

To cope with this observation we 
\begin{itemize}
\item
employ a mutual potential interaction of the $c\bar c$ pairs with a potential, which is adjusted to lattice data. It is active between all $c\bar c$ pairs in singlet states.
\item
modify below $T_{diss}$ the Wigner density of the \J in the QGP medium by introducing a temperature dependent width in the Gaussian parametrization of the \J Wigner density. This width reproduces the \J in medium radius given by a potential model based on lattice data. 
\end{itemize}

For the collisions of the $c$ or $\bar c$ with the QGP constituents, the quarks and gluons, we use the MC@sHQ model which has already been successfully applied for the studies of open charm mesons ref.\cite{Gossiaux:2008jv,Gossiaux:2009mk}. They change the Wigner density of the $c$ or $\bar c$ quark which was involved in these collisions and therefore the convolution of the $c\bar c$ Wigner density with that of the \J changes as well.

In central heavy ion collisions one would expect a strong \emph{enhancement} of the \J multiplicity as compared to the multiplicity in a single pp collision multiplied by the number of these collisions in the heavy ion reaction. The reason is that in heavy ion collisions \cc from different elementary vertices can form a \J. This enhancement is, however, (over)compensated by the late production of \Js. \Js~ can only be produced when the QGP temperature has fallen below $T_{diss}$. There the average distance between the \cc is considerable larger and therefore the convolution with the Wigner density is smaller. 

We observe furthermore that the collisions between QGP partons as well as the potential interaction between \cc quarks enhance the \J yield as compared to a free streaming scenario. The collisions shift the $p_T$ distribution of heavy quarks towards lower values, the potential keeps $c\bar c$ pairs closer together. Both processes shift therefore the two-body Wigner density of the $c\bar c$ pairs to regions where the \J Wigner density is large. 

We find in our approach reasonable agreement with the experimental data at low $p_T$, where the
enhancement has been observed, simultaneously for $R_{AA}$ and for $v_2$.
The latter is created due to the collisions of \cc with the QGP partons.

We have employed in our approach a simple model for the color degrees of freedom and we have concentrated on \J mesons. It should be noted that the gluon-dissociation mechanism $g+\Phi \to Q+\bar{Q}$ as well as its detailed balance counterpart -- expected to become significant for deeply bound states, thus around $T_c$ -- were not included in our dynamical treatment. \JA{At the highest $p_T$ values, considered in our work, one observes deviations between the experimental values for $R_{AA}$ and $v_2$ and our predictions. They come from several shortcomings of our approach in this kinematic regime. There, due to the large $\gamma_{cm}$ of the $Q\bar Q$ cm system, the calculation of the potential interaction between the $Q$ and the $\bar Q$ in their center of mass system has large systematic errors and has to be improved. Also \J from excited charmonium states and from B meson decay have to be included for a quantitative description and \Js~, which are produced in the corona, and do not pass the QGP, have to be consistently added. Last but not least, if a Q is close by, which forms with the considered $\bar Q$ a color neutral state, the interaction of a heavy quark with the QGP partons has to be modified to take the dipole character of this interaction into account. }
To improve on these aspects will be the subject of an upcoming publication. The new EPOS4 approach will also allow to treat correctly the correlations between the initially formed $c\bar c$ pairs which is as well an important ingredient of the microscopic modelling.

%% file: acknowledgement.tex
The authors thank Taesoo Song for providing the program for the calculation of \J in pp collisions as well as Taesoo Song and Elena Bratkovskaya for inspiring discussions. This study is part of a project that has received funding from the European Union’s Horizon 2020 research and innovation program under grant agreement STRONG – 2020 - No 824093. We are also pleased to acknowledge the support from the Region Pays de la Loire, under contract n° 2015-08473.

%% file: appendix.tex
\appendix
The HQ are created in the early phase of the HIC through hard scattering processes and can have relativistic energies in the computational frame, the nucleus-nucleus center of mass frame. It is crucial that the Wigner function, used in our coalescence approach, is adequate to describe HQ under these conditions.

To obtain such a Wigner density we need first to build an orthogonal set of states that represents the bound states of $Q\bar Q$ and to ensure the correct normalization and relativistic invariance.  From that we can then proceed to derive the Wigner function associated with the $Q\bar Q$ states, expressed in this orthogonal basis. 

\section{Finding Orthogonal Two-Particle Basis for non-relativistic states}
Given a longitudinal direction and a transverse plane, a quantum state for a two-particle non relativistic  system (two scalar particles) can be written as 
\begin{equation}
\vert \Phi \rangle =\int \frac{dp_{L,1}d^{2}\pv_{T,1}}{2E_{1}}\frac{dp_{L,2}d^{2}\pv_{T,2}}{2E_{2}}f(\pv_{1},\pv_{2}) \vert 1 \rangle \vert 2 \rangle \label{eq:QQstates}
\end{equation}
where the single particle states are normalized to
\be
\langle 1'|1 \rangle = 2 E_1\delta(\pv_1-\pv\,'_1). 
\ee
States with a defined center of mass momentum $\Pv$, $|\Psi_{\Pv} \rangle $, should ideally be  normalized as 
\begin{equation}\nonumber
\langle \Phi_{\Pv^{'}}\vert \Phi_{\Pv}\rangle = 2E\delta(\Pv-\Pv^{'}) 
\label{eq:norm}
\ee
For eq. \ref{eq:QQstates} this imposes 
\be\int \frac{d^{3}p_{1}}{2E_{1}}\frac{d^{3}p_{2}}{2E_{2}} f_{P}^*(\pv_{1},\pv_{2})f_{P'}(\pv_{1},\pv_{2}) =2E\delta(\Pv-\Pv^{'})  
\end{equation}
We now consider the correlation function
\begin{equation}
f_{i,\Pv}(\pv_{1},\pv_{2})=\delta(\Pv-\pv_{1}-\pv_{2})f_{i}(\frac{\pv_{1}-\pv_{2}}{2}) 
\label{eq:reststate}
\end{equation}
where the index $i$ refers to possible internal states. Then, with $\vec q= \frac{\pv_1-\pv_2}{2}$, the normalization reads:
\bea
\langle \Phi_{i,P^{''}}\vert \Phi_{j,P'}\rangle &=&\int \frac{d^3p_{1}}{2E_{1}}\frac{d^3p_{2}}{2E_{2}}\delta({\Pv}-{\pv}_{1}-{\pv}_{2})\delta({\Pv}^{'}-{\pv}_{1}-{\pv}_{2})\nonumber \\
&\times&f_{i}^*(\frac{{\pv}_{1}-{\pv}_{2}}{2})f_{j}(\frac{{\pv}_{1}-{\pv}_{2}}{2})\n
&=& \delta({\Pv'}-{\Pv''})\int\frac{d^3q}{4E_1E_2}f^*_j(\vec q)f_i(\vec q)\n
&=& \delta({\Pv'}-{\Pv''})\int\frac{d^3q}{4(q^2+m^2)}f^*_j(\vec q)f_i(\vec q),
\label{eq:norm1}
\eea
where $m$ is the mass of the individual particles.
In the last expression we have assumed that the CM momentum is small compared to the mass of the heavy quarks. Provided that the integral equals $\delta_{ij}E_P$ we obtain the proper normalization. At small momentum $P$, one has 
$E_P\approx M_i$ -- the mass of the two-particle state --, so that it is always possible to impose such constrain on the integral\ldots but thus this ceases to be so for finite values of $P$. To further discuss our approach we temporary neglect the transverse degrees of freedom. This reduces our approach to one momentum space and one energy dimension. 

\subsection{Variable Transformation}
To advance towards fully relativistic definitions let us consider the invariant measure
\bea
I&=&\int\frac{dp_1}{2e_1}\frac{dp_1}{2e_1}=\int dp_1de_1dp_2de_2\delta(p_1^2-m^2)\nonumber \\ &\times&\delta(p_2^2-m^2)\theta(e_1)\theta(e_2).
\label{invme}
\eea
We introduce the relative momentum $\Delta=\frac{{p}_{1}-{p}_{2}}{2}=(q^0,q)$ and the center of mass momentum $\Sigma=p_1+p_2= (P^0,P)$. With 
$ d^2p_1d^2p_2=d^2\Sigma d^2\Delta $,
we obtain
\begin{equation}
I=\int \frac{d^2\Sigma d^2\Delta}{2}\delta(\frac{\Sigma^2}{4}+\Delta^2-m^2)\delta(\Sigma\cdot \Delta)\theta(\frac{P^0}{2}-\vert q^{0} \vert).
\end{equation}
$\delta(\Sigma\cdot \Delta)$ implies 
\begin{equation}
\vert q^{0}\vert=\sqrt{\frac{(P^0)^2-4(m^2+q^2)}{(P^0)^2-4q^2}}\vert q\vert \leq \vert q \vert
\label{eq:onshellmassqrel}    
\end{equation}
for $\vert q \vert < P^0/2$. The $\delta$ functions takes therefore care that the condition $|q^0|<P^0/2$ is always fulfilled.
$\Sigma$ is a time-like vector, therefore $\delta(\Sigma\cdot \Delta)$ implies that $\Delta$ has to be a space-like vector. Introducing $Q^2=-\Delta^2$, we now perform a second variable transformation
\bea
P^0&=&\sqrt{s}\ \cosh Y\ , \ P=\sqrt{s}\ \sinh Y\nonumber \\ q^0&=&Q\ \sinh y\ , \ q= \pm Q\ \cosh y
\eea
with $d^2\Sigma d^2\Delta=\frac{ds}{2} Q dQ dY dy$, including both, the $q=+Q \cosh y$ and the $q=-Q \cosh y$ sector. In these variables we find
\begin{equation}\nonumber
\delta(\Sigma\cdot \Delta)=\delta(\sqrt{s}Q \sinh(y\pm Y))= \frac{\delta(y\pm Y)}{\sqrt{s}Q }   \end{equation}
where the $\pm$ stand for the $q=\pm Q \cosh y$ sectors, respectively. The 2nd distribution $\delta(\frac{\Sigma^2}{4}+\Delta^2-m^2)$ simply writes 
\begin{equation}
\delta\left(\frac{\Sigma^2}{4}+\Delta^2-m^2\right)= \delta\left(\frac{s}{4}-Q^2-m^2\right),   
\label{eq:onshell}
\end{equation}
implying that the condition $|q|<P^0/2$ is always satisfied. The integral I (eq. \ref{invme}) can thus be rewritten as
\begin{eqnarray}
I=I_++I_- &=& \int dY
\frac{d|q|}{\sqrt{s}}(\Theta(q>0)+\Theta(q<0))
\nonumber\\
&=& \int dY
\int_{-\infty}^{+\infty}\frac{dq}{\sqrt{s}},
\end{eqnarray}
where the 2 sectors have been merged.

\subsection{Relativistic Two-Particles States}
These new variables allow for reformulating our relativistic state. For any lab frame $S'$, we assume that there exists a frame $S$ where the two-particle state CM is nearly at rest and define $y_\Phi$ as the rapidity of $S$ in $S'$. In $S$, the state is defined as
\bea
\vert \Phi_{i,P\approx 0}\rangle &=& \int \frac{dp_1 dp_2}{2e_{1}2e_{2}}
\delta({P}-p_1-p_2)f_{i}(p_1-p_2)\vert p_{1}\rangle\vert p_{2}\rangle \nonumber\\
&=&
\int \frac{dq}{\sqrt{s}} d Y  \delta({P}-p_1-p_2) f_i(q)
\vert p_{1}\rangle\vert p_{2}\rangle. 
\end{eqnarray}
Let $f_i$ be defined as a boost invariant wave function depending on the relative momentum $q$ measured in the rest frame. Proceeding to a Lorentz transform to the lab frame $S'$, the two-particle CM state reads
\bea
\vert \Phi_{i,y_{\Phi}}\rangle &=&\int \frac{dq}{\sqrt{s}}dY\delta(\sqrt{s} \sinh(y_{\Phi}-Y)) f_{i}(q)\vert p_{1}'\rangle\vert p_{2}'\rangle \n 
&=&\int \frac{dq}{s}dY\delta(y_{\Phi}-Y)f_{i}(q)\vert p_{1}'\rangle\vert p_{2}'\rangle   
\label{eq:state_relat_def}
\eea
where the momenta $p_{1}'$ and $p_{2}'$ are taken as $(\sqrt{m^{2}+q^{2}},q)= m(\cosh(\hat{y}),\sinh(\hat{y}))$ in $S$ and then boosted with a rapidity shift $+y_{\Phi}$, leading to

\bea\nonumber
p_{1}'&=&m(\cosh(y_{\Phi}+\hat{y}),\sinh(y_{\Phi}+\hat{y}))  \n  
p_{2}'&=&m(\cosh(y_{\Phi}-\hat{y}),\sinh(y_{\Phi}-\hat{y})).
\eea
It remains to be checked that the orthogonality conditions of the state  (\ref{eq:norm}) is fulfilled. After some trivial calculation, one finds
\bea
\langle \Phi_{i,y^{'}_{\Phi}}\vert \Phi_{j,y_{\Phi}}\rangle &=&\int \int \frac{dp_{1}}{2E_{1}}\frac{dp_{2}}{2E_{2}} \times \frac{\delta(y^{'}_{\Phi}-Y)}{\sqrt{s}}
\frac{\delta(y_{\Phi}-Y)}{\sqrt{s}}\n &\times&  f_{i}^\star(q)f_{j}(q)\nonumber \\
&=&\int\frac{dq  dY}{\sqrt{s}}\frac{\delta(y^{'}_{\Phi}-Y)\delta(y_{\Phi}-Y)}{s} \nonumber \\ &\times& f^{\star}_{i}(q)f_{j}(q)\nonumber \\
&=&\delta(y^{'}_{\Phi}-y_{\Phi})\int\frac{dq}{s^{3/2}}f^{\star}_{i}(q)f_{j}(q)
\label{eq:checkorth}
\eea
and  we can derive the invariant orthogonality relations provided we require
\begin{equation}\nonumber
\int \frac{dq}{s^{3/2}}f^{\star}_{i}(q)f_{j}(q)=\delta_{ij}.
\label{eq:orthogonality}
\end{equation}
By introducing the non relativistic wave function
\begin{equation}
\psi_{i}(q)=\frac{f_{i}(q)}{s^{\frac{3}{4}}(q) }    
\label{eq:wfnon}
\end{equation}
we obtain the orthogonality relation for non relativistic wave functions $\int dq \psi_i^*(q)\psi_j(q)=\delta_{ij}$.

Although we have been able to develop a prescription for the construction of an orthogonal invariant base, there are still some comments to be made: first, we do not strictly recover the anticipated normalization  $\langle \Phi_{y^{'}_{\Phi},i}\vert \Phi_{y_{\Phi},i}\rangle =2E\delta (p-p^{'})\delta_{ij}$. This can be explained due to the fact that for a given state $i$ and a given rapidity $y_{\Phi}$, the total energy depends on the relative momentum $q$. This has consequences for the orthogonality relation \ref{eq:orthogonality}, which slightly differs from \ref{eq:norm1}. However, these states admit proper transformation laws under Lorentz boosts and the relationship \ref{eq:checkorth} allows to create a completely orthogonal set of two-particle particle states 
\begin{equation}
\vert \Phi_{i}\rangle :=\int dy_{\Phi}g_{i}(y_{\Phi})\vert \Phi_{y_{\Phi},i}\rangle,  
\label{eq:basisform}
\end{equation}
where $g$ is an arbitrary function which satisfies the  relation 
\begin{equation}
\langle\Phi_{i}\vert \Phi{'}_{j}\rangle =\delta_{ij}\int dy_{\Phi}g^{*}_{i}(y_{\Phi})g_{j}^{'}(y_{\Phi}),  \end{equation}
with an easy connection to the rapidity spectrum:
 \begin{equation}
\frac{dN_{i}}{dy_{\Phi}}=\vert \langle \Phi_{y_{\Phi},i}\vert\Phi\rangle\vert^{2}=\vert g_{i}\vert^{2}. 
\label{eq:basisesprec}
\end{equation}
After having developed a method to build an orthogonal boost invariant basis on which we can project our two-particle states, we turn to the building of the Wigner function on this basis. 

\section{Relativistic  Wigner Function from a Given Basis in (1+1)D Case}

The main difficulty for finding a relativistic Wigner function from an orthogonal basis of the form (\ref{eq:basisform}), lies in the difficulty of defining a conjugate variable of the rapidity. Only the option to take the relative momentum $q$ in the center of mass as the conjugate variable allowed to arrive at a satisfactory conclusion and at the same time to obtain  results which have a clear physical significance. Here we discuss this option $[Y,q]$. 

To obtain a Wigner density in $[Y,q]$ for the basis of the form (\ref{eq:basisform}) it is convenient to rewrite (\ref{eq:basisesprec}) in the form 
\begin{equation}
\frac{dN_{i}}{dy_{\Phi}} = Tr(\widehat{\rho}_{\Phi}\widehat{\rho}_{i,y_{\Phi}})
\label{traJA}
\end{equation}
where the trace  is performed over the phase space variables ($[Y,q]$), $ y_{\Phi}$ being the rapidity of the two-particle system (quarkonium). The density operators $ \widehat{\rho}_{\Phi} $ and $\widehat {\rho}_{i,y_{\Phi}}$ have the form
\begin{equation}\label{A}
\begin{split}
\widehat \rho_{\Phi} =\vert \Phi \rangle \times  \langle \Phi \vert \quad ; \quad 
\widehat \rho_{i,y_{\Phi}} =\vert \Phi_{i,y_{\Phi}} \rangle \times  \langle \Phi_{i,y_{\Phi}} \vert
\end{split}
\end{equation}
in which $\vert \Phi\rangle $ and $\vert \Phi_{i,y_{\Phi}}\rangle$ represents a generic $(Y,q)$ two-particle wave function and the two-particle wave function for a state i with the rapidity $y_{\Phi}$ respectively. Inserting the identity operator 
\begin{equation}
\widehat{I}= \int dY \frac{d{q}}{\sqrt{s}} \vert 1,2\rangle \langle 1,2\vert
\end{equation}
in the expression for the spectrum (\ref{traJA}) we obtain
\begin{equation}
\frac{dN_{i}}{dy_{\Phi}} = \int dY dY^{'}\frac{d{q}d{q'}}{\sqrt{ss^{'}}}\langle Y,q\vert \widehat{\rho}\vert Y^{'},{q'} \rangle \langle Y^{'},{q'}\vert \widehat{\rho}_{i,y_{\Phi}}\vert Y,q\rangle  
\end{equation}
where $\vert Y,p_r \rangle$ refers to the $\vert 1,2 \rangle$ state with a total rapidity Y and a relative momentum $p_r$ and $\sqrt{s}$ respectively $\sqrt{s^{'}}$ are the center of mass energy of the states.

Defining 
\begin{equation}
\overline{\rho}({Y},q;Y^{'},{q'})= \frac{\langle {Y}^{'},{q'} \vert\widehat{\rho} \vert {Y},q \rangle}{(ss^{'})^{1/4}},
\label{densmat}
\end{equation}
we get
\begin{equation}
\frac{dN_{i}}{dy_{\Phi}}= \int dY dY^{'}d{q}d{q'}\overline{\rho}({Y'},{q'};{Y},q)\bar{\rho}_{i,y_{\Phi}}({Y},q;{Y}^{'},{q'})
\label{eq:spectdens}
\end{equation}
We introduce now, in preparation of the Wigner transformation, the auxiliary variables $\overline{{Y}}=\frac{{Y}+ {Y}^{'}}{2}$, $\overline{q}=\frac{q+{q'}}{2}$, $\Delta {Y}={Y}-{Y}^{'}$ and $\Delta q=q- {q'}$,  which transform (\ref{eq:spectdens}) to
\bea
\frac{dN_{i}}{dy_{\Phi}}&=&\int d\overline{Y}d\overline{q}d\Delta Y d\Delta Y^{'} d\Delta {q}d\Delta {q'}\n &\times& \overline{\rho}(\overline{{Y}} - \frac{\Delta {Y}}{2},\overline{{p}}_{r} - \frac{\Delta q}{2};\overline{{Y}} + \frac{\Delta {Y}}{2},\overline{q} + \frac{\Delta q}{2}) \n &\times & \overline{\rho}_{i,y_{\Phi}}(\overline{{Y}} + \frac{\Delta {Y}'}{2},\overline{q} + \frac{\Delta q'}{2};\overline{Y} - \frac{\Delta {Y}'}{2},\overline{q} - \frac{\Delta q'}{2})\n &\times&\delta(\Delta {Y} -\Delta {Y}^{'})\delta(\Delta q-\Delta {q'}).
\label{eq:spectrap1D}
\eea
The relationship between conjugate variables is given by
\bea
\delta(\Delta q-\Delta {q'}) &=&\frac{1}{2\pi\hbar}\int dx_{r}e^{i\frac{{x}_{r}(\Delta {q}-\Delta {q'})}{\hbar}}
\n
\delta(\Delta{Y} -\Delta {Y}^{'})&=&\frac{1}{2\pi}\int d{k_3} e^{i{k_3}(\Delta {Y} -\Delta {Y}^{'})} 
\label{eq:deltarelation}
\eea
where $ {x}_{r}$ is the relative position measured in the $Q\bar Q$ center of mass system and $k_3$ corresponds to the dimensionless eigenvalues of the boost operator \cite{Durand:1976}.
We recall as well the definition of the Wigner function of a density operator $\rho({r},{r}^{'})$, for instance associated to some wave function $\psi$ through $\rho({r},{r}^{'})=\psi({r}) \psi^\star(r') $: 
\begin{equation}
W(r,p)=\frac{1}{2\pi\hbar}\int dy e^{-i\frac{{py}}{\hbar}}\rho({r}+\frac{{y}}{2},{r}-\frac{{y}}{2}).
\label{eq:Wignerdeff}    
\end{equation}
 Substituting the delta distributions relations (\ref{eq:deltarelation}) and comparing the definition of the Wigner function with the factors in the equation (\ref{eq:spectrap1D}) we obtain the Wigner function for the two-particle density operator 
\bea
W_{i,y_{\Phi}}(\overline{{Y}},{k_3};q,{x}_{r})&=&\frac{1}{(2\pi)^2\hbar }\int d\Delta Y^{'}d\Delta {q'}\n 
&&\hspace{-2cm}\times \overline{\rho}_{i,y_{\Phi}}
(\overline{{Y}}+\frac{\Delta {Y}^{'}}{2},\overline{{q}}+\frac{\Delta {q'}}{2};\overline{{Y}}-\frac{\Delta {Y}^{'}}{2},\overline{q}-\frac{\Delta {q'}}{2})\n &&\hspace{-2cm}\times e^{-i({k_3}\Delta {Y}^{'}+{x}_{r}\frac{\Delta q'}{\hbar})}
\label{eq:wignerstateyphi}
\eea \\
and 
\bea
W(\overline{{Y}},{k_3};{q},{x}_{r})&=&\frac{1}{(2\pi)^2\hbar}\int d\Delta Y d\Delta {q}\n && \hspace{-2cm}\times  \overline{\rho}(\overline{Y}-
\frac{\Delta Y}{2},\overline{q}-\frac{\Delta q}{2};\overline{{Y}}+\frac{\Delta Y}{2},\overline{{q}}+\frac{\Delta {q}}{2})\n &&\hspace{-2cm}\times e^{i({k_3}\Delta {Y} +{x}_{r}\frac{\Delta{q}}{\hbar})}.
\label{wdja}
\eea
Substituting the equations for the Wigner functions, \ref{wdja} and \ref{eq:wignerstateyphi}
in equation \ref{eq:spectrap1D}  we obtain
\bea
\frac{dN_{i}}{dy_{\Phi}}&=& (2\pi)^2\hbar \int d\overline{Y}d\overline{q}dk_3 dx_{r} W(\overline{{Y}},{k_3};\overline{q},{x}_{r})\n &\times&
W_{i,y_{\Phi}}(\overline{{Y}},{k_3};\overline{q},{x}_{r})
\label{eq:dNdy1DfromWigner}
\eea
so we can, as in the non relativistic case, consider $\frac{dN_{i}}{dy_{\Phi}}$ as a convolution of the two Wigner densities.
Next we need to evaluate $W_{i,y_{\Phi}}(\overline{{Y}},{k_3};\overline{q},{x}_{r})$. We start from eq. \ref{densmat} and obtain:
\bea
\overline{\rho}_{i,y_{\Phi}}({Y},{q};{Y}^{'},{q}^{'})&=&\frac{f^\star_{i,y_{\Phi}}(q)f_{i,y_{\Phi}}(q')}{(ss^{'})^{\frac{1}{4}}}\\
&&\hspace{-2cm}=\frac{\delta(\sqrt{s}(Y-y_{\Phi}))\delta(\sqrt{s^{'}}(Y^{'}-y_{\Phi}))f^{'\star}_{i}({q})f_{i}({q}^{'})}{(ss^{'})^{\frac{1}{4}}}\n
&&\hspace{-2cm}=\frac{\delta(Y-y_{\Phi})\delta(Y^{'}-y_{\Phi})f^{'\star}_{i}({q})f_{i}({q}^{'})}{(ss^{'})^{\frac{3}{4}}}\nonumber.
\eea
with $f_{i}(q)$,  the "wave function" in eq. \ref{eq:state_relat_def}. With this density matrix
we calculate now the Wigner density, eq. \ref{eq:wignerstateyphi}
\bea
W_{i,y_{\Phi}}(\overline{{Y}},{k_3};\overline{q},{x}_{r})&=&\frac{1}{(2\pi)^2\hbar}\int \frac{d\Delta Y d\Delta {q}}{(ss^{'})^{3/4}} e^{-i{x}_{r}\frac{\Delta {q}}{\hbar}}\times \n &&
\hspace{-2 cm} \delta(\overline{{Y}}-{y}_{\Phi})\delta (\Delta Y) f^{*}_{i}(\overline{{p}}_{r} + \frac{\Delta {q}}{2})
f_{i}(\overline{{p}}_{r} - \frac{\Delta {q}}{2}) \nonumber \\
&&\hspace{-2 cm} =\frac{\delta(\overline{{Y}}-{y}_{\Phi})}{(2\pi)^2\hbar}
\int d\Delta {q}\, \psi^{\star}_{i}(\overline{{p}}_{r} + \frac{\Delta {q}}{2})
\psi_{i}(\overline{q} - \frac{\Delta {q}}{2})\n &&\hspace{0.5cm}\times e^{-i{x}_{r}\frac{\Delta {q}}{\hbar}} 
\eea
where we have employed $\delta(\overline{{Y}} +\frac{\Delta {Y}}{2}-{y}_{\Phi})\delta(\overline{{Y}}-\frac{\Delta {Y}}{2}-{y}_{\phi})=\delta(\overline{{Y}}-{y}_{\phi})\delta (\Delta {Y})$ and eq. \ref{eq:wfnon}.
We realize that 
 $\frac{1}{2\pi\hbar}\int d\Delta {q}\psi^{\star}_{i}(\overline{{p}}_{r} + \frac{\Delta {q}}{2})\psi_{i}(\overline{{p}}_{r} - \frac{\Delta {q}}{2})e^{-i{x}_{r}\frac{\Delta {q}}{\hbar} }$ is nothing than the non relativistic Wigner function $ W_{i,{\rm NR}}$ expressed in the coordinates of the center of mass. So we can finally establish 
\bea
W_{i,y_{\Phi}}(\overline{{Y}},{k_3};\overline{q},{x}_{r})&=&\frac{\delta(\overline{Y}-{y}_{\Phi})}{2\pi}
W_{i,{\rm NR}}(\overline{q},{x}_{r})
\label{eq:boostWigner}
\eea
The equation for $ W_ {i, y_{\Phi}} $ does not depend on the boost operator eigenvalue $k_3$, whose meaning in physical terms is that the states used in the definition of $\bar{\rho}_{i,y_{\Phi}}$ are plane waves wrt the center of mass motion and come with a single rapidity, leading to a $\delta(\Delta Y)$ in the Wigner calculation.   
$W_{i,y_{\Phi}}$ depends on the rapidity of the center of mass of the quarkonia state in the computational frame. Equation (\ref{eq:boostWigner}) describes how we can evaluate the Wigner function for a
$Q\bar Q$ pair in the bound state $i$ with $y_{\Phi}$.  It is also important to  realize that the boosted Wigner function (\ref{eq:boostWigner}) inherits some properties of the non relativistic Wigner function $W_{\rm NR}$, including the normalization. \\ 
Substituting equation (\ref{eq:boostWigner}) in equation (\ref{eq:dNdy1DfromWigner}), we obtain the rapidity distribution of the $Q\bar Q$ pairs which are bound in a state $i$:
\bea
\frac{dN_{i}}{dy_{\Phi}}&=&(2\pi)^2\hbar \int d\overline{Y}d\overline{q}dk_3 dx_{r}W(\overline{{Y}},k_3;\overline{{q}},x_{r})W_{i,y_{\Phi}}(\overline{{Y}},\overline{{q}},{x}_{r})\n
&=&2\pi\hbar \int d\overline{Y}\delta(\overline{{Y}}-y_{\Phi}) \int d\overline{q}dx_{r}W_{i,{\rm NR}}(\overline{{q}},{x}_{r})\n &&\times \underbrace{\int dk_3 W(\overline{{Y}},{k_3};\overline{{q}},{x}_{r})}_{\bar{W}(\overline{{Y}},\overline{{q}},{x}_{r})}
\n
&=& 
2\pi\hbar \int d\overline{q}dx_{r}W_{i,{\rm NR}}(\overline{{q}},{x}_{r})  \overline{W}(y_\Phi,\overline{{q}},{x}_{r})
\label{eq:coalescen1Dform}
\eea
In equation (\ref{eq:coalescen1Dform}) the Wigner function $W_{i,\rm NR}(\overline{{q}},{x}_{r})$ represents the probability density (non-relativistic) of formation of a quarkonium state $i$. $\int dk_3 W(\overline{{Y}},{k_3};\overline{{q}},{x}_{r})$ represents the probability density of finding a $Q\bar Q$ pair  with relative momentum and position $q$ and $x_{r}$ and with a center of mass rapidity $y_\Phi$. Since $W$ references to free $Q$ and $\bar Q$ before they form a bound state, the relationship with the two-particle probability density operator $\rho_{Q\bar Q}$, can be traced back by using the Wigner function definition
\bea
\overline{W}(\overline{Y},{q},{x}_{r})&=&\int dk_3 W(\overline{Y},{k_3};\overline{{q}},{x}_{r})\\ &&\hspace{-1cm}=\frac{1}{(2\pi)^2\hbar}\int dk_3 \int d\Delta Y d\Delta {q} e^{-i{k_3}\Delta {Y} - i{x}_{r}\frac{\Delta {q}}{\hbar}}\n && \hspace{-1cm}\times 
\overline{\rho}(\overline{{Y}} + \frac{\Delta {Y}}{2},\overline{q} + \frac{\Delta {q}}{2};\overline{{Y}} - \frac{\Delta {Y}}{2},\overline{{q}} - \frac{\Delta {q}}{2}) \n 
&&\hspace{-1cm} = \frac{1}{2\pi\hbar}\int d\Delta {q}\overline{\rho}(\overline{{Y}},\overline{{q}} + \frac{\Delta {q}}{2};\overline{{Y}},\overline{{q}}-\frac{\Delta {q}}{2})e^{ - i{x}_{r}\frac{\Delta {q}}{\hbar}}\nonumber. 
\eea
The integral of $\overline{W}$ over $x_{r}$ has the form : 
\begin{equation}
\int dx_{r}\overline{W}(\overline{{Y}},\overline{q},{x}_{r})= \overline{\rho}(\overline{{Y}},\overline{q};\overline{{Y}},\overline{q})=\frac{\langle \overline{{Y}},\overline{q}\vert \overline{\rho}_{Q\bar Q}\vert \overline{{Y}},\overline{q} \rangle}{\sqrt{s}}    
\end{equation}
which satisfies the following relationship:
\bea
&\int& d{q}d \overline{{Y}} \int dx_{r}\overline{W}(\overline{{Y}},{q},{x}_{r}) =\int \frac{d\overline{Y} d{q}}{\sqrt{s}}\rho_{Q\bar Q}(\overline{{Y}} ,{q})\n
&=&\int
 \frac{d^{3}p_{1}d^{3}p_{2}}{2e_{1}e_{2}}\rho_{Q\bar Q}({p}_{1},{p}_{2}) =N_{Q\bar Q}
\label{eq:Wignerbar}
\eea  
where $N_{Q\bar Q}=N_{Q}\times N_{\bar Q}$, is the total number of different $Q\bar Q$ pairs that are present in the system. One can thus write the following relationship for  $\overline{W}(\overline{{Y}},\overline{q},{x}_{r})$ 
\begin{equation}
\overline{W}(\overline{{Y}},\overline{q},{x}_{r})=\frac{d^{3}N_{Q\bar Q}}{d\overline{{Y}} d\overline{q}dx_{r}}
\end{equation}
\\
which is simply the differential form of equation (\ref{eq:Wignerbar}). 
The rapidity distribution, eq. \ref{eq:coalescen1Dform} can now be expressed as
\begin{equation}
\frac{dN_{i}}{dy_{\Phi}}=2\pi\hbar \int d\overline{q}dx_{r}\frac{d^{3}N_{Q\bar Q}}{dy_\Phi d\overline{q}dx_{r}}W_{i,\rm NR}(\overline{q},{x}_{r})
\label{eq:spectrapidty}
\end{equation}  

\section{Generalizing for the (3+1)D Case}
  The results derived in equation (\ref{eq:spectrapidty}) can be generalized to the case (3 +1)D if we start from a state generated according to equation (\ref{eq:QQstates})
\begin{equation}
\vert \Phi_{i,\mathbf{u}_{\Phi}} \rangle = \int \frac{d^3u}{u^0}\frac{d^3q^{\rm cm}}{\sqrt{s}} f_{i,\mathbf{u}_{\Phi}}(\mathbf{u},\mathbf{q}^{\rm cm})\vert \mathbf{p}_{1},\mathbf{p}_{2} \rangle
\end{equation}
where $(u^{0},\mathbf{u})$ is the 4-velocity of the quarkonium state $\Phi$ and $\mathbf{q}^{\rm cm}$ is the relative momentum evaluated in the quarkonium state CM (we use an explicit "cm" superscript anticipating a similar formulation in the lab-frame). It is also possible to express these quarkonium states in the basis $(\mathbf{u_T},Y)$, if we separate  $\mathbf{u}$ into its transverse $\mathbf{u}_{T}$ and longitudinal $u_{z}$ components, where $Y={\rm atanh}(\frac{u_{z}}{u^{0}})$, leading to
the analogous of equation (\ref{eq:state_relat_def}):
\bea
\vert \Phi_{i,y_{\Phi},{\bf u}_{T,\phi}}\rangle 
&=&\int \frac{d^3 q^{\rm cm}}{s}dY d^2{u}_{T}\delta(y_{\Phi}-Y)
\delta({\bf u}_{T\phi}-{\bf u}_{T}) \n
&&  f_{i}(q^{\rm cm}) \vert p_{1}\rangle   \vert p_{2}\rangle   
\eea
where ${\bf u}_{T,\phi}$ is the transverse component of the quarkonium 4-velocity. One can then generalize equation (\ref{eq:spectrapidty}) by extending this formula to the transverse component (3D case), we arrive at 
\bea
&&\frac{d^3N_{i}}{dy_{\Phi}d^2u_{T,\Phi}}= (2\pi)^6(\hbar)^{3}  \int dY d^2u_{T}d^3q^{\rm cm}d^3x^{\rm cm}_{r}\n  &\times& \frac{d^9N_{Q\bar Q}}{dY d^2u_{T}d^3q^{\rm cm} d^3x^{\rm cm}_{r}}
W_{i,y_{\Phi},\mathbf{u}_{T,\Phi}}(Y,\mathbf{u}_{T},\mathbf{q}^{\rm cm},\mathbf{x}^{\rm cm}_{r})\n
\eea  
where 
\bea
&&W_{i,y_{\Phi},\mathbf{u}_{T,\Phi}}(Y,\mathbf{u}_{T},\mathbf{q}^{\rm cm},\mathbf{x}^{\rm cm}_{r})
\label{eq:boosterWigner3D}\\ &=& \frac{1}{(2\pi)^3}\delta(y-y_{\Phi})\delta^{(2)}(\mathbf{u}_{T,\Phi}-\mathbf{u}_{T})
W_{\rm NR}(\mathbf{q}^{\rm cm},\mathbf{x}^{\rm cm}_{r})\nonumber.
\eea
Equation (\ref{eq:boosterWigner3D}) is the 3D generalization of equation (\ref{eq:boostWigner}). Inspection shows that we have just to multiply by the factor $\frac{\delta^{(2)}(\mathbf{u}_{T,\Phi}-\mathbf{u}_{T})}{(2\pi)^{2}}$. 
After performing the integrals over the delta distributions we obtain:
\bea
\frac{d^3N_{i}}{dy_{\Phi}d^2u_{T,\Phi}}&=&(2\pi\hbar)^3\int d^3q^{\rm cm}d^3x^{\rm cm}_{r}\frac{d^9N_{Q\bar Q}}{dYd^2u_{T}d^3q^{\rm cm}d^3x^{\rm cm}_{r}}\n &\times& W_{i,\rm NR}(\mathbf{q}^{\rm cm},\mathbf{x}^{\rm cm}_{r})
\label{eqntransversevelocity}
\eea
in which the nine-fold distribution in the integral is taken at $Y=y_\Phi$ and ${\bf u}_T={\bf u}_{T,\Phi}$.
This expression can be also expressed in the coordinates of the computational frame
\bea
\frac{d^3N_{i}}{dy_{\Phi}d^2u_{T,\Phi}}&=& h^3 \int d^3q^{\rm lab}d^3x^{\rm lab}_{r}\frac{d^9N_{Q\bar Q}}{dY d^2u_{T}d^3q^{\rm lab}d^3x^{\rm lab}_{r}}\n &\times&
W_{i,\rm NR}(\mathbf{q}^{\rm cm}(\mathbf{q}^{\rm lab}),\mathbf{x}^{\rm cm}_{r}(\mathbf{x}^{\rm lab}_{r}))
\label{eq:densitylabcm}
\eea
in which the coordinates $\mathbf{q}^{\rm cm}$ and $\mathbf{x}^{\rm cm}_{r}$ have to be expressed as a function of the  $\mathbf{q}^{\rm lab}$ and  $\mathbf{x}^{\rm lab}_{r}$.
From the previous equation, integrating by the variables $u_{T,\Phi}$ and $y_{\Phi}$ we can obtain the absolute number of states of quarkonium $i$-th
\bea
N_{i}&=&h^3\int dy_{\Phi}d^{2}u_{T}d^{3}q^{\rm lab}d^{3}x^{\rm lab}_{r}\frac{d^9N_{Q\bar Q}}{dY d^2u_{T}d^3q^{\rm lab}d^3x^{\rm lab}_{r}}\n &\times&
W_{i,\rm NR}(\mathbf{q}^{\rm cm}(\mathbf{q}^{\rm lab}),\mathbf{x}^{\rm cm}_{r}(\mathbf{x}^{\rm lab}_{r})) 
\eea
\\
The nine-fold distribution has to be considered according to the physical situation\ldots In nucleus-nucleus collisions, we take all possible $(Q,\bar Q)$ combinations into account in order to form $i$-th quarkonium state. In Monte Carlo Implementation, equation
\ref{eq:densitylabcm} becomes
\bea
\frac{dN_{i}}{d y_\Phi d^2 u_{T,\Phi}} &=& h^3 \sum^{N_{Q}\times N_{\bar Q}}_{l=1} W_{i,y_{\Phi},\mathbf{u}_{T,\Phi}}(Y,\mathbf{u}_{T},\mathbf{q},\mathbf{x}_{r})
\label{eq:densitylabcmMC} \\ 
&&\hspace{-1cm}= h^3 \sum^{N_{Q}\times N_{\bar Q}}_{l=1} \delta(Y-y_{\Phi})
\delta(\mathbf{u}_{T,\Phi}-\mathbf{u}_{T})
W_{i,\rm NR}(\mathbf{q}^{\rm cm},\mathbf{x}^{\rm cm}_{r})\nonumber
\eea
where the sum runs over all $N_{Q}\times N_{\bar Q}$ possible combinations and where the $\{\mathbf{q}^{\rm cm},\mathbf{x}^{\rm cm}_{r},\ldots\}$ are constructed for each pair. So the total number of states $ \Phi_{i} $ is given by the expression
\begin{equation}
N_{i}= h^3 \sum^{N_{Q}\times N_{\bar Q}}_{l=1} W_{i,\rm NR}(\mathbf{q}^{\rm cm},\mathbf{x}^{\rm cm}_{r}).
\end{equation}
The Wigner distribution function $W_{i,\rm NR}$ has to be multiplied by the factor $g=\frac{2s_{\Psi}+1}{8(2s_{Q}+1)^2}$ to take into account the spin of the quarkonium state and of the quarks. The factor 1/8 is due to the fact that in our approach only color neutral $Q\bar Q$ combinations can form a quarkonium.